\documentstyle[12pt,a4,epsf]{article}

\begin{document}

\makeatletter
\renewcommand{\theequation}{\thesection.\arabic{equation}}
\@addtoreset{equation}{section}
\makeatother

\begin{titlepage}
\title{
\vspace*{-10ex}
\hfill
\parbox{40mm}
{\normalsize UT-816\\KUCP-0113\\OHU-9808\\hep-th/9808034}\\
\vspace{1cm}
{\Large \bf Valley Views:\\
\Large \bf Instantons, Large Order Behaviors, and Supersymmetry}}
\author{Hideaki {\sc Aoyama}$^{\dagger}$,
Hisashi {\sc Kikuchi}$^{\dagger\dagger}$, 
Ikuo {\sc Okouchi}$^{\ddagger}$,
\\
Masatoshi {\sc Sato}$^{\star}$ 
and Shinya {\sc Wada}$^{\ddagger}$
\vspace{0.5cm}
\\
$^{\dagger}${\normalsize\sl Faculty of Integrated Human Studies,}\\
{\normalsize\sl Kyoto University, Kyoto 606-8501, Japan}\\
{\normalsize\tt aoyama@phys.h.kyoto-u.ac.jp}
\vspace{0.2cm}\\
$^{\dagger\dagger}${\normalsize\sl Ohu University, Koriyama 963-8611, 
Japan}\\
{\normalsize\tt kikuchi@yukawa.kyoto-u.ac.jp}
\vspace{0.2cm}\\
$^{\ddagger}${\normalsize\sl Graduate School of Human and Environmental 
Studies,}\\
{\normalsize\sl Kyoto University, Kyoto 606-8501, Japan}\\
{\normalsize\tt dai@phys.h.kyoto-u.ac.jp, shinya@phys.h.kyoto-u.ac.jp}
\vspace{0.2cm}\\
$^{\star}${\normalsize\sl Department of Physics, 
University of Tokyo, Tokyo 113-0033, Japan}\\
{\normalsize\tt msato@hep-th.phys.s.u-tokyo.ac.jp}}
\date{\small August, 1998}

\setcounter{page}{1} 
\maketitle
\thispagestyle{empty}
\begin{abstract} 
The elucidation of the properties of the
instantons in the topologically trivial sector
has been a long-standing puzzle.
Here we claim that the properties can be summarized in terms of the
geometrical structure in the configuration space, the valley.      
The evidence for this claim is presented in various ways. 
The conventional perturbation theory and the non-perturbative
calculation are unified, and the ambiguity of the Borel transform of the
perturbation series is removed.
A `proof' of Bogomolny's ``trick'' is presented, 
which enables us to go beyond the dilute-gas approximation.
The prediction of the large order behavior of the perturbation theory is
confirmed by explicit calculations, in some cases to the 
478-th order. 
A new type of supersymmetry is found as a by-product, and our result is 
shown to be consistent with the non-renormalization theorem. 
The prediction of the energy levels is confirmed with numerical
solutions of the Schr\"{o}dinger equation.
\end{abstract}
\end{titlepage}

\newpage
\setcounter{page}{1}
\tableofcontents
\newpage

\section{Introduction} 

Elucidation of the 
properties of the instantons in the topologically trivial sector
has been a long-standing puzzle.
Although there is no doubt as to the physical importance of the instantons,
it is not trivial to define them in this sector.
When they are separated by infinite distance they satisfy the classical
equation.
But once the distance between them becomes finite
there are no classical solutions. 
Due to the attractive force between the instanton and the anti-instanton
they can easily collapse to a vacuum.
Thus, it is impossible to distinguish them from the
fluctuations around the vacuum. 
Of course, no pragmatic problem arises in the dilute-gas
approximation, but the lack of a precise definition has prevented us
from going beyond this approximation.  
The main purpose of the present article is to clarify the structure of the
topologically trivial sector in theories with tunneling,
and to provide a method which goes beyond the dilute-gas approximation.

From the early stages of the study of the instanton \cite{CDG}, 
it was evident that this difficulty is connected with the fact that the
perturbation theory in the presence of tunneling is not Borel summable
\cite{folk,BPZtwo,review}. 
Both problems come from the non-separability of the tunneling and the
perturbative effects. 
And if we change the sign of the square of the coupling constant so that
the force between instantons becomes repulsive, they disappear; the
instanton configuration
is now free from the collapse and at the same time the perturbation series
becomes Borel summable.  
Bogomolny \cite{Bog} was the first to point out that the above
``analytic continuation'' of the coupling constant is the key to going
beyond the dilute-gas approximation,
and our work gives a precise realization of his suggestion
in the context of the structures in the configuration space.

In what follows, we will present novel structures for the configuration space
in the topologically trivial sector.
We will perform an explicit analysis in one-dimensional quantum
mechanics, with asymmetric double-well potential, but
our analysis may apply to quantum field theories with weak coupling, 
though additional complications will be introduced in such cases. 

Our starting point is a geometrical definition of the instantons
in the topologically trivial sector.
As was indicated by Balitsky and Yung \cite{BY2}, the instantons in
the topologically trivial sector form valleys in the configuration space. 
In Section 2, following Refs.\cite{AK,HS,AKHSW,AKHOSW}, 
we review a method to define
the valleys in the configuration space and present the construction
of the valley configurations and their basic constituents, the 
valley-instantons. 
Section 3 is the core of this paper; 
we will give a detailed analysis of a valley that contains a pair 
consisting of
an instanton and an anti-instanton, and show that 
the interplay between the perturbative effects and the tunneling effects
can be summarized in a unified way in terms of the collective 
coordinate of the valley. 
The summation over the multi-instantons is carried out in
Section 4, which leads to the expression for the
non-perturbative part of the energy spectrum.
In Sections 5 and 7, we summarize various tests of our results.
In Section 5, the prediction of the large order behavior is checked
against numerical and exact calculation of the perturbative series
from the 200-th to the 478-th order for a wide range of the 
parameters of our model.
This result shows an apparent lack of
Borel singularities at certain values of the parameters,
which suggests the existence of some non-renormalization theorem.
In Section 6, we show that the model we analyze has a new type of
supersymmetry, which we dub  ``${\cal N}$-fold supersymmetry".
The non-renormalization theorem makes it possible to test the prediction of
the energy spectrum. 
These predictions are checked against the numerical values of 
the energy spectrum in Section 7.
We conclude with some additional remarks in Section 8.
The four appendices contain some of the supporting calculations.

\section{The valley\label{valcon;sec}} 

\begin{figure}
\centerline{\epsfxsize=7cm\epsfbox{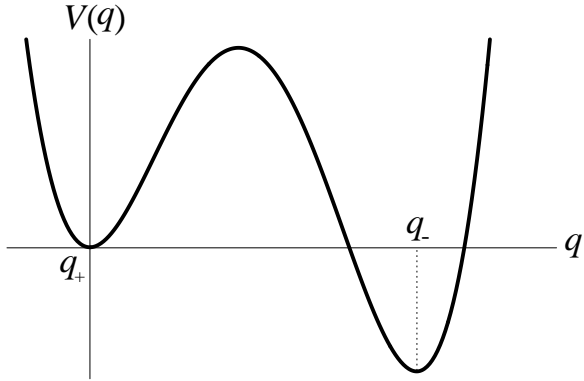}}
\caption{The asymmetric double-well potential $V(q)$ 
defined by Eq.(\protect{\ref{eqn:pot}}) for $\epsilon=5$
and $g=0.1$.}
\label{fig:pot}
\end{figure}

As our object of study we will take
a one-dimensional quantum mechanical system with
an asymmetric double-well potential.
We denote the coordinate by $q$,
and write the Euclidean action of this system as follows:
\begin{eqnarray}
S[q]&=& \int d\tau \left[
\frac{1}{2}\, \left(\frac{d q}{d \tau}\right)^2+V(q)\right],
\label{eqn:action}\\
V(q)&=&\frac{1}{2}\, q^2(1-g q)^2-\epsilon g q.
\label{eqn:pot}
\end{eqnarray}
This potential is plotted in Fig.\ref{fig:pot}
for  $\epsilon=5$ and $g=0.1$.
For small $\epsilon g^2$, the points $q \approx 0$ and $q \approx 1/g$
are the local minima of the potential, where $V(q) \approx 0$ and 
$-\epsilon$, respectively.
Thus $\epsilon$ is essentially the difference between the energies
of the two minima.
The parameter $g$ plays the role of the coupling constant, 
since the $q^3$ and $q^4$-terms have coefficients $g$ and $g^2$,
respectively.
The height of the potential barrier is approximately
$V(1/2g) \approx 1/(32 g^2)$, and
throughout this paper we will work in the small coupling regime $g \ll 1$.
In general, the existence of the two minima is limited to the range
$\epsilon g^2 < \sqrt{3}/18$, to which we restrict ourselves.
We will denote the coordinate of the upper and lower local minima 
by $q_+$ and $q_-$, respectively (see Fig.\ref{fig:pot}).
For the sake of definiteness, we assume that $g > 0$ and
$\epsilon \ge 0$, which is convenient for calculation
and entails no loss of generality.

Note that other types of quartic potential for the asymmetric double-well type
can be made in the form (\ref{eqn:pot}) by a
suitable linear transformation on the coordinate $q$ and a scale change
of the Euclidean time $\tau$.
Therefore, with this form of the potential we are covering
a wide range of potentials.
Furthermore, as we will see later, this form is essential
for the argument discussing the supersymmetry.\footnote{In 
an earlier paper \cite{AKHSW}, we have used a $(4g^3q^3 - 3g^4q^4)$ term
in place of $gq$ in the $\epsilon$ term of the potential.
This was for the sake of the exact relations
$q_+=0$ and $q_-=1/g$.  But this choice obscures the
supersymmetry we present in Section 6 and is thus avoided.}

Let us first describe the qualitative features of the
valleys.  The actual construction and calculations will
be given in the subsequent sections.

For $\epsilon=0$ this model has
the instanton and anti-instanton solutions of this system, 
given by the following:
\begin{eqnarray}
q^{(I)}_0(\tau; \tau_{I}) &=&\frac{1}{g} \frac{1}{1 + e^{-(\tau -\tau_{I})}},
\\
q^{(\bar{I})}_0(\tau; \tau_{\bar{I}})&=&
\frac{1}{g} \frac{1}{1 + e^{(\tau -\tau_{\bar{I}})}}.
\end{eqnarray}
The parameter $\tau_I$ ($\tau_{\bar I}$) denotes the
positions of the (anti-)instanton.

In the topologically trivial sector, the simplest valley
in the functional space of $q(\tau)$ starts
from the vacuum $q(\tau)\equiv 0$.
On the outskirts of the valley, the configuration 
has an almost-flat region of $q \simeq 1/g$.
When the width $R$ of this region 
becomes infinity, the transient regions become the exact instanton and 
the exact anti-instanton solutions.
We will call this valley the $I \bar I$ valley.
Correspondingly, there is also an $\bar I I$ valley, 
which starts from $q(\tau)\equiv 1/g$
and develops to a pair of $\bar I$ and $I$.
For both valleys the action is a monotonically increasing function of $R$,
approaching the sum of the instanton action and the anti-instanton action
as $R\rightarrow \infty$.
There are more complicated valleys, which start either from 
$q(\tau)\equiv 0$ or $1/g$ and develop to
configurations with multiple $I$ and $\bar I$.
At the leading order (for small coupling), this part of the valley 
is relevant for the dilute-instanton-gas calculus.

For $\epsilon \ne 0$, the valley configurations were found to
behave in a qualitatively similar manner \cite{AKHSW,AKOSW}.
There are configurations that smoothly connect one
minimum to the other minimum, which we call
``valley-instanton'' and ``anti-valley-instanton'',
and configurations that are similar to the pairs of 
these valley-instantons and anti-valley-instantons
exist as solutions of the valley equation.
For a one-pair configuration with distance $R (\gg 1)$ 
(for preview, see Fig.\ref{fig:iibar1}),
its action behaves as $S^{(I \bar I)} \approx 
2S^{(I)} - \epsilon R -(2/g^2)e^{-R}$ (Fig.\ref{fig:iibar2}). 
The first term of this action comes from
the valley-instanton and the anti-valley-instanton, 
while the second term is the ``volume energy'' from the region between 
them, where $q(\tau)=q_-$ (and $V(q_-)=-\epsilon$), 
and the third term is the ``interaction" between
the instanton and the anti-instanton.
When $R$ decreases, this configuration reaches the ``bounce''
solution where the action peaks, and then reduces to $q(\tau)=q_+$ for $R=0$.
At the bounce point, the valley line 
corresponds to the direction of the negative
eigenvalue of the bounce solution.
We will also call this an $I \bar I$ valley.
There is also an $\bar I I$ valley (Fig.\ref{fig:ibari1}), 
in which $q(\tau) = q_-$ at
$R=0$, and the action $S^{(\bar I I)}(R)$ 
is a monotonically increasing function of $R$
for the whole range of $R = 0 \sim \infty$.  Specifically,
$S^{(\bar I I)}(R) \approx 2S^{(I)} + \epsilon R-(2/g^2)e^{-R}$ for $R \gg 1$
(Fig.\ref{fig:ibari2}). 

All these can be seen explicitly and in detail by solving the valley equation.
In the rest of this section, 
we briefly explain the valley method used,
and analytically construct the valley configurations.
Furthermore, quantities needed for later calculations,
such as Jacobians and determinants, are also obtained.

\subsection{The valley method}
\label{valm;sec}

\begin{figure}
\centerline{\epsfxsize=7cm\epsfbox{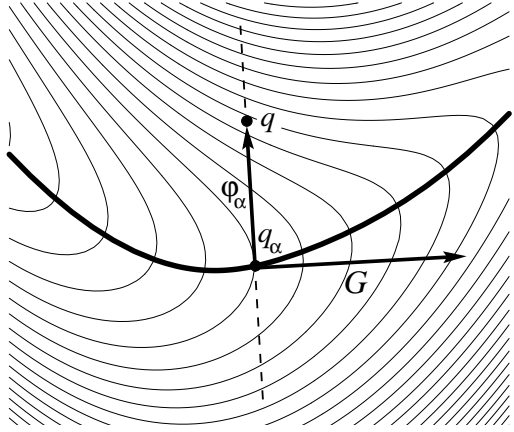}}
\caption{A schematic drawing of a valley in a two-dimensional
space. The thin curved lines are the contour lines of the action.
The thick line denotes the valley line.
The horizontal arrow denotes the gradient vector $G$ at a point
$q_\alpha$ and the broken vertical line is the 
integration subspace for $\varphi_\alpha$, which is perpendicular to $G$.
Note that the vector $G$ is not tangential to the valley line, which
is a general property of the valley equation.}
\label{fig:valpic}
\end{figure}

We use the following equation as 
a precise definition of the valley\cite{AK}:
\begin{eqnarray}
\int d\tau'\frac{\delta^2 S[q]}{\delta q(\tau)\delta q(\tau')}
\frac{\delta S[q]}{\delta q(\tau')}= 
\lambda\frac{\delta S[q]}{\delta q(\tau)},
\label{eqn:valley}
\end{eqnarray}
This equation can be rewritten in the following form:
\begin{equation}
\frac{\delta}{\delta q(\tau)} \left[ \frac{1}{2}\int 
\left(\frac{\delta S[q]}{\delta q(\tau')} \right)^2 d\tau'
-\lambda S[q] \right] =0. 
\end{equation}
This implies that a solution of the valley equation 
(\ref{eqn:valley}) extremizes the norm of the gradient vector 
$\int d\tau (\delta S[q]/\delta q)^2$ in the functional subspace 
of $q(\tau)$ with a fixed value of $S$, with $\lambda$
playing the role of the Lagrange multiplier.
This makes this equation suitable for a geometrical definition of the 
valley (see Fig.\ref{fig:valpic}).

More important is the fact that with this definition, the 
eigenvalue $\lambda$ of the second derivative of $S$
is removed from the loop integrations.
Initially, we separate the integration along the valley line
from the whole functional integration.
After this step, we are left with the
integrations over the directions perpendicular to the gradient ``vector", 
$\delta S / \delta q(\tau)$, which are the loop-integrations.
This is done in the following manner.

Firstly, we parameterize the valley by a parameter $\alpha$ and
denote the valley configuration by $q_\alpha(\tau)$,
and then take a functional integral,
\begin{equation}
Z = {\cal J}\int {\cal D}q \,e^{-S[q]},
\label{eqn:calz}
\end{equation}
where we are adopting the natural unit, $\hbar=c=1$.
(Inserting an operator ${\cal O}$ in the above does not change
the following argument.)
The normalization of the path-integral measure is the same as in
Ref.\cite{Coleman}.
We define the Faddeev-Popov determinant $\Delta[\varphi_\alpha]$,
which is actually the Jacobian for introducing $\alpha$ as an
integration variable, by the following:
\begin{equation}
\int d\alpha \, \delta \left(\int \varphi_\alpha (\tau) G(\tau)d\tau\right)
\Delta[\varphi_\alpha] =1,
\label{eqn:fpone}
\end{equation}
where $\varphi_{\alpha}(\tau) \equiv q(\tau)-q_\alpha(\tau)$ are the
fluctuations
over which we will be doing loop-integrations, and $G(\tau)$ is the 
normalized gradient vector,
\begin{equation}
G(\tau) = \frac{\delta S}{\delta q_\alpha(\tau)}\Biggl/\sqrt{\int
\left(\frac{\delta S}{\delta q_\alpha(\tau')} \right)^2 d\tau'}. 
\end{equation}
By inserting the factor (\ref{eqn:fpone}) into the functional integral
(\ref{eqn:calz}), we obtain the following:
\begin{equation}
Z = {\cal J} \int d\alpha \int {\cal D} q \,
\delta \left(\int \varphi_\alpha(\tau) G(\tau)d\tau\right)
\Delta[\varphi_\alpha] e^{-S[q]}.
\label{eqn:calz2}
\end{equation}
We expand the action $S[q]$  around $\varphi_\alpha(\tau)=0$:
\begin{equation}
S[q] = S[q_\alpha] + \int \frac{\delta S[q_\alpha]}{\delta q_\alpha(\tau)}
\varphi_\alpha(\tau) d\tau +
\frac{1}{2}\int \frac{\delta^2 S[q_\alpha]}
{\delta q_\alpha(\tau)\delta q_\alpha(\tau')}
\varphi_\alpha(\tau) \varphi_\alpha(\tau') d\tau d\tau' + \cdots .
\end{equation}
The first order term in $\varphi_\alpha(\tau)$ vanishes due to the
$\delta$-function.  The integration of the second order term
in  $\varphi_\alpha(\tau)$ yields the determinant of the
second functional derivative $D(\tau, \tau')\equiv
\delta^2 S[q_\alpha]/\delta q_\alpha(\tau) \delta q_\alpha(\tau')$,
{\it in the subspace defined by the $\delta$-function}.
At this order, the Jacobian $\Delta[\varphi_\alpha]$
is approximated as follows,
\begin{eqnarray}
\Delta[\varphi_\alpha] = 
\int \frac{d q_\alpha(\tau)}{d\alpha} G(\tau)d\tau 
= \frac{\displaystyle\frac{d S[q_\alpha]}{d\alpha}}
{\sqrt{\displaystyle\int d\tau \left(\frac{\delta S[q_\alpha]}
{\delta q(\tau)}\right)^2}}
(\equiv\Delta).
\label{eqn:jacob}
\end{eqnarray}
With this, the integral (\ref{eqn:calz2}) reduces to,
\begin{eqnarray}
Z = {\cal J} \int \frac{d\alpha}{\sqrt{2 \pi \det'D}}
\Delta \,e^{-S[q_\alpha]}.
\label{eq:nprime}
\end{eqnarray}
at the one-loop order.  In the above, $\det'$ denotes the determinant
in the subspace described above. 
Note the apparent reparametrization invariance
of $\alpha$, which should exist, as we have not specified the
choice of the valley parameter $\alpha$.
The factor $1/\sqrt{2\pi}$ is induced by the fact that
the $\delta$-function reduces the number of the integrations over 
$q(\tau)$ by one.

The valley equation (\ref{eqn:valley}) dictates that the 
subspace $\int \varphi_\alpha(\tau) G(\tau)d\tau =0$
does not contain the eigenvector
of the eigenvalue $\lambda$.  Therefore $\det' D$
is simply the product of all the eigenvalues of $D$
less $\lambda$.
This means that the valley equation (\ref{eqn:valley}) converts
the eigenvalue $\lambda$ completely to  
the collective coordinate $\alpha$.
This is the key property of the valley equation (\ref{eqn:valley}), 
since in actual
applications one encounters negative, zero, or very small eigenvalues,
which render the one-loop approximation useless, or at least dangerous.
By choosing $\lambda$ to be the undesired eigenvalue, we remove it 
completely from the Gaussian (and higher order) integration.

The valley equation can be made more perspicuous by
introducing an auxiliary coordinate.
First we will rewrite the valley equation (\ref{eqn:valley})
as follows:
\begin{equation}
\frac{\delta S_{\rm V}[q]}{\delta q(\tau)} =0,  
\end{equation}
where $S_{\rm V}$ is defined by,
\begin{equation}
S_{\rm V}[q] =  S[q] - \frac{1}{2\lambda}\int d\tau 
\left(\frac{\delta S[q]}{\delta q(\tau)} \right)^2. 
\end{equation}
The second term of the above $S_{\rm V}[q]$
contains a fourth derivative term of $q(\tau)$ with respect to
$\tau$. It can be removed by introducing an auxiliary variable 
$F(\tau)$ as follows:
\begin{eqnarray}
\bar S_{\rm V}[q] &=& S_{\rm V}[q]  + \frac{1}{2\lambda}\int d\tau 
\left(F(\tau) - \frac{\delta S[q]}{\delta q(\tau)} \right)^2
\nonumber\\
&=& S[q] +  \frac{1}{2\lambda}\int d\tau 
\left(F(\tau)^2 - 2F(\tau)\frac{\delta S[q]}{\delta q(\tau)} \right).
\label{eqn:bars}
\end{eqnarray}
Taking the functional derivative of $\bar S_{\rm V}[q]$ with
respect to $q(\tau)$ and $F(\tau)$ we obtain the following equations:
\begin{eqnarray}
\frac{\delta S[q]}{\delta q(\tau)} &=& F(\tau), 
\label{eqn:valleysf}\\
\int d\tau'\frac{\delta^2 S[q]}{\delta q(\tau)\delta q(\tau')}
F(\tau')&=&\lambda F(\tau),
\label{eqn:valleyf}
\end{eqnarray}
which are apparently equivalent 
to the original valley equation (\ref{eqn:valley}) upon elimination
of $F(\tau)$.
It is evident that any solution of the equation of motion
is also a solution of the valley equation (\ref{eqn:valleyf}) with
$F(\tau) \equiv 0$.
Some of the other important properties of the valley method are
noted in Appendix A.

Like any classical solution, 
the solution of the valley equation (\ref{eqn:valley}) may break some of
the symmetries of the system.  
To restore them, we must introduce additional collective coordinates.
The conventional collective coordinate method cannot be applied,
because, in contrast to the classical solution, the existence of the
zero modes of the symmetries is not guaranteed for the solution of the
valley equation. 
We will introduce them in the manner developed in Ref.\cite{HS}.
In order to achieve definiteness, we will assume 
that the translational symmetry is broken
by the valley configuration, which
is indeed the case in the model we treat in this paper.

Under translation $\tau_0$, the quantum fluctuation
$\varphi_{\alpha}(\tau)\equiv q(\tau)-q_{\alpha}(\tau)$ transforms as
\begin{eqnarray}
\varphi_\alpha(\tau)\rightarrow
\varphi^{\tau_0}_{\alpha}(\tau)&\equiv&q(\tau+\tau_0)-q_{\alpha}(\tau)
\nonumber\\
&=&\varphi_\alpha(\tau+\tau_0)+q_{\alpha}(\tau+\tau_0)-q_{\alpha}(\tau).
\label{gt;eqn}
\end{eqnarray}
This transformation induces an analogue of the gauge
transformation on $S[q_{\alpha}+\varphi_{\alpha}] $\cite{HK}.
The basic strategy of our collective coordinate method is to introduce
the translation $\tau_0$ into the path-integral as the ``gauge
orbit''. 
This procedure automatically guarantees the manifest invariance under the
translation through the following deformation, and does not require the
zero mode.

To extract the gauge orbit, we will use the
Faddeev-Popov technique.
In place of Eq.(\ref{eqn:fpone}),
we introduce the following Faddeev-Popov
determinant $\Delta[\varphi_{\alpha}]$: 
\begin{eqnarray}
\int d\tau_0 \int d\alpha\,
\delta\left(\int\varphi_{\alpha}^{\tau_0}(\tau)G(\tau)d\tau\right)\, 
\delta\left(\int\varphi_{\alpha}^{\tau_0}(\tau){\cal G}(\tau)d\tau\right)
\Delta[\varphi_{\alpha}]=1,
\label{fpone2;eqn}
\end{eqnarray}
where ${\cal G}(\tau)$ is a suitable
function independent from $\tau_0$, and gives the gauge fixing
condition of the translation:   
\begin{eqnarray}
\int \varphi_{\alpha}(\tau){\cal G}(\tau)d\tau=0.
\end{eqnarray}
By inserting the factor (\ref{fpone2;eqn}) into the functional integral,
we obtain
\begin{eqnarray}
Z&=&{\cal J}\int d\tau_{0} \int d\alpha \int {\cal D}q\,
\delta\left(\int\varphi_{\alpha}^{\tau_0}(\tau)G(\tau)d\tau\right)
\nonumber\\
&&\times
\delta\left(\int\varphi_{\alpha}^{\tau_0}(\tau){\cal G}(\tau)d\tau\right)
\Delta[\varphi_{\alpha}]e^{-S[q]}.
\end{eqnarray}
As well as the usual gauge symmetry, the following relations hold.
\begin{eqnarray}
\Delta[\varphi_{\alpha}]=\Delta[\varphi_{\alpha}^{\tau_0}],\quad 
S[q_{\alpha}+\varphi_{\alpha}^{\tau_0}]=S[q_{\alpha}+\varphi_{\alpha}].
\end{eqnarray}
Therefore, the integration of $\tau_0$ is trivially factored out.
\begin{eqnarray}
Z&=&{\cal J}\int d\tau_{0}\int d\alpha 
\int {\cal D}\varphi_{\alpha}\,
\delta\left(\int\varphi_{\alpha}(\tau)G(\tau)d\tau\right)
\nonumber\\
&&\times
\delta\left(\int\varphi_{\alpha}(\tau){\cal G}(\tau)d\tau\right)
\Delta[\varphi_{\alpha}]e^{-S[q_{\alpha}+\varphi_{\alpha}]}.
\end{eqnarray}
At the one-loop level, we find that the Faddeev-Popov determinant becomes
as follows:
\begin{eqnarray}
\Delta[\varphi_{\alpha}]&=&
\left|\det\left(
\begin{array}{cc}
\displaystyle
\int \frac{\partial q_{\alpha}(\tau)}{\partial \tau}{\cal G}(\tau)d\tau 
&
\displaystyle
\int \frac{\partial q_{\alpha}(\tau)}{\partial \alpha}{\cal G}(\tau)d\tau \\
\noalign{\vspace{2mm}}
\displaystyle
\int \frac{\partial q_{\alpha}(\tau)}{\partial \tau}G(\tau)d\tau
&
\displaystyle
\int \frac{\partial q_{\alpha}(\tau)}{\partial \alpha}G(\tau)d\tau
\end{array}
\right)\right|.
\end{eqnarray}
Because of the following equation
\begin{eqnarray}
\int \frac{\partial q_{\alpha}(\tau)}{\partial \tau}G(\tau)d\tau
=\int \frac{\partial q_{\alpha}(\tau)}{\partial \tau}
\frac{\delta S}{\delta q_{\alpha}(\tau)}d\tau
\left/\sqrt{\int\left(\frac{\delta S}{\delta
        q_{\alpha}(\tau')}\right)^2d\tau'}\right.=0,
\end{eqnarray}
it can be simplified as
\begin{eqnarray}
\Delta[\varphi_{\alpha}]
=\left|\int \frac{\partial q_{\alpha}(\tau)}{\partial
    \tau}G(\tau)d\tau \right|
\left|\int \frac{\partial q_{\alpha}(\tau)}{\partial \alpha}{\cal G}(\tau)
d\tau\right|. 
\label{eqn:delsimp}
\end{eqnarray}

Some of the valleys may contain classical solutions,
and in the neighborhood of such a solution, 
the valley has a quasi-zero mode which is different from $\lambda$
because of the broken symmetry. 
To remove this dangerous mode $\eta$ from the path-integral, we 
will designate ${\cal
G}(\tau)$ as the normalized eigenfunction of $\eta$:
\begin{eqnarray}
\int d\tau'\frac{\delta^2 S[q]}{\delta q(\tau)\delta q(\tau')}{\cal G}(\tau')
=\eta{\cal G}(\tau).
\end{eqnarray}
With this gauge fixing, we obtain
\begin{eqnarray}
Z={\cal J}\int\int \frac{d\tau_0d\alpha}{2\pi\sqrt{\det''D}}
\Delta[\varphi_{\alpha}]e^{-S[q_{\alpha}]},
\end{eqnarray}
where 
\begin{eqnarray}
{\rm det}''D=\frac{\det D}{\lambda\eta}.
\label{eqn:detppdsimp}
\end{eqnarray}

\subsection{Valley configurations}
For the action (\ref{eqn:action}), 
the valley equations (\ref{eqn:valleysf}) and (\ref{eqn:valleyf}) 
yield the following partial differential equations,
\begin{eqnarray}
-\partial_\tau^2 q + V'(q) &=& F 
\label{eqn:valleysf2}\\
\left( - \partial_\tau^2 + V^{\prime\prime}(q)\right) F &=& \lambda F.
\label{eqn:valleyf2}
\end{eqnarray}
From Eq.(\ref{eqn:valleysf2}) we see that the auxiliary coordinate $F$
acts as an external force introduced into the usual equation of motion
for $q$. Eq.(\ref{eqn:valleyf2}) is a self-consistent equation for $F$.

Since the terms in the action $\bar S_{\rm V}$ in (\ref{eqn:bars}) 
contain two time-derivatives at most, the canonical formalism can
be applied.
Denoting the ``Lagrangian'' of $\bar S_{\rm V}$ as $L_{\rm V}$, 
the canonical momentum $p_q$ and $p_F$ for the coordinates $q$ and $F$
are given as follows,
\begin{eqnarray}
p_q &=& \frac{\partial L_{\rm V}}{\partial \dot q} =
\dot q - \frac{1}{\lambda}\dot F, \\
p_F &=&  \frac{\partial L_{\rm V}}{\partial \dot F} =
- \frac{1}{\lambda}\dot q.
\end{eqnarray}
This leads to the following conserved ``Hamiltonian" of the system,
\begin{eqnarray}
H_V &=& p_q \dot q + p_F \dot F -L_{\rm V} \nonumber\\ 
&=& \frac{1}{2}\dot q^2 - V(q) 
- \frac{1}{\lambda}\left( \frac{1}{2}F^2
 + \dot F \dot q - F V'(q) \right).
\label{eqn:hami}
\end{eqnarray}
The conservation of this quantity can be directly confirmed by
the use of the valley equations (\ref{eqn:valleysf2}) and
(\ref{eqn:valleyf2}).

The valley Hamiltonian $H_V$ in (\ref{eqn:hami}) is useful
for numerical calculations of the valley configurations. 
Any $I \bar I$ or $\bar I I$ valley configuration
is symmetric around the middle point between the instantons.
When solving the differential equations (\ref{eqn:valleysf2}) and 
(\ref{eqn:valleyf2}) for a given $\lambda$,
we choose initial values at this symmetric point so that
the solutions converge as $\tau \rightarrow \infty$.
This shooting method is somewhat complicated in its operation
if the initial data is multi-dimensional.
At the symmetric point, say $\tau=0$, we have $\dot q (0)=\dot F(0)=0$. 
The convergence conditions, $q(\infty)=q_\pm, F(\infty)=0$, are 
to be satisfied choosing $q(0)$ and $F(0)$ as appropriate.  
Therefore at this stage the search has to be made in two-dimensional
space. The valley Hamiltonian (\ref{eqn:hami}) reduces the degrees of
freedom in the initial data by one, since $H_V=-V(q_{\pm})$ by the 
asymptotic condition for $\tau \rightarrow \infty$.
Therefore we only need to choose one initial datum, say $q(0)$,
to find the valley configuration.
Having only one parameter to determine, this task is straightforward.

The numerical calculations were carried out for
$g=0.1$, $\epsilon=5$, which corresponds to the potential 
in Fig.\ref{fig:pot}.
Fig.\ref{fig:iibar1} is the plot of $q_\alpha(\tau)$ and 
$F_\alpha(\tau)$ of the configurations in the $I \bar I$ valley.
The values of the action $S$ and the eigenvalue $\lambda$
for this valley are plotted in Fig.\ref{fig:iibar2}, where the
horizontal coordinate is $||q||\simeq |q_{-}-q_{+}|R$ for $R\gg 1$.
The configuration drawn by the broken line Fig.\ref{fig:iibar1} is
the bounce solution, which has $F(\tau)=0$.
The action peaks at this point in the valley as seen in 
Fig.\ref{fig:iibar2}.
The corresponding plots for the $\bar I I$ valley are given in 
Fig.\ref{fig:ibari1} and Fig.\ref{fig:ibari2}.
Corresponding to the fact that there are no solutions 
of the equation of motion with the boundary condition $q(\pm\infty)=q_-$, 
the action behaves monotonically, as seen in Fig.\ref{fig:ibari2}.
\begin{figure}
\centerline{\epsfxsize=7.0cm\epsfbox{qfconfig.eps}}
\caption{Configurations in $I \bar I$ valley.
The symmetric point is chosen to be at $\tau=0$, and only the half
of the configuration, $\tau > 0$, is shown.
The upper plot shows the behavior of $q(\tau)$,
where the broken line denotes the bounce solution.
The lower plot is $F(\tau)$. 
The labels $a$, $b$, and $c$ show
the correspondence between $q(\tau)$ and $F(\tau)$. }
\label{fig:iibar1}
\vspace{5mm}
\centerline{\epsfxsize=7.0cm\epsfbox{actlambda.eps}}
\caption{The action $S$ (upper plot) and the eigenvalue $\lambda$
(lower plot) of the $I \bar I$ valley.
The labels show the action of the corresponding solutions
in Fig.\protect{\ref{fig:iibar1}}.
The horizontal axis is a valley parameter defined as
$\alpha=||q||\equiv\int_0^\infty |q(\tau)-q_+| d\tau$.
For $||q||\gg 1$, $||q||\simeq |q_{-}-q_{+}|R$.
The peak of the action is given by the bounce solution 
shown in Fig.\protect{\ref{fig:iibar1}}.}
\label{fig:iibar2}
\end{figure}
\begin{figure}
\centerline{\epsfxsize=7.0cm\epsfbox{qfconfig2.eps}}
\caption{Configurations in $\bar I I$ valley.}
\label{fig:ibari1}
\vspace{5mm}
\centerline{\epsfxsize=7.0cm\epsfbox{actlambda2.eps}}
\caption{The action $S$ (upper plot) and the eigenvalue $\lambda$
(lower plot) of the $\bar I  I$ valley.
There are no solutions of the equation of motion under the
boundary condition $q(\pm\infty)=q_-$. Therefore, the action behaves
monotonically.}
\label{fig:ibari2}
\end{figure}

In Fig.\ref{fig:iibar1} and Fig.\ref{fig:ibari1}, 
we readily see that these valleys
are qualitatively similar to the valleys for the symmetric double-well
case $\epsilon=0$.
Particularly, for large distances $R \gg 1$,
the valley configurations 
are in fact made of flat inside regions of $q=q_\mp$ and 
flat outside regions where $q=q_\pm$.
The transient regions are of fixed shape, which simply moves
outward as $R$ increases (see the curves a,b,c of 
Figs.\ref{fig:iibar1},\ref{fig:ibari1}).
It is apparent that for $R \rightarrow \infty$, 
each transient configuration with boundary conditions
$q(-\infty)=q_\pm, q(\infty)=q_\mp$ becomes
an independent solution of the valley equation.
We call the left transient region of the $I \bar I$ valley,
where $q$ changes from $q_+$ to $q_-$, as valley-instanton, while the other
as the anti-valley-instanton, and vice versa for $\bar I I$ valley.
The behaviour of the eigenvalue $\lambda$ in Fig.\ref{fig:iibar2} and 
Fig.\ref{fig:ibari2} suggests that
these valley-instantons have an eigenvalue of exactly zero.
We note that this property is not trivial for $\epsilon\ne 0$,
since in that case the valley-instanton is not a solution 
of the equation of motion.
In fact, taking the time-derivative of Eq.(\ref{eqn:valleysf2}),
we obtain,
\begin{equation}
\left( - \partial_\tau^2 + V^{\prime\prime}(q)\right) \dot q = \dot F,
\label{eqn:notzero}
\end{equation}
which shows that $\dot q$ is {\it not} a zero mode.
This is as it should be, since simple translation of the
valley-instanton {\it does} change the action by the corresponding 
volume energy.

Eq.(\ref{eqn:notzero}), however, can be combined with
Eq.(\ref{eqn:valleyf2}) to show that $\lambda=0$. For this purpose, 
we multiply $\dot q$ on Eq.(\ref{eqn:valleyf2}),
integrate over $\tau$ from $-\infty$ to $\infty$, and 
carry out partial-integration
twice to move $\partial_\tau^2$ to act on $\dot q$.
At $\tau=\pm\infty$, $q$ of the valley-instanton 
is at the extrema $q_\pm$ of the 
potential. Therefore the external force $F$ goes to zero for $\tau 
\rightarrow \pm\infty$ (see Fig.\ref{fig:iibar1} and Fig.\ref{fig:ibari1}).  
The surface terms of the partial integrations vanish
due to this property.
We thus obtain the following:
\begin{equation}
\int_{-\infty}^\infty \dot F F d\tau =
\lambda \int_{-\infty}^\infty \dot q F d\tau.
\end{equation}
The left-hand side of this equation is evidently zero according to 
the boundary conditions.
On the right-hand side, the integral
$\int_{-\infty}^\infty \dot q F d\tau$ 
is not zero, as is seen in Fig.\ref{fig:iibar1} and in Fig.\ref{fig:ibari1}.
Therefore, for the valley-instanton we find that $\lambda=0$ and the
auxiliary coordinate $F$ is the zero mode. 
Note that this proof does not apply to the 
$I\bar I$ and $\bar I I$ configurations.
This is because for those configurations the integral
$\int_{-\infty}^\infty \dot q F d\tau$ 
is zero due to the fact that $q(\tau)$ and $F(\tau)$ are
even functions around the symmetric point.
And in fact, $\lambda$ is not identically zero for
the $I\bar I$ and $\bar I I$ valley.

Although the global features are evident from the numerical
results presented in this section, we need analytic expressions of
the action and other related quantities.
Of special importance are the interaction terms
of the actions of the $I \bar I$ and $\bar I I$ configurations.
In the following we carry out an analytic construction
for these configurations, using the fact that
the valley-instanton possesses an exact zero eigenvalue.

\subsection{Valley-instanton}
The existence of the zero eigenvalue $\lambda=0$ 
and the zero mode $F$ enables us to carry out 
the construction of the valley-instanton.
We will first carry out the analytic construction of the valley-instanton.
For this purpose it is convenient to use the rescaled coordinate 
$\tilde q = g q$.
The action (\ref{eqn:action}) is written in terms of this 
$\tilde q $ as follows:
\begin{equation}
S[q] = \frac{1}{g^2}\tilde S[\tilde q],\quad
\tilde S[\tilde q] = \int d\tau \left[
\frac{1}{2}\, \left(\frac{d \tilde q}{d \tau}\right)^2+ 
\tilde V(\tilde q)\right],
\end{equation}
\begin{equation}
\tilde V(\tilde q) = 
\tilde V_0(\tilde q) + \epsilon g^2 \tilde V_1(\tilde q),\quad
\tilde V_0(\tilde q) = \frac{1}{2}\tilde q^2(1-\tilde q)^2, \quad
\tilde V_1(\tilde q) = -\tilde q.
\label{eqn:rescaled}
\end{equation}
This implies that in the following expansion in $\epsilon$, 
the actual perturbation parameter is $\epsilon g^2$, 
which is natural considering that this is a dimensionless quantity.
We can also define a rescaled auxiliary coordinate $\tilde F = g F$
and rewrite the valley equations (\ref{eqn:valleysf2}) and (\ref{eqn:valleyf2})
as follows:
\begin{eqnarray}
-\partial_\tau^2 \tilde q + \tilde V'(\tilde q) &=& \tilde F 
\label{eqn:valleysf3}\\
\left( - \partial_\tau^2 + \tilde V^{\prime\prime}(\tilde q)\right) \tilde F 
&=& 0,
\label{eqn:valleyf3}
\end{eqnarray}
where we used the fact that $\lambda=0$.
(In the above the prime denotes the derivative with respect to $\tilde q$.)
Let us obtain the solution of this set of equations as
perturbative series in $\epsilon g^2$,
\begin{eqnarray}
\tilde q &=& \tilde q_0 + \epsilon g^2 \tilde q_1 + (\epsilon g^2)^2
\tilde q_2 + \cdots ,\\
\tilde F &=& \epsilon g^2 \tilde F_1 + (\epsilon g^2)^2 \tilde F_2 
+ \cdots .
\end{eqnarray}
As the zeroth order, we have only the following equation:
\begin{equation}
-\partial_\tau^2 \tilde q_0 + \tilde V_0'(\tilde q_0) =0,
\label{eqn:zeroth}
\end{equation}
whose solution is the (scaled) ordinary instanton 
$\tilde q_0 = gq^{(I)}_0(\tau; 0)$.
(We choose $\tau_I=0$ for the ease of notation in this analysis.
It may be reintroduced by replacing $\tau$ by $\tau - \tau_I$
at the end.)

At the first order in $\epsilon g^2$, Eqs.(\ref{eqn:valleysf3}) and
(\ref{eqn:valleyf3}) leads to the following:
\begin{eqnarray}
\left( -\partial_\tau^2 + \tilde V_0^{\prime\prime}(\tilde q_0) \right) 
\tilde q_1
&=& \tilde F_1 - \tilde V_1'(\tilde q_0),
\label{eqn:valleysf-p1}\\
\left(-\partial_\tau^2 + \tilde V_0^{\prime\prime}(\tilde q_0)\right) 
\tilde F_1 
&=& 0.
\label{eqn:valleyf-p1}
\end{eqnarray}
Eq.(\ref{eqn:valleyf-p1}) dictates that $\tilde F_1$ be proportional to
the zero mode of $\tilde q_0$, which is $\dot{\tilde q}_0$:
\begin{equation}
\tilde F_1 = c_1 \dot{\tilde q}_0.
\end{equation}
The proportionality constant $c_1$ can be fixed by the following
consideration:
For Eq.(\ref{eqn:valleysf-p1}) to have a solution, 
the zero mode of the operator 
$-\partial_\tau^2 + \tilde V_0^{\prime\prime}(\tilde q_0)$
should not be contained in the right-hand side.
Therefore the constant $c_1$ should be such that $\tilde F_1$ cancels 
the zero mode in $\tilde V_1'(\tilde q_0)$.
In fact, multiplying $\dot{\tilde q}_0$ on both sides of 
Eq.(\ref{eqn:valleysf-p1}), integrating over $\tau$ from
$-\infty$ to $\infty$ and performing partial integration twice,
we find that
\begin{eqnarray}
0 &=& \int_{-\infty}^\infty \dot{\tilde q}_0 \tilde F_1 d\tau 
- \int_{-\infty}^\infty \dot{\tilde q}_0 \tilde V_1'(\tilde q_0) d\tau 
\nonumber\\
&=& 
c_1 \int_0^1 \dot{\tilde q}_0 d\tilde q_0 
- \int_0^1 \tilde V_1'(\tilde q_0) d\tilde q_0 
= \frac{1}{6}c_1 +1,
\end{eqnarray}
which fixes $c_1$ as $c_1=-6$.
(It is useful to note
that $\dot{\tilde q}_0 = \tilde q_0 (1-\tilde q_0)$
in the calculation of the first term.)
The expression for $\tilde q_1$ is obtained by the same procedure
as above, but integrating only from $-\infty$ to $\tau$, which leads to
the following:
\begin{eqnarray}
\ddot{\tilde q}_0 {\tilde q}_1 -\dot{\tilde q}_0 \dot{\tilde q}_1
=-6\left( \frac{\tilde q_0^2}{2} - \frac{\tilde q_0^3}{3}\right) + \tilde q_0
= \tilde V_0'(\tilde q_0) = \ddot{\tilde q_0}.
\end{eqnarray}
The general solution of this equation is 
$\tilde q_1 = 1 + c_2 \dot{\tilde q}_0$,
where $c_2$ is an integration constant, which remains arbitrary at this
order.

At the second order in $\epsilon g^2$, we have the following:
\begin{eqnarray}
\left(-\partial_\tau^2 + \tilde V_0^{\prime\prime}(\tilde q_0) \right) 
\tilde q_2
&=& \tilde F_2 
- \frac{1}{2}\tilde V_0^{\prime\prime\prime}(\tilde q_0) \tilde q_1^2
- \tilde V_1^{\prime\prime}(\tilde q_0) \tilde q_1, 
\label{eqn:valleysf-p2}\\
\left(-\partial_\tau^2 + \tilde V_0^{\prime\prime}(\tilde q_0)\right) 
\tilde F_2 
&=& -\tilde V_0^{\prime\prime\prime}(\tilde q_0) \tilde q_1 \tilde F_1.
\label{eqn:valleyf-p2}
\end{eqnarray}
The parameter $c_2$ in $\tilde q_1$ is fixed by the condition that
zero mode be absent from the right-hand side of Eq.(\ref{eqn:valleyf-p2}).
Straightforward calculation shows that this leads to $c_2=0$.
Proceeding in a manner similar to the first-order calculation,
we find that $\tilde F_2$ satisfies the following equation:
\begin{equation}
\ddot{\tilde q}_0 \tilde F_2 - \dot{\tilde q}_0 \dot{\tilde F}_2
=-18 \dot{\tilde q}_0^2.
\end{equation}
The general solution of this equation is $\tilde F_2 =
(18\tau + c_3) \dot{\tilde q}_0$, where the constant $c_3$ is a free
parameter.
It is straightforward to show that 
the condition for the absence of the zero mode 
$\dot{\tilde q}_0$ on the right-hand side
of Eq.(\ref{eqn:valleysf-p2}) fixes the parameter $c_3$ as $c_3=0$, using 
the fact that 
the integral $\int_{-\infty}^\infty \dot{\tilde q}_0^2 \tau d\tau$ vanishes
since the integrand is an odd function of $\tau$.

Summarizing the perturbative solution obtained so far, we have
\begin{eqnarray}
\tilde q (\tau) &=& \tilde q_0 (\tau) + \epsilon g^2 + 
O\left((\epsilon g^2)^2\right),
\label{eqn:smalltau}
\\
\tilde F(\tau) &=& -6 \epsilon g^2 \dot{\tilde q}_0 (\tau)
+ 18 (\epsilon g^2)^2 \tau \dot{\tilde q}_0 (\tau)
+ O\left((\epsilon g^2)^3\right).
\label{eqn:smalltau2}
\end{eqnarray}
The $(\epsilon g^2)^2$ term in $\tilde F$ indicates that
these perturbative series are valid only for $|\tau| \ll 1/\epsilon g^2$.

For $\tau \rightarrow \pm\infty$, we may solve the linearized
valley equation,
\begin{eqnarray}
\left(-\partial_\tau^2 + \tilde V^{\prime\prime}(\tilde q_\pm) \right) 
\delta \tilde q
&=& \tilde F, 
\label{eqn:valleysf-asym}\\
\left(-\partial_\tau^2 + \tilde V^{\prime\prime}(\tilde q_\pm) \right) 
\tilde F 
&=& 0,
\label{eqn:valleyf-asym}
\end{eqnarray}
where $\delta \tilde q = \tilde q - \tilde q_\pm$,
The general solution of these equations are,
\begin{eqnarray}
\tilde q(\tau) &=& \tilde q_\mp + 
C_\mp e^{\mp\omega_\mp\tau} + \frac{F_\mp}{2 \omega_\mp}\tau 
e^{\mp\omega_\mp\tau},
\label{eqn:asymp}\\
\tilde F(\tau) &=& F_\mp e^{\mp\omega_\mp\tau},
\label{eqn:asymp2}
\end{eqnarray}
for $\tau \rightarrow \pm\infty$, respectively.
The coefficients $F_\pm, C_\pm$ are arbitrary constants and
$\omega_\pm \equiv \sqrt{\tilde V^{\prime\prime}(\tilde q_\pm)}$.

For small $\epsilon g^2$, 
$\tilde q_+ = \epsilon g^2 + O\left((\epsilon g^2)^2\right)$ and
$\tilde q_- = 1 + \epsilon g^2 + O\left((\epsilon g^2)^2\right)$.
Also, $\omega_\pm = 1 \mp 3 \epsilon g^2 + O\left((\epsilon g^2)^2\right)$,
which means that the linearized solution 
(\ref{eqn:asymp})--(\ref{eqn:asymp2}) is valid for $1 \ll |\tau|$.
Therefore, we find 
an overlapping region $1 \ll |\tau| \ll 1/\epsilon g^2$, where both
Eqs.(\ref{eqn:smalltau}), (\ref{eqn:smalltau2})
and Eqs.(\ref{eqn:asymp}), (\ref{eqn:asymp2}) are valid.
In this region the exponentials in the 
asymptotic solutions Eq.(\ref{eqn:asymp})
can be expanded in terms of $\epsilon g^2 \tau$ and be matched with
Eq.(\ref{eqn:smalltau})--(\ref{eqn:smalltau2}).
This matching indeed works out and fixes the constants $F_\pm, C_\pm$ as 
$F_\pm = -6\epsilon g^2$, $C_\pm = \pm 1$.

In summary of the analytic construction of the valley-instanton 
configuration, we have the following:
\begin{eqnarray}
q(\tau)=\left\{
\begin{array}{ll}
\displaystyle
\epsilon g+\frac{1}{g}e^{\omega_{+}\tau}
+\frac{3\epsilon g}{\omega_{+}}\tau e^{\omega_{+}\tau}+\cdots
& \mbox{if $\tau\ll-1$ ;}\\  
\noalign{\vspace{2mm}} \displaystyle
\frac{1}{g}\frac{1}{1+e^{-\tau}}+\epsilon g+\cdots 
& 
\mbox{if $-1/\epsilon g^2 \ll \tau\ll 1/\epsilon g^2 $ ;}\hspace*{-15mm}\\
\noalign{\vspace{2mm}} \displaystyle
\frac{1}{g}+\epsilon g
-\frac{1}{g}e^{-\omega_{-}\tau}-\frac{3\epsilon g}{\omega_{-}}\tau 
e^{-\omega_{-}\tau}+\cdots
& \mbox{if $\tau\gg 1$ ,}
\end{array}
\right.\nonumber\\
\label{eqn:approxq}
\end{eqnarray}
\begin{eqnarray}
F(\tau)=\left\{
\begin{array}{ll}
\displaystyle
-6 \epsilon ge^{\omega_{+}\tau}+\cdots
& \mbox{if $\tau\ll-1$ ;}\\  
\noalign{\vspace{2mm}} \displaystyle
-6 \epsilon g  \frac{1 - 3\epsilon g^2 \tau}{(1 + e^{-\tau})(1 + e^{\tau})}
+\cdots & \mbox{if $-1/\epsilon g^2 \ll \tau\ll 1/\epsilon g^2 $ ;}\\
\noalign{\vspace{2mm}} \displaystyle
-6 \epsilon g e^{-\omega_{-}\tau}+\cdots
& \mbox{if $\tau\gg 1$ .}
\end{array}
\right.\nonumber\\
\label{eqn:approxF}
\end{eqnarray}

From this solution we can evaluate the action $S$. 
First let us calculate the contribution of 
the solution (\ref{eqn:approxq}) for $|\tau| \ll 1/\epsilon g^2$.
\begin{equation}
S[q_0 + \epsilon g^2 q_1] =
S_0[q_0] + \epsilon g^2 S_1[q_0] 
+ \epsilon g^2 \int \frac{\delta S_0[q_0]}{\delta q_0(\tau)}
q_1(\tau) d\tau + \cdots,
\end{equation}
where we used unscaled quantities $q_i \equiv \tilde q_i /g$, etc.
And $S_1$ is the contribution of the $V_1$-term in
Eq.(\ref{eqn:rescaled}), while 
$S_0$ denotes the rest of the terms in the action.
We find that the second term vanishes.
The third term is also zero due to Eq.(\ref{eqn:zeroth}).
In the asymptotic region, $|\tau| \gg 1$, the linearized solutions should be
used, which simply yields the volume energy plus the higher order
terms in $\epsilon g^2$.
Therefore, the action of the configuration (\ref{eqn:approxq}) 
restricted in region $\tau \in [-T/2, T/2]$ ($T \gg 1$) is given by
$S \simeq (1/6g^2) - \epsilon T/2 + \cdots$.
The second term of this expression is the ``volume energy''
due to the energy difference between $q=q_+$ and $q_-$,
while the first term is the
leading contribution to the valley-instanton action $S^{(I),(\bar I)}$:
\begin{equation}
S^{(I),(\bar I)} = \frac{1}{6g^2} 
\left[ 1+ O\left( (\epsilon g^2)^2 \right)\right].
\end{equation}

We will now turn to the evaluation of the Jacobian (\ref{eqn:jacob}).
The numerator is the squared norm of $F(\tau)$,
the leading contribution to which comes from the integration 
near $\tau \sim 0$. Substituting the expression
(\ref{eqn:approxF}) in this region, we find that the
leading contribution is,
\begin{equation}
\int_{-\infty}^\infty F^2 d\tau = 6 \epsilon^2 g^2 
+ \cdots .
\end{equation}
Therefore, we find that when we choose the position coordinate of 
the valley-instanton as the valley parameter $\alpha$,
the Jacobian (\ref{eqn:jacob}) is given by the following:
\begin{equation}
\Delta = \frac{\epsilon}
{\sqrt{\displaystyle\int_{-\infty}^\infty F^2 d\tau}} =
\frac{1}{\sqrt{6g^2}}\left[ 1+ O( \epsilon g^2 )\right].
\label{eqn:delta}
\end{equation}
This leading term is the same as that of the ordinary instanton
for $\epsilon =0$.  

The factor $\det'D$ can also be evaluated by the use of the fact
that $\lambda=0$.
The derivation is given in Appendix B.
The result for the valley-instanton located at $\tau=\tau_I$ in
the range $\tau \in [-T/2, T/2]$ ($T \gg 1$) is the following
(see Eq.(\ref{eq:appa})):
\begin{equation}
\frac{\det'(-\partial_\tau^2 + V^{\prime\prime})}
{\det(-\partial_\tau^2 + \omega_+^2)}
=\frac{e^{(\omega_- - \omega_+)(T/2 - \tau_I)}}
{2\omega_-F_+F_-}
\int_{-\infty}^{\infty}d\tau F^2(\tau).
\label{rdetvi;eqn}
\end{equation}
Note that the exponential factor in the ratio of the determinant
gives the right quantity for the zero-point energy contribution. 
In fact, the normalization factor ${\cal J}$ of the functional integral 
(\ref{eq:nprime}) is given by the following:
\begin{equation}
\frac{{\cal J}}{\sqrt{\det(-\partial_\tau^2 + \omega_+^2)}}
= e^{-\omega_+T/2} \Upsilon_{+},
\label{vinormalization;eqn}
\end{equation}
where $\Upsilon_{+}$ is a constant dependent on the definition of $Z$.
For the partition function $Z=\lim_{T\rightarrow\infty}{\rm Tr}(e^{-HT})$, 
\begin{eqnarray}
\Upsilon_{+}=1,
\end{eqnarray}
and for the amplitude $Z=\lim_{T\rightarrow\infty}\langle
q_{+}|e^{-HT}|q_{+}\rangle$,  
\begin{eqnarray}
\Upsilon_{+}=|\Psi(0 ;\omega_{+})|^2,
\end{eqnarray}
where $\Psi(q;\omega)$ is the normalized Schr\"odinger wave function
of the ground state for the harmonic oscillator of frequency $\omega$
and therefore 
$\Psi(0;\omega) = (\omega/\pi)^{1/4}$. 
Thus we find that
\begin{equation}
\frac{{\cal J}}{\sqrt{\det'(-\partial_\tau^2 + V^{\prime\prime})}}
=\frac{\Upsilon_{+}}{\sqrt{\kappa^{(I)}}}\exp\left\{
-\frac{\omega_+}{2}\left(\frac{T}{2}+\tau_I\right)
-\frac{\omega_-}{2}\left(\frac{T}{2}-\tau_I\right)\right\},
\label{eqn:videt}
\end{equation}
where
\begin{equation}
\kappa^{(I)} \equiv \frac{1}{2 \omega_- F_+ F_-} 
\int_{-\infty}^\infty F^2 d\tau.
\end{equation}

For the anti-valley-instanton, $\det'D$ is evaluated in a similar
manner, but slightly modified due to the
difference of the boundary condition at $\tau=\pm\infty$.
Eqs.(\ref{rdetvi;eqn}), (\ref{vinormalization;eqn}) and
(\ref{eqn:videt}) become as follows, 
\begin{equation}
\frac{\det'(-\partial_\tau^2 + V^{\prime\prime})}
{\det(-\partial_\tau^2 + \omega_-^2)}
=\frac{e^{(\omega_+ - \omega_-)(T/2 - \tau_{\bar{I}})}}
{2\omega_+F_+F_-}
\int_{-\infty}^{\infty}d\tau F^2(\tau),
\label{rdetavi;eqn}
\end{equation}
\begin{equation}
\frac{{\cal J}}{\sqrt{\det(-\partial_\tau^2 + \omega_-^2)}}
= e^{-\omega_-T/2} \Upsilon_{-},
\label{avinormalization;eqn}
\end{equation}
\begin{equation}
\frac{{\cal J}}{\sqrt{\det'(-\partial_\tau^2 + V^{\prime\prime})}}
=\frac{\Upsilon_{-}}{\sqrt{\kappa^{(\bar{I})}}}\exp\left\{
-\frac{\omega_-}{2}\left(\frac{T}{2}+\tau_{\bar{I}}\right)
-\frac{\omega_+}{2}\left(\frac{T}{2}-\tau_{{\bar I}}\right)\right\}.
\label{eqn:avidet}
\end{equation}
Here
\begin{equation}
\kappa^{(\bar{I})} \equiv \frac{1}{2 \omega_+ F_+ F_-} 
\int_{-\infty}^\infty F^2 d\tau,
\end{equation}
and for $Z=\lim_{T\rightarrow\infty}{\rm Tr}(e^{-HT})$, $\Upsilon_{-}$ is
\begin{eqnarray}
\Upsilon_{-}=1,
\end{eqnarray}
and for $Z=\lim_{T\rightarrow\infty}\langle q_{-}|e^{-HT}|q_{-}\rangle$
\begin{eqnarray}
\Upsilon_{-}=|\Psi(0;\omega_-)|^2.
\end{eqnarray}

The results (\ref{eqn:videt}) and (\ref{eqn:avidet}) 
properly represent the zero-point energies $\omega_\pm /2$ at
$q=q_\pm$.
At the leading order in $\epsilon g^2$, we find that
$\omega_{+}=\omega_{-}=1$
and 
\begin{equation}
\kappa^{(I)}=\kappa^{(\bar{I})}
=\kappa\equiv\frac{1}{12},
\label{eqn:kappa}
\end{equation}
thus the difference between 
the results (\ref{eqn:videt}) and (\ref{eqn:avidet}) 
disappears.

\subsection{$I \bar I$ valley configurations with large separations}

The rescaled valley equations can be solved for
the  $I \bar I$ valley configurations
in much the same manner as above.
The calculation is especially straightforward for the leading terms
in the $\epsilon g^2$-expansion, which we will carry out 
in the following.\footnote{This interaction term 
coincides with the one obtained in Ref.\cite{Bog}.
However our derivation is based on the valley method.}

\begin{figure}
\centerline{\epsfxsize=8cm\epsfbox{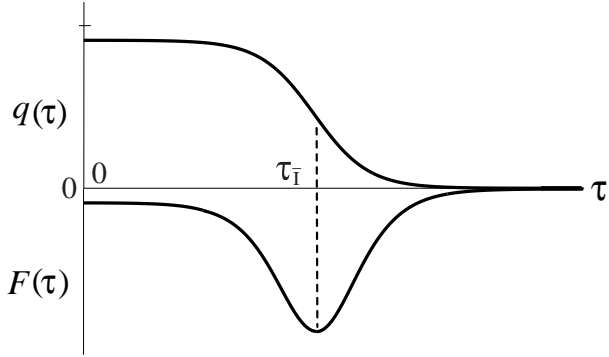}}
\caption{The $I \bar I$ valley configuration with a large separation. 
The symmetric point is chosen
to be at $\tau = 0$, so that $\tau_I=-\tau_{\bar I}$.}
\label{fig:iibareps}
\end{figure}

For this purpose, we will neglect the ${\tilde V}_1$ term
in Eq.(\ref{eqn:rescaled}).
Furthermore, we will confine ourselves to the outskirts of the
valley, where the eigenvalue $\lambda \ll 1$.
This allows us to solve the valley equation in a perturbation 
by $\lambda$:
\begin{eqnarray}
\tilde q(\tau) &=& \tilde q_0(\tau) + \lambda \tilde q_1(\tau)+ \cdots,
\\
\tilde F(\tau) &=& 
\lambda \tilde F_1(\tau)+ \lambda^2 \tilde F_2(\tau)+ \cdots.
\end{eqnarray}
We will nominate the symmetric point of the configuration 
as \(\tau = 0\), and solve the equation for the region $\tau \ge 0$ 
(see Fig.\ref{fig:iibareps}).
Therefore we choose 
 $\tilde q_0(\tau)$ as the ordinary anti-instanton solution at
$\tau=\tau_{\bar{I}} (\gg 1)$; 
$\tilde q_0 (\tau) = g q_0^{(\bar{I})}(\tau; \tau_{\bar{I}})$.
The parameter $\tau_{\bar{I}}$ will be related to $\lambda$ through the
boundary conditions,
\begin{eqnarray}
\dot{\tilde q}(0) &=\dot{\tilde F}(0)=0,
\label{eqn:bc1}
\\
\tilde q(\infty) &=\tilde F(\infty)=0,
\label{eqn:bc2}
\end{eqnarray}
for $\lambda \rightarrow 0$ as $\tau_{\bar{I}} \rightarrow \infty$.
The equations we obtain at each order of $\lambda$ are as
follows:
\begin{eqnarray}
O(\lambda^0): &&
-\partial_\tau^2 \tilde q_0+ \tilde V_0^\prime(\tilde q_0)= 0, 
\label{eqn:v1}\\
O(\lambda^1): &&
\left(-\partial_\tau^2 + \tilde V_0^{\prime\prime}(\tilde q_0)\right) 
\tilde q_1 = \tilde F_1,
\label{eqn:v2}\\
&&
\left(-\partial_\tau^2 + \tilde V_0^{\prime\prime}(\tilde q_0)\right) 
\tilde F_1 = 0,
\label{eqn:v3}\\
O(\lambda^2): &&
\left(-\partial_\tau^2 + \tilde V_0^{\prime\prime}(\tilde q_0)\right) 
\tilde F_2 
+\tilde V_0^{\prime\prime\prime}(\tilde q_0)\tilde q_1 \tilde F_1
= \tilde F_1.
\label{eqn:v4}
\end{eqnarray}
The first equation (\ref{eqn:v1}) is satisfied by the anti-instanton solution.
In the following we first solve Eq.(\ref{eqn:v3}) for $\tilde F_1$, 
then Eq.(\ref{eqn:v2}) for $\tilde q_1$,
and subsequently Eq.(\ref{eqn:v4}) for $\tilde F_2$, satisfying the 
boundary conditions (\ref{eqn:bc2}) at $\tau=\infty$.
At the conclusion of the process
we require the boundary conditions (\ref{eqn:bc1}) at
$\tau=0$, which read;
\begin{eqnarray}
\dot{\tilde q}_0(0) + \lambda \dot{\tilde q}_1(0) =0,
\label{eq:q12bc}\\
\dot{\tilde F}_1(0) + \lambda \dot{\tilde F}_2(0) =0.
\label{eq:f12bc}
\end{eqnarray}

The solution of Eq.(\ref{eqn:v3}) is simply given by the
translation mode of the anti-instanton 
$\dot{\tilde q}_0(\tau) (\equiv -\psi_1(\tau))$:
\begin{equation}
\tilde F_1(\tau) = c_1 \psi_1(\tau).
\end{equation}
This evidently satisfies $\tilde F_1(\infty)=0$.
The constant $c_1$ is to be determined by the boundary condition
at $\tau=0$ later.

In order to solve Eq.(\ref{eqn:v2}), it is convenient to define
a Green function,
\begin{equation}
\bar G(\tau, \tau') \equiv \frac12
\left( 
\psi_1(\tau)\psi_2(\tau')\theta(\tau-\tau')
+\psi_1(\tau')\psi_2(\tau)\theta(\tau'-\tau)
\right),
\end{equation}
where $\psi_2(\tau)$ is the other zero mode solution of 
Eq.(\ref{eqn:v3}), which is chosen so that
the Wronskian is,
\begin{equation}
W = \psi_1 \dot\psi_2 - \dot\psi_1 \psi_2 = 2,
\end{equation}
and therefore has the following asymptotic behaviour:
\begin{equation}
\psi_2(\tau) \simeq \left\{
\begin{array}{ll}
e^{\tau-\tau_{\bar{I}}} & \mbox{for $\tau-\tau_{\bar{I}} \gg 1$;}\\
-e^{-(\tau-\tau_{\bar{I}})} & \mbox{for $\tau-\tau_{\bar{I}} \ll -1$.}\\
\end{array}\right.
\end{equation}
The solution of Eq.(\ref{eqn:v2}) is then written as follows,
\begin{equation}
\tilde q_1(\tau) = \int_0^\infty d\tau'
\bar G(\tau, \tau')\tilde F_1(\tau').
\end{equation}
It is straightforward to show that $\tilde q_1(\infty)=0$. 
Further, 
\begin{equation}
\dot{\tilde q}_1(0) = \frac{c_1}{2}\dot\psi_2(0) \int_0^\infty
\psi_1(\tau)^2 d\tau \simeq 
\frac{c_1 e^{\tau_{\bar{I}}}}{12}.
\end{equation}
From this, we find that the boundary condition (\ref{eq:q12bc})
leads to,
\begin{equation}
c_1 = \frac{12}{\lambda g} e^{-2\tau_{\bar{I}}}.
\end{equation}

Eq.(\ref{eqn:v4}) can be solved as follows:
\begin{equation}
\tilde F_2 (\tau) =  \int_0^\infty d\tau'
\bar G(\tau, \tau') B(\tau')\tilde F_1(\tau'),
\end{equation}
where $B(\tau) \equiv (1 - 
\tilde V_0^{\prime\prime\prime}(\tilde q_0)\tilde q_1 )$.
Just as before, this satisfies $\tilde F_2 (\infty) =0$. 
At $\tau=0$, we find that
\begin{equation}
\dot{\tilde F}_2 (0) =  
\frac{c_1}{2}\dot\psi_2(0)B,
\end{equation}
where
\begin{equation}
B \equiv \int_0^\infty B(\tau)\psi_1^2(\tau) d\tau,
\end{equation}
is the quantity of $O(1)$.
From this, we find that the boundary condition (\ref{eq:f12bc})
leads to
\begin{equation}
\lambda = -\frac{2}{B} e^{-2\tau_{\bar{I}}}.
\label{lambda;eqn}
\end{equation}
The factor $B$ is evaluated to be $1/12$ in Appendix C.

Next we will evaluate the contribution of the region 
$\tau \in [0, \infty]$ to the action as follows:
\begin{eqnarray}
S[q] &=&
S[q_0] + \frac{\delta S[q_0]}{\delta q_0} (\lambda q_1 + \cdots)
+ \frac12 \frac{\delta^2 S[q_0]}{\delta q_0^2}
 (\lambda q_1 + \cdots)^2 + \cdots \nonumber\\
&=&
S[q_0] + \frac{\lambda^2}{2}\int_0^\infty q_1 F_1 d\tau 
-\frac{\lambda}{2}\dot q_0(0) q_1(0) + \cdots.
\end{eqnarray}
where we used the fact that 
the linear term in $\lambda q_1 + \cdots$ vanishes due to 
the equation of motion
and the last term is the sum of the surface terms.
The first term on the left side is given by the
following:
\begin{equation}
S[q_0] = \frac{1}{g^2}\left(\frac{1}{6} - \frac12 e^{-2\tau_{\bar{I}}}
\right).
\end{equation}
The second term is of order of $\lambda^2 \sim e^{-4\tau_{\bar{I}}}$,
and is of next-to-leading order.
The last term is evaluated as follows,
\begin{equation}
-\frac{\lambda}{2}\dot q_0(0) q_1(0)
= -\frac{\lambda}{2} \left(-\frac{1}{g}e^{-\tau_{\bar{I}}}\right)
\frac{c_1}{2}\psi_2(0)\int_0^\infty \psi_1(\tau)^2 d\tau
\simeq - \frac{e^{-2\tau_{\bar{I}}}}{2g^2}.
\end{equation}
Combining all contributions, we find that 
at the leading order of $\epsilon g^2$ ,
\begin{equation}
S^{(I \bar I)} = 2S^{(I)} - \epsilon R + S_{\rm int} ^{(I \bar I)},
\end{equation}
where $S_{\rm int} ^{(I \bar I)}$ is the interaction term,
\begin{equation}
S_{\rm int}^{(I \bar I)} = - \frac{2}{g^2} e^{- R}.
\label{eq:interact1}
\end{equation}
We have denoted the distance between $I$ and $\bar I$ by
$R=|\tau_I-\tau_{\bar{I}}|=2\tau_{\bar{I}}$. 
In the same way, the $\bar I I$ configuration with a large separation $R$
between the anti-valley-instanton and the valley-instanton 
is given by the following at the leading order of $\epsilon g^2$:
\begin{equation}
S^{(\bar I I)} = 2S^{(I)} + \epsilon R + S_{\rm int} ^{(\bar I I)},
\end{equation}
where $S_{\rm int} ^{(\bar I I)}$ is the interaction term,
\begin{equation}
S_{\rm int} ^{(\bar I I)} = - \frac{2}{g^2} e^{- R}.
\label{eq:interact2}
\end{equation}

Finally, let us evaluate $\Delta[\varphi_{R}]$ in Eq.(\ref{eqn:delsimp}) 
and $\det''D$ in Eq.(\ref{eqn:detppdsimp}).
(Here we denote $\varphi_\alpha$ as $\varphi_R$.)
We choose the eigenfunction of the second lowest eigenvalue of
$D(\tau,\tau')$ as the gauge fixing function ${\cal G}(\tau)$. 
This choice enables us to be  completely free from the difficulties that
come from the quasi zero mode.
The configuration on the $I\bar{I}$ valley constructed above is
\begin{equation}
q_{R}^{(I\bar{I})}(\tau)
\mathop{=}^{R\rightarrow\infty}_{\epsilon g^2\rightarrow 0}
\theta(-\tau)q^{(I)}_0(\tau;\tau_{I})
+\theta(\tau)q^{(\bar{I})}_0(\tau;\tau_{\bar I}),
\end{equation}
\begin{equation}
F^{(I\bar{I})}(\tau)
\mathop{=}^{R\rightarrow\infty}_{\epsilon g^2\rightarrow 0}
\theta(-\tau)\lambda\dot{q}^{(I)}_0(\tau;\tau_{I})
-\theta(\tau)\lambda\dot{q}^{(\bar{I})}_0(\tau;\tau_{\bar{I}}),
\end{equation}
where $\tau_{\bar{I}}=-\tau_{I}$ and $\lambda$ is given by
Eq.(\ref{lambda;eqn}).
Therefore, we find
\begin{eqnarray}
\int\frac{\partial q_{R}^{(I\bar{I})}(\tau)}{\partial R}G(\tau)d\tau
\mathop{=}^{R\rightarrow\infty}_{\epsilon g^2\rightarrow 0}
\frac{1}{\sqrt{12g^2}}.
\label{fp1;eqn}
\end{eqnarray}
The gauge fixing function behaves as follows:
\begin{eqnarray}\\
{\cal G(\tau)}
\mathop{=}^{R\rightarrow\infty}_{\epsilon g^2\rightarrow 0}
\frac{\theta(-\tau)\dot{q}^{(I)}_0(\tau;\tau_I)
+\theta(\tau)\dot{q}^{\bar{(I)}}_0(\tau;\tau_{\bar I})}
{\sqrt{\displaystyle\int(\theta(-\tau)\dot{q}^{(I)}_0(\tau;\tau_I)
+\theta(\tau)\dot{q} ^{(\bar{I})}_0(\tau;\tau_{\bar I}))^2 d\tau }}.
\end{eqnarray}
Consequently, we find 
\begin{eqnarray}
\int\frac{\partial q_{R}^{(I\bar{I})}(\tau)}{\partial \tau}
{\cal G}(\tau)d\tau
\mathop{=}^{R\rightarrow\infty}_{\epsilon g^2\rightarrow 0}
\sqrt{\frac{1}{3g^2}}.
\label{fp2;eqn}
\end{eqnarray}
Combining Eq.(\ref{fp1;eqn}) and Eq.(\ref{fp2;eqn}), 
$\Delta[\varphi_{R}]$ is evaluated as
\begin{eqnarray}
\Delta[\varphi_{R}]
\mathop{=}^{R\rightarrow\infty}_{\epsilon g^2\rightarrow 0}
\frac{1}{6g^2}.
\end{eqnarray}
Note that the factorization property at $R\rightarrow\infty$ holds:
The above $\Delta[\varphi_{R}]$ coincides with the product of  
the Jacobian (\ref{eqn:delta})
of the valley-instanton and that of the anti-valley-instanton.

The one-loop determinant can be evaluated from the following equation:
\begin{eqnarray}
\frac{\cal J}{\sqrt{\det''(-\partial_{\tau}^2+V''_{I\bar{I}})}}
=\frac{\cal J}{\sqrt{\det(-\partial_{\tau}^2+\omega_{+}^{2})}}
\sqrt{\frac{\det(-\partial_{\tau}^2+\omega_{+})}
{\det''(-\partial_{\tau}^2+V''_{I\bar{I}})}},
\end{eqnarray}
where $V_{I\bar{I}}=V(q_R^{(I\bar{I})})$.
Under the above choice of ${\cal G}(\tau)$, the following relation holds:
\begin{eqnarray}
\frac{\det''(-\partial_{\tau}^2+V_{I\bar{I}}'')}
{\det(-\partial_{\tau}^2+\omega_{+}^2)}
\stackrel{R\rightarrow\infty}{=}
\frac{\det'(-\partial_{\tau}^2+V_{I}'')}{\det(-\partial_{\tau}^2+\omega_{+}^2)}
\cdot 
\frac{\det'(-\partial_{\tau}^2+V_{\bar{I}}'')}
{\det(-\partial_{\tau}^2+\omega_{-}^2)},
\end{eqnarray}
where $V_I(V_{\bar{I}})$ is the potential of the (anti-)valley-instanton
background.
From the valley-instanton results (\ref{rdetvi;eqn}) and
(\ref{vinormalization;eqn}), we find 
\begin{eqnarray}
\frac{{\cal J}}{2\pi\sqrt{\det''(-\partial^2_{\tau}+V''_{I\bar{I}})}}
\Delta[\varphi_{R}]
\mathop{=}^{R\rightarrow\infty}_{\epsilon g^2\rightarrow 0}
\Upsilon\times\frac{e^{-T/2}}{\pi g^2},
\label{ibarif;eqn}
\end{eqnarray}
where $\Upsilon\equiv\lim_{\epsilon g^2\rightarrow 0}\Upsilon_{\pm}$.

For the $\bar{I}I$ valley, we find the following in the similar manner:
\begin{eqnarray}
\Delta[\varphi_{R}]
\mathop{=}^{R\rightarrow\infty}_{\epsilon g^2\rightarrow 0}
\frac{1}{6g^2}.
\end{eqnarray}
\begin{eqnarray}
\frac{{\cal J}}{2\pi\sqrt{\det''(-\partial^2_{\tau}+V''_{\bar{I}I})}}
\Delta[\varphi_{R}]
\mathop{=}^{R\rightarrow\infty}_{\epsilon g^2\rightarrow 0}
\Upsilon\times\frac{e^{-T/2}}{\pi g^2}.
\label{bariif;eqn}
\end{eqnarray}

\section{$I\bar{I}$ valley and $\bar{I}I$ valley}
\label{ibari;sec}

Consider the transition amplitudes between one of the local minima
of the potential and itself:
\begin{eqnarray}
Z=\left\{
\begin{array}{c}
\displaystyle
\lim_{T\rightarrow\infty}\langle q_{+}|e^{-HT}|q_{+}\rangle ,\\
\noalign{\vspace{2mm}} \displaystyle
\lim_{T\rightarrow\infty}\langle q_{-}|e^{-HT}|q_{-}\rangle .
\end{array}
\right. 
\label{Zamplitude;eqn}
\end{eqnarray}
In general, all configurations made of the same number of $I$ and
$\bar{I}$ are relevant for the amplitude, but we will first
evaluate this against the background of
the $I\bar{I}$ and $\bar{I}I$ valleys.
Due to the restriction of the background, only a few states in the
Hilbert space can be treated in an appropriate way, but  
the essence of the physics in the topologically trivial sector already
appears in this background. 
The general configurations will be taken into account 
in Section \ref{multi;sec}.
Due to the boundary condition, the $I\bar{I}$ valley is relevant for the
former amplitude of Eq.(\ref{Zamplitude;eqn}), and the $\bar{I}I$ valley
for the latter one. 

As was shown in Section \ref{valm;sec}, except for the collective
coordinates of the valleys, the path integral 
in the background of these valleys can be performed with  
Gaussian integrals. 
Then we obtain
\begin{eqnarray}
Z&=&\lim_{T\rightarrow\infty}
{\cal J}\int\int \frac{d\tau_0dR}{2\pi\sqrt{\det''D}}
\Delta[\varphi_{R}]e^{-S(R)}
\nonumber\\
&=&\lim_{T\rightarrow\infty}|\Psi(0)|^2\times \frac{e^{-T/2}}{\pi g^2}
\int_{0}^{T}dR(T-R)e^{-\tilde{S}(R)/g^2},
\label{valley-amplitude;eqn}
\end{eqnarray}
where $R$ is the collective coordinate of the valley corresponding to
the relative distance between the valley-instanton and the
anti-valley-instanton,\footnote{Strictly
speaking, only when $R\gg 1$, $R$ coincides with the relative distance
between the valley-instanton and anti-valley-instanton.  
For other values of $R$, we define $R$ so as to satisfy 
Eqs.(\ref{valley-amplitude;eqn})-(\ref{valley-action2;eqn}).} 
and we have abbreviated $\Psi(0;\omega)$ with $\omega=1$ as $\Psi(0)$.
(Namely, $\Psi(0)=(1/\pi)^{1/4}$.)
The factor before the integration of $R$ comes from
Eqs.(\ref{ibarif;eqn}) and (\ref{bariif;eqn}).
Here we have also performed the integration of the collective coordinate of the
translational symmetry. 
Since the configurations in the valleys must be one of the minima of
the potential at $\tau=\pm T/2$, a non-trivial factor $(T-R)$ results 
from the integration. 
As was shown in the previous section, the action of the valley behaves as
follows:
\begin{eqnarray}
\tilde{S}(R)=\left\{
\begin{array}{ll}
\displaystyle 
\sigma R^2& \mbox{at $R\rightarrow 0$ ;}\\
\noalign{\vspace{2mm}} \displaystyle
\frac{1}{3}-2e^{-R}-\epsilon g^2 R& \mbox{at $R\rightarrow\infty$,}
\end{array}
\right.
\label{valley-action;eqn}
\end{eqnarray}
for the $I\bar{I}$ valley, and 
\begin{eqnarray}
\tilde{S}(R)=\left\{
\begin{array}{ll}
\displaystyle 
\sigma R^2& \mbox{at $R\rightarrow 0$ ;}\\
\noalign{\vspace{2mm}} \displaystyle
\frac{1}{3}-2e^{-R}+\epsilon g^2 R& \mbox{at $R\rightarrow\infty$,}
\end{array}
\right.
\label{valley-action2;eqn}
\end{eqnarray}
for the $\bar{I}I$ valley.
Here $\sigma$ is a constant which will be determined later. 

For both of the valleys, the configuration at $R=0$ is the vacuum and at
$R\sim\infty$ the
well-separated valley-instanton--anti-valley-instanton pair.
These configurations are smoothly connected along the valleys.  
In the following, we will show that these geometrical structures 
in the configuration space lead to
the nontrivial structure of the quantum physics in the topological
trivial sector.

\subsection{The $\epsilon =0$ case}

First, let us consider the case of $\epsilon=0$.
We will now carry out 
the Borel transformation of the path-integral, which was
first suggested in Ref.\cite{thooft} 
and later developed on the basis of the valley
method \cite{Kik}.    
Changing the integration variable $R$
to $t=\tilde{S}(R)$, the transition amplitude becomes
\begin{eqnarray}
Z=\lim_{x\rightarrow 1/3}|\Psi(0)|^2\times\frac{e^{-T/2}}{\pi g^2}
\int_{C_V}dtF(t)e^{-t/g^2},
\label{t-amplitude;eqn}
\end{eqnarray}
where $x=\tilde{S}(T)=1/3-2e^{-T}$, $C_V=[0,x]$. From
Eq.(\ref{valley-action;eqn}), it is found that
$F(t)$ behaves as follows:
\begin{eqnarray}
F(t)=\left\{
\begin{array}{ll}
\displaystyle 
\frac{T}{\sqrt{4\sigma t}}& \mbox{for $t\rightarrow 0$ ;}\\
\noalign{\vspace{2mm}} \displaystyle
\frac{1}{1/3-t}\ln\left(\frac{1/3-t}{1/3-x}\right)
& \mbox{for $t\rightarrow 1/3$.}
\end{array}
\right.
\end{eqnarray}
To obtain the full form of $F(t)$, a detailed analysis of the valley
is needed. 
Instead, we will simply assume that the form of $F(t)$ is as follows:
\begin{eqnarray}
F(t)=\frac{f(t)}{\sqrt{4\sigma t}(1/3-t)}
\ln\left(\frac{1/3-t}{1/3-x}\right),
\label{f1;eqn}
\end{eqnarray}
where $f(t)$ is a function that satisfies
\begin{eqnarray}
f(0)=-\frac{T}{3\ln(1-3x)}, \quad 
f(1/3)=\sqrt{\frac{4\sigma}{3}}.
\end{eqnarray}

\begin{figure}
\centerline{\epsfxsize=9cm\epsfbox{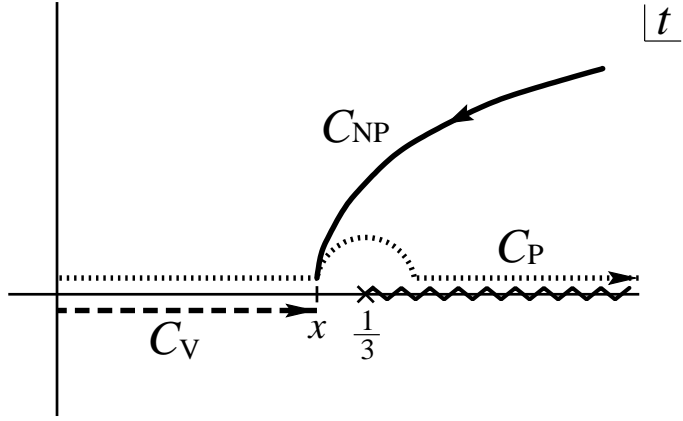}}
\caption{Deformation of  the contour $C_V$ 
to the sum of $C_{\rm P}$ and $C_{\rm NP}$.}
\label{fig:cv}
\end{figure}

The integral (\ref{t-amplitude;eqn}) contains both the perturbative
contribution at $t\sim 0$ 
and the non-perturbative one at $t\sim 1/3$.
To separate the perturbative and non-perturbative contributions, we
can deform the contour $C_{\rm V}$ to the sum of $C_{\rm P}$ and 
$C_{\rm NP}$ as is shown in
Fig.\ref{fig:cv}.
Then the amplitude becomes
\begin{eqnarray}
Z&=&\lim_{x\rightarrow 1/3}|\Psi(0)|^2\times\frac{e^{-T/2}}{\pi g^2}
\int_{C_{\rm P}}dtF(t)e^{-t/g^2}
\nonumber\\
&&+\lim_{x\rightarrow 1/3}|\Psi(0)|^2\times\frac{e^{-T/2}}{\pi g^2}
\int_{C_{\rm NP}}dt F(t)e^{-t/g^2}.
\label{Borel-amplitude;eqn}
\end{eqnarray}
Note that there is a significant resemblance between the first term on
the right-hand side and the formal Borel summation of the perturbation series.
{\em Therefore, we identify the first term of
Eq.(\ref{Borel-amplitude;eqn}) as the formal Borel-summation of the
perturbative series, and the second term as the non-\linebreak
perturbative contribution.}
As we shall see, this decomposition naturally explains the interplay 
between them.
For simplicity, we denote the first term as $Z_{\rm P}$ and the second as
$Z_{\rm NP}$.

According to the above assumption, the first term
of Eq.(\ref{Borel-amplitude;eqn}) is given by the ordinary perturbation
theory. 
Schematically, it is given by 
\begin{eqnarray}
Z_{\rm P}=\lim_{T\rightarrow\infty}
|\Psi(0)|^2\times
e^{-T/2}\left(1+P_0(g^2)+P_1(g^2)T+P_2(g^2)T^2+\cdots\right), 
\end{eqnarray}
where $P_i(g^2)\, (i=0,1,2,\cdots)$ are polynomials starting with $g^2$-terms.
In order to avoid contradicting 
the assumption, the parameter $\sigma$ is determined as
\begin{eqnarray}
\sigma=\frac{1}{4\pi g^2 T^2}.
\end{eqnarray}
The second term of Eq.(\ref{Borel-amplitude;eqn}) can be evaluated as follows,
\begin{eqnarray}
Z_{\rm NP}&=&\lim_{x\rightarrow 1/3}|\Psi(0)|^2\times\frac{e^{-T/2}}{\pi g^2}
\int_{\infty e^{i\chi}}^{x}
dtF(t)e^{-t/g^2}\nonumber\\
&=&\lim_{x\rightarrow 1/3}|\Psi(0)|^2\times 
\frac{e^{-T/2-1/3g^2}}{\pi g^2}
\nonumber\\
&&\hspace{3ex}\times\int_{x-1/3}^{\infty e^{i\chi}}
ds\frac{f(1/3+s)}{\sqrt{4\sigma(1/3+s)}s}\ln\left(\frac{-s}{1/3-x}\right)
e^{-s/g^2}\nonumber\\
&=&\lim_{x\rightarrow 1/3}|\Psi(0)|^2\times
\frac{e^{-T/2-1/3g^2}}{\pi g^2}\int_{x-1/3}^{\infty e^{i\chi}}ds\frac{1}{s}
\ln\left(\frac{-s}{1/3-x}\right)e^{-s/g^2}+\cdots
\nonumber\\
&=&\lim_{T\rightarrow\infty}|\Psi(0)|^2\times 
\frac{e^{-T/2-1/3g^2}}{\pi g^2}
\left[\frac{T^2}{2}-T\left\{\gamma+\ln\left(-\frac{2}{g^2}\right)\right\}
\right.\nonumber\\
&&\left.\hspace{10ex}+\frac{\Gamma^{''}(1)}{2}+
\gamma\ln\left(-\frac{2}{g^2}\right)
+\frac{1}{2}\left\{\ln\left(-\frac{2}{g^2}\right)\right\}^2+\cdots
\right],
\nonumber\\
\end{eqnarray}
where $s=t-1/3$, and $\chi$ is the angle of the path ${\rm C}_{\rm NP}$
at infinity.
In the above we have retained only the leading contribution in $g^2$.
Therefore, we obtain the following equation.
\begin{eqnarray}
Z&=&Z_{\rm P}+Z_{\rm NP}\nonumber\\
&=&
\lim_{T\rightarrow\infty}
\sum_{\varsigma=\pm 1}
\frac{|\Psi(0)|^2}{2}
\Biggl[ 1+P_0(g^2)\nonumber\\
&&\left.+\alpha^2
\left\{\frac{\Gamma''(1)}{2}+\gamma\ln\left(-\frac{2}{g^2}\right)
+\frac{1}{2}\left\{\ln\left(-\frac{2}{g^2}\right)\right\}^2+\cdots\right\}
\right]
\nonumber\\
&&\times e^{-T/2}
\Biggl[1+T\left\{\bar{P}_1(g^2)+\varsigma\alpha
-\alpha^2\left\{\gamma+\ln\left(-\frac{2}{g^2}\right)
+\cdots\right\}\right\}
\nonumber\\
&&+\frac{T^2}{2}\left\{\bar{P}_2(g^2)+\alpha^2\left\{1+\cdots\right\}\right\}
+\cdots\Biggr], 
\end{eqnarray}
where $\alpha=e^{-1/6g^2}/\sqrt{\pi g^2}$ and $
\bar{P}_i(g^2)=P_i(g^2)/(1+P_0(g^2))$ $(i=1,2)$.

To obtain the non-perturbative corrections, we can compare
this with the following:  
\begin{eqnarray}
Z&=&\lim_{T\rightarrow\infty}\langle q_+|e^{-HT}|q_+\rangle
\nonumber\\
&=&\lim_{T\rightarrow\infty}\sum_{\varsigma=\pm 1}
|\Psi_{\varsigma}(q_+)|^2 e^{-E_{\varsigma}(\epsilon=0,N_+=0)T}+\cdots,
\end{eqnarray}
where $E_{\varsigma}(\epsilon=0,N_+=0), \varsigma=\pm 1$ are the energies
of the first two lowest states and  
$\Psi_{\varsigma}(q_+)$ are the values of the normalized wave
functions of the states at $q=q_+$.
These states are perturbatively degenerated as,
\begin{eqnarray}
E_{\varsigma}(\epsilon=0, N_{+}=0)=\frac{1}{2}+O(g^2),
\end{eqnarray}
and the degeneracy is
resolved by the tunneling effect. 
The non-perturbative corrections for these energies are thus found as
\begin{eqnarray}
E_{\varsigma,{\rm NP}}(\epsilon=0,N_+=0)
=\varsigma\alpha+\alpha^2\left\{\gamma+\ln\left(-\frac{2}{g^2}\right)\right\},
\label{ge-double;eqn}
\end{eqnarray}
where $\varsigma=\pm 1$.
In addition, we obtain the non-perturbative correction for the values of
the wave functions at $q=q_+$:
\begin{eqnarray}
&&|\Psi_{\varsigma}(q_+)|^2_{\rm NP}
\nonumber\\
&&=\frac{\alpha^2}{2}|\Psi(0)|^2
\left\{\frac{\Gamma''(1)}{2}+\gamma\ln\left(-\frac{2}{g^2}\right)
+\frac{1}{2}\left\{\ln\left(-\frac{2}{g^2}\right)\right\}^2\right\}.
\label{gw-double;eqn}
\end{eqnarray}
Note that there are imaginary terms in Eqs.(\ref{ge-double;eqn}) and
(\ref{gw-double;eqn}).
As will be explained in Section \ref{dis;sec}, 
this is because that the perturbative corrections for
them are non-Borel-summable. 

\subsection{The $\epsilon\neq 0$ case}

\subsubsection{$I\bar{I}$ valley}

As was seen in Fig.\ref{fig:iibar2}, 
the action of the $I\bar{I}$ valley $\tilde{S}(R)$ is not
an invertible function of $R$.
Then, if we change the integral variable $R$ to $t=\tilde{S}(R)$ in
Eq.(\ref{valley-amplitude;eqn}), there arises an apparent singularity. 
To avoid this, we add $\epsilon g^2\tilde{S}_1(R)$ 
\begin{eqnarray}
\epsilon g^2 \tilde{S}_1=\epsilon g^2\int d\tau \tilde{q}_R(\tau)
\end{eqnarray}
from the action and define the new integral variable $t$ as
\begin{eqnarray}
t=\tilde{S}(R)+\epsilon g^2 \tilde{S}_1(R)=\tilde{S}_0(R).
\end{eqnarray}
Here $\tilde{q}_R(\tau)$ is the $I\bar{I}$ valley with $R$.
The function $\tilde{S}_0(R)$ 
is invertible, so this causes no apparent singularity.
When $\epsilon\rightarrow 0$, this definition of $t$ smoothly coincides
with that in the case of $\epsilon=0$, since $\epsilon g^2\tilde{S}_1(R)$ is
zero when $\epsilon=0$.   

The remaining analysis is an extension of the case of $\epsilon=0$. 
Using $t$, the transition amplitude is rewritten as  
\begin{eqnarray}
Z=\lim_{x\rightarrow 1/3}|\Psi(0)|^2\times\frac{e^{-T/2}}{\pi g^2}
\int_{C_V}dtF(t)e^{-t/g^2},
\end{eqnarray}
where $x=\tilde{S}_0(T)=1/3-2e^{-T}$ and $C_V=[0,x]$.
As $\tilde{S}_0(R)$ behaves as
\begin{eqnarray}
\tilde{S}_0(R)=\left\{
\begin{array}{ll}
\displaystyle 
\sigma R^2& \mbox{for $R\rightarrow 0$ ;}\\
\noalign{\vspace{2mm}} \displaystyle
\frac{1}{3}-2e^{-R}& \mbox{for $R\rightarrow\infty$,}
\end{array}
\right.
\label{action:eqn}
\end{eqnarray}
then $F(t)$ satisfies
\begin{eqnarray}
F(t)=\left\{
\begin{array}{ll}
\displaystyle 
\frac{T}{\sqrt{4\sigma t}}& \mbox{for $t\rightarrow 0$ ;}\\
\noalign{\vspace{2mm}} \displaystyle
\frac{1}{1/3-t}\left(\frac{1/3-t}{2}\right)^{-\epsilon}
\ln\left(\frac{1/3-t}{1/3-x}\right)& \mbox{for $t\rightarrow 1/3$.}
\end{array}
\right.
\end{eqnarray}
We assume the form of $F(t)$ to be
\begin{eqnarray}
F(t)=\frac{f(t)}{\sqrt{4\sigma t}(1/3-t)}
\left(\frac{1/3-t}{2}\right)^{-\epsilon}
\ln\left(\frac{1/3-t}{1/3-x}\right),
\label{f2;eqn}
\end{eqnarray}
where $f(t)$ is a function which satisfies
\begin{eqnarray}
f(0)=-\frac{6^{-\epsilon}T}{3\ln(1-3x)},
\quad 
f(1/3)=\sqrt{4\sigma/3}.
\end{eqnarray}
We can 
identify the perturbative contribution and non-perturbative contribution 
in the same manner of the case of $\epsilon=0$, and
we find the non-perturbative contribution to be as follows.
\begin{itemize}
 \item $\epsilon\notin{\bf Z}$
\begin{eqnarray}
Z_{\rm NP}&=&\lim_{T\rightarrow\infty}|\Psi(0)|^2\times 
\frac{e^{-T/2-1/3g^2}}{\pi g^2}
\left[\left(-\frac{2}{g^2}\right)^{\epsilon}
\Gamma(-\epsilon)T
\right.\nonumber\\
&&\left.\hspace{3ex}
-\left(-\frac{2}{g^2}\right)^{\epsilon}
\left\{\ln\left(-\frac{2}{g^2}\right)\Gamma(-\epsilon)
-\Gamma '(-\epsilon)\right\}
\right.\nonumber\\
&&\left.\hspace{3ex}
+\sum_{N_{-}=0}^{\infty}
\left(\frac{2}{g^2}\right)^{N_{-}}\frac{e^{-(N_{-}-\epsilon)T}}
{N_{-}!(N_{-}-\epsilon)^2}
\right]. 
\label{znp-ibari1;eqn}
\end{eqnarray}
\item$\epsilon={\cal N}\geq 0, \quad {\cal N}\in{\bf Z}$
\begin{eqnarray}
Z_{\rm NP}&=&\lim_{T\rightarrow\infty}|\Psi(0)|^2\times 
\frac{e^{-T/2-1/3g^2}}{\pi g^2}
\left[\left(\frac{2}{g^2}\right)^{\cal N}\frac{1}{{\cal N}!}\frac{T^2}{2}
\right.\nonumber\\
&&\left.
-\left(\frac{2}{g^2}\right)^{\cal N}
\left\{\ln\left(-\frac{2}{g^2}\right)
-\psi({\cal N}+1)\right\}\frac{T}{{\cal N}!}\right.
\nonumber\\
&&\left.+\left(\frac{2}{g^2}\right)^{\cal N}\frac{1}{{\cal N}!}
\left\{
\frac{1}{2}\left\{\ln\left(-\frac{2}{g^2}\right)\right\}^2
-\psi({\cal N}+1)\ln\left(-\frac{2}{g^2}\right)
\right.\right.\nonumber\\
&&\left.\left.
+\frac{\Gamma^{''}(1)}{2}
-\gamma\sum_{k=1}^{\cal N}\frac{1}{k}
+\sum_{k=1,l=1,k\ge l}^{{\cal N}}\frac{1}{k\,l}\right\}
\right.\nonumber\\
&&\left.
+\sum_{N_{-}=0,N_{-}\neq {\cal N}}^{\infty}\left(\frac{2}{g^2}\right)^{N_{-}}
\frac{e^{-(N_{-}-{\cal N})T}}{N_{-}!(N_{-}-{\cal N})^2}
\right],
\label{znp-ibari2;eqn}
\end{eqnarray}
where $\psi(z)\equiv d\log\Gamma(z)/dz$ is the psi function.
\end{itemize}
The result (\ref{znp-ibari1;eqn}) leads to the following for the
non-perturbative contribution to the perturbative ground state at $q=q_+$, 
\begin{eqnarray}
E^{(+)}_{\rm NP}(\epsilon,N_+=0)&=&
-\alpha^2\left(-\frac{2}{g^2}\right)^{\epsilon}\Gamma(-\epsilon).
\label{enpn+01;eqn}
\end{eqnarray}
Furthermore, the non-perturbative contribution to the normalized wave
function of this state is found to be the following
\begin{eqnarray}
|\Psi(q_+)|_{\rm NP}^2
&=&-\alpha^2|\Psi(0)|^2\left(-\frac{2}{g^2}\right)^{\epsilon}
\nonumber\\
&&\times
\left\{\ln\left(-\frac{2}{g^2}\right)\Gamma(-\epsilon)
-\Gamma'(-\epsilon)\right\}.
\label{wfnp+01;eqn}
\end{eqnarray}
The case where $\epsilon={\cal N}\geq 0,\, {\cal N}\in{\bf Z}$ requires 
further discussion, which will be given in the next section. 

\subsubsection{$\bar{I}I$ valley}
The evaluation of the amplitude in the background of the $\bar{I}I$ valley
can be performed in the same manner as that of the $I\bar{I}$ valley.
The non-perturbative part of the amplitude is the following:
\begin{eqnarray}
Z_{\rm NP}&=&\lim_{T\rightarrow\infty}|\Psi(0)|^2\times 
\frac{e^{(\epsilon-1/2)T-1/3g^2}}{\pi g^2}
\left[\left(-\frac{2}{g^2}\right)^{-\epsilon}
\Gamma(\epsilon)T
\right.\nonumber\\
&&\left.\hspace{3ex}
-\left(-\frac{2}{g^2}\right)^{-\epsilon}
\left\{\ln\left(-\frac{2}{g^2}\right)\Gamma(\epsilon)
-\Gamma '(\epsilon)\right\}
\right.\nonumber\\
&&\left.\hspace{3ex}
+\sum_{N_{+}=0}^{\infty}
\left(\frac{2}{g^2}\right)^{N_{+}}\frac{e^{-(N_{+}+\epsilon)T}}
{N_{+}!(N_{+}+\epsilon)^2}
\right]. 
\end{eqnarray}
From the above equation, we find the following non-perturbative
contributions for the perturbative ground state at $q=q_{-}$:
\begin{eqnarray}
E^{(-)}_{\rm NP}(\epsilon,N_{-}=0)=
-\alpha^2\left(-\frac{2}{g^2}\right)^{-\epsilon}\Gamma(\epsilon),
\label{enpn-0;eqn}
\end{eqnarray}
\begin{eqnarray}
|\Psi(q_-)|_{\rm NP}^2&=&-\alpha^2|\Psi(0)|^2
\left(-\frac{2}{g^2}\right)^{-\epsilon}
\nonumber\\
&&\times\left\{\ln\left(-\frac{2}{g^2}\right)\Gamma(\epsilon)
-\Gamma'(\epsilon)\right\}. 
\label{wfnp-0;eqn}
\end{eqnarray}

\subsection{Singularity of the Borel plane}\label{dis;sec}

An immediate consequence of our decomposition of the perturbative and
non-perturbative contribution is  
\begin{eqnarray}
{\rm Im}Z_{\rm P}+{\rm Im}Z_{\rm NP}=0.
\end{eqnarray}
This is because $Z=Z_{\rm P}+Z_{\rm NP}$ is real.
Thus, this simple equation explains why the imaginary part of the formal Borel
summation of the perturbation series is canceled by that of the
non-perturbative contribution. 
At the same time, it also shows that the nonzero imaginary part of the
non-perturbative contribution is a necessary and sufficient condition
for the non-Borel-summability of the perturbative expansion.

\begin{figure}
\centerline{\epsfxsize=10cm\epsfbox{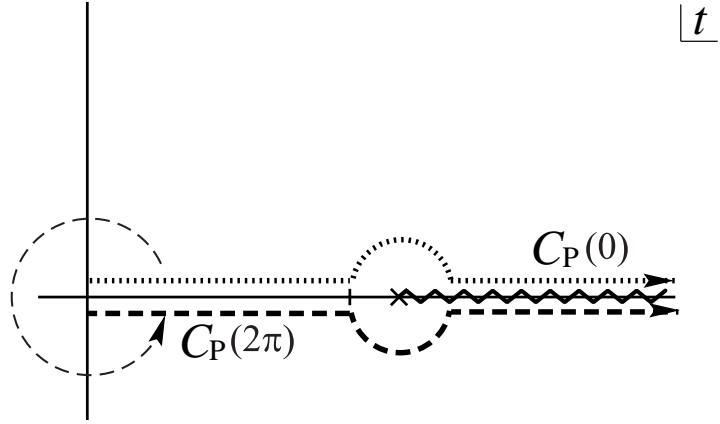}}
\caption{The change of the contour 
$C_{\rm P}(\theta)$ of $Z_{\rm P}(|g^2|e^{i\theta})$
as $\theta$ is changed from zero to $2\pi$.}
\label{fig:cprotation}
\end{figure}

\begin{figure}
\centerline{\epsfxsize=9cm\epsfbox{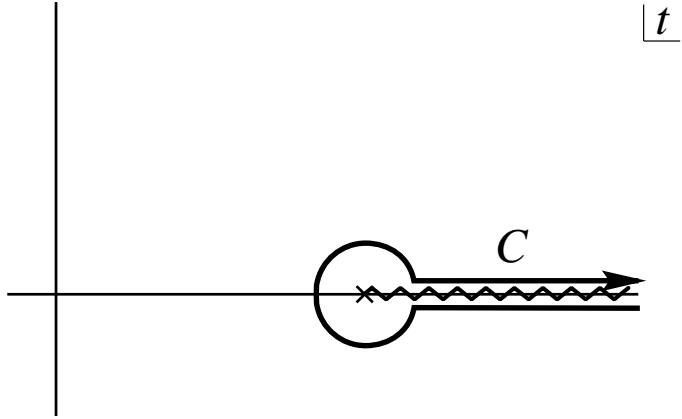}}
\caption{The contour of integral}
\label{fig:imaginary}
\end{figure}

Furthermore, assuming that $f(t)$ in 
Eqs.(\ref{f1;eqn}) and (\ref{f2;eqn}) has the following form:
\begin{eqnarray}
f(t)=\tilde{f}(t,g^2/t),
\end{eqnarray}
where $\tilde{f}(t,s)$ is an analytic function of $t$ 
for small $s$, we can predict the large order behavior of the
perturbative contribution.
To this end, we examine the analyticity of $Z_{\rm P}(g^2)$ in the complex
$g^2$-plane.
Let $\theta$ denote the phase of $g^2$, $g^2=|g^2|e^{i\theta}$.
When the phase of $g^2$ changes to $2\pi$, 
the contour of $Z_{\rm P}(g^2)$ changes as Fig.\ref{fig:cprotation}. 
Thus we obtain
\begin{eqnarray}
Z_{\rm P}(g^2e^{2\pi i})
=Z_{\rm P}(g^2)+\lim_{x\rightarrow 1/3}|\Psi(0)|^2
\times \frac{e^{-T/2}}{\pi g^2}\int_{C}dtF(t)e^{-t/g^2},
\end{eqnarray} 
where the contour $C$ is given in Fig.\ref{fig:imaginary}.
Therefore, $Z_{\rm P}(g^2)$ has a cut on the real axis in the 
complex $g^2$-plane.
This is the only singularity near the origin. 
Thus the dispersion relation becomes
\begin{eqnarray}
Z_{\rm P}(g^2)&=&\frac{1}{2\pi i}\oint_{C_{g^2}} dz\frac{Z_{\rm P}(z)}{z-g^2}
\nonumber\\
&=&\frac{1}{\pi}\int_{0}^{\infty}dz\frac{{\rm Im}Z_{\rm P}(z)}{z-g^2}
+\cdots
\nonumber\\
&=&-\frac{1}{\pi}\int_{0}^{\infty}dz\frac{{\rm Im}Z_{\rm NP}(z)}{z-g^2}
+\cdots
\nonumber\\
&=&-\frac{1}{\pi}\sum_{n=0}^{\infty}\int_{0}^{\infty}dz
\frac{{\rm Im}Z_{\rm NP}(z)}{z^{n+1}}g^{2n}
+\cdots,
\end{eqnarray}
where $C_{g^2}$ is the contour that circles around $z=g^2$,
and we have neglected the
contribution from the singularity far from the origin.
Thus, the following large order behavior of $Z_P(g^2)$ is predicted:
\begin{eqnarray}
&&Z_{\rm P}(g^2)=\sum_{m=1}^{\infty}c_m g^{2m},\nonumber\\
&&c_m\stackrel{m\rightarrow\infty}{=} 
-\frac{1}{\pi}\int_0^{\infty}dg^2 \frac{{\rm Im}Z_{\rm NP}(g^2)}{g^{2m+2}}.
\end{eqnarray}
Rewriting this in terms of the energies and the wave functions, we find
the perturbative contribution to the energy $E_{\rm P}$
and to the wave function $|\Psi(q_{\pm})|_{\rm P}^2$ as follows,
\begin{eqnarray}
E_{\rm P}(g^2)&=&\sum_{m=0}^{\infty}e_{m}g^{2m},\nonumber\\
e_{m}&\stackrel{m\rightarrow\infty}{=}&
-\frac{1}{\pi}\int_0^{\infty}dg^2 \frac{{\rm Im}E_{\rm NP}(g^2)}{g^{2m+2}}.
\end{eqnarray}
and 
\begin{eqnarray}
|\Psi(q_{\pm})|_{\rm P}^2&=&\sum_{m=0}^{\infty}w_{m}g^{2m},\nonumber\\
w_{m}&\stackrel{m\rightarrow\infty}{=}&
-\frac{1}{\pi}\int_0^{\infty}dg^2 \frac{{\rm Im}|\Psi(q_{\pm})|_{\rm NP}^2}
{g^{2m+2}}. \label{eqn:psicorr}
\end{eqnarray}
Detailed tests of these predictions are performed in Section \ref{lob;sec}.
 
\subsection{Bogomolny's trick}
\label{bog;sec}

In the early eighties, Bogomolny \cite{Bog} suggested 
a procedure to 
calculate the tunneling effect beyond the dilute-gas approximation. 
He suggested that by the formal analytic continuation of the coupling
constant which changes the force between instantons from
attractive to repulsive,
the instantons in the topologically trivial sector become meaningful and their
contributions can be separated from the perturbation theory. 
The same trick was used by Zinn-Justin \cite{Zin} to obtain the energy
levels of the excited states.
Our identification of the non-perturbative contribution is a precise
realization of Bogomolny's suggestion, and it provides an explicit
justification for this trick. 

\begin{figure}
\centerline{\epsfxsize=9cm\epsfbox{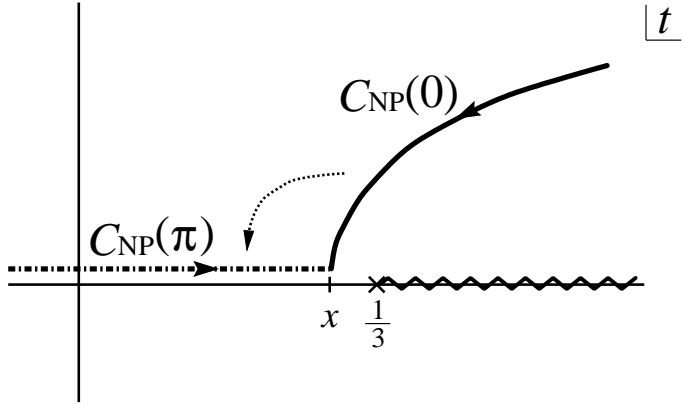}}
\caption{The change of the contour 
$C_{\rm NP}(\theta)$ of $Z_{\rm NP}(|g^2|e^{i\theta})$
as $\theta$ is changed from zero to $\pi$.}
\label{fig:cvrotation}
\end{figure}

To see this, let us examine $Z_{\rm NP}(g^2)$ in the complex
$g^2$-plane. 
As is shown in Fig.\ref{fig:cvrotation},
if we perform the analytic continuation of $Z_{\rm NP}(|g^2|e^{i\theta})$ 
from $\theta=0$ to $\theta=\pi$, 
the contour for $Z_{\rm NP}(|g^2|e^{i\theta})$
changes from $C_{\rm NP}(0)$ to $C_{\rm NP}(\pi)$.
And in the weak coupling limit 
the integral of $C_{\rm NP}(\pi)$ can be
well-approximated by that of $C_{\rm V}$, because
the dominant contribution of the integral comes from $t\sim x$.
Therefore, in the case of $g^2=|g^2|e^{i\pi}$ when the interaction between
instantons is repulsive, 
$Z_{\rm NP}(|g^2|e^{i\pi})$
coincides approximately with 
what Bogomolny suggested as a method of evaluation:
\begin{eqnarray}
Z_{\rm NP}(|g^2|e^{i\pi})=Z(|g^2|e^{i\pi}).
\label{coincide;eqn}
\end{eqnarray}
Note that the coincidence $Z_{\rm NP}(g^2)=Z(g^2)$ does not hold for
$\theta=0$;
while $Z_{\rm NP}(g^2)$ could be complex, $Z(g^2)$ never
possesses the imaginary part in this case.
This is because a so-called Stokes phenomena occurs for $Z(g^2)$; the
terms in $Z(g^2)$ which are negligible for $\theta=\pi$ become relevant
for $\theta=0$.  
Since the imaginary part of $Z_{\rm NP}(g^2)$ is necessary to cancel the
ambiguity of the Borel transform of the perturbation theory,
the formal analytic continuation of $Z(g^2)$
which Bogomolny originally suggested should be replaced  
with the {\em real} analytic continuation of $Z_{\rm NP}(g^2)$.

One can easily find that $Z_{\rm NP}(g^2)$ has the same asymptotic expansion
for $\theta=0$ and $\theta=\pi$.
Therefore, the asymptotic expansion of $Z_{\rm NP}(g^2)$ with $\theta=0$ can be
obtained by the analytic continuation of the asymptotic expansion of
$Z(g^2)$ with $\theta=\pi$.  
We will use this in the next section. 

\section{The multi valley} 
\subsection{Multi valley calculus}\label{multi;sec}

In this section we will evaluate the partition function
$Z=\lim_{T\rightarrow\infty} {\rm Tr} (e^{-HT})$,
by summing over 
those configurations made of several valley-instantons,
by utilizing the knowledge of the valley-instantons and the interactions 
among them in Section 2, and applying the 
analytic continuation discussed in Section \ref{bog;sec}.
This enables us to evaluate non-perturbative contributions
to excited states as well as the ground state.

\begin{figure}
\centerline{\epsfxsize=12cm\epsfbox{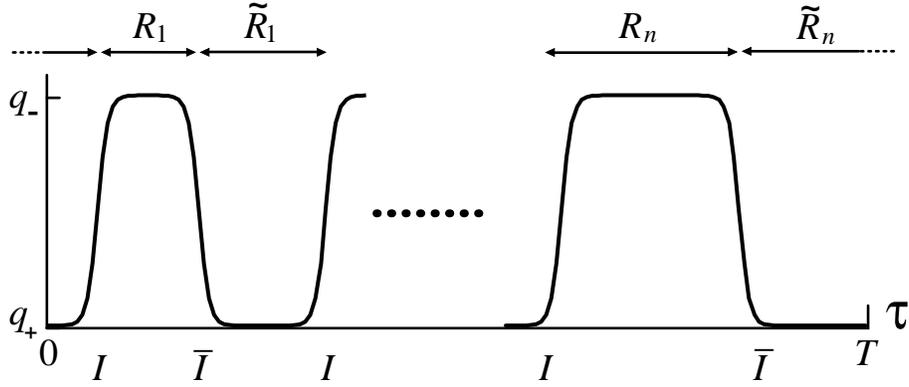}}
\caption{The collective coordinates $R_i$ and $\tilde R_i$ for 
$n$-pair configuration of valley-instanton and anti-valley-instanton.}
\label{fig:r-tilder}
\end{figure}

We will take a valley made of  $n$-pairs of the valley-instanton
and the anti-valley-instanton with periodic boundary condition.
Since we perform 
the calculation for $g^2=|g^2|e^{i\pi}<0$, the force between the
valley-instanton and the anti-valley-instanton becomes repulsive.
Therefore, the configurations with large separations between
valley-instantons dominate.
From the calculation of the action for one pair in Section 2, we
find that the action of this $n$-pair configuration is given by the following:
\begin{eqnarray}
S=\frac{n}{3g^2}-\epsilon \sum_{i=1}^{n}R_i
-\frac{2}{g^2}\sum_{i=1}^{n}e^{-R_i}
-\frac{2}{g^2}\sum_{i=1}^{n}e^{-\tilde{R_i}},
\label{eqn:slr}
\end{eqnarray}
where $R_i$ is the distance between the $i$-th 
valley-instanton and the $i$-th anti-valley-instanton 
and $\tilde{R_i}$ is the distance between
the $i$-th anti-valley-instanton and the $(i+1)$-th valley-instanton (mod $n$).
(See Fig.\ref{fig:r-tilder}.)
The expression (\ref{eqn:slr}) is valid when all the valley-instanton
and the anti-valley-instantons are well-separated, that is, when
$R_i, \tilde R_i \gg 1$.

We will 
write the sum of the contributions of the $n$-pairs of valley-instantons 
to the partition function $Z$
for $n=1 \sim \infty$ as follows,
\begin{eqnarray}
Z_{\rm NP}=\lim_{T\rightarrow\infty}
\sum_{n=1}^{\infty}\alpha^{2n}J_{n},   
\label{eqn:Z}
\end{eqnarray}
where the constant $\alpha$ contains the contributions of the
Jacobian (\ref{eqn:delta}) 
and the $R$-independent part of the determinant (\ref{eqn:kappa}),
and is at the leading order in $\epsilon g^2$,
\begin{equation}
\alpha= \frac{\Delta}{\sqrt{2\pi\kappa}} e^{-S^{(I)}} 
=\frac{e^{-1/6g^2}}{g\pi^{1/2}}.
\end{equation}
The term $J_{n}$ is defined as follows:
\begin{eqnarray}
J_{n}&=&\frac{T}{n}\int_{0}^{\infty}\prod_{i=1}^{n}dR_i 
d\tilde{R}_i
\, \delta\left(\sum_{i=1}^{n}(R_{i}+\tilde{R}_i)-T\right)\nonumber\\ 
&& \hspace*{-15mm} \times \exp\left(
\left(\epsilon - \frac{1}{2}\right)\sum_{i=1}^{n}R_i
- \frac{1}{2}\sum_{i=1}^{n}\tilde R_i 
+\frac{2}{g^2}\sum_{i=1}^{n}e^{-R_i}
+\frac{2}{g^2}\sum_{i=1}^{n}e^{-\tilde{R_i}}\right).
\label{eqn:22}
\end{eqnarray}

Eq.(\ref{eqn:Z}) can be evaluated in a manner similar to that performed
by Zinn-Justin for the ordinary instanton case \cite{Zin}.
First, we rewrite the $\delta$-function as follows:
\begin{eqnarray}
\delta\left(\sum_{i=1}^{n}(R_i+\tilde{R}_i)-T\right)=\frac{1}{2\pi i}
\int_{-i\infty-\eta}^{i\infty-\eta}ds 
\exp\left(-sT+s\sum_{i=1}^{n}(R_i+\tilde{R}_i)\right),
\end{eqnarray}
where $\eta$ is a positive real number.
This allows factorization of the integrals over $R_i$ and
$\tilde{R}_i$ in the following manner.
\begin{eqnarray}
J_n = \frac{1}{2\pi i}
\int_{-i\infty-\eta}^{i\infty-\eta}ds \,
K_+(s)^n K_-(s)^n,
\end{eqnarray}
where $K_\pm(s)$ are,
\begin{eqnarray}
K_\pm(s) &\equiv& \int_0^\infty d R \, \exp\left(
s_\pm R + \frac{2}{g^2}e^{- R} \right) \nonumber\\
&=& 
\left(-\frac{2}{g^2}\right)^{s_\pm}
\Gamma\left( -s_\pm\right),
\label{eqn:kpm}
\end{eqnarray}
\begin{equation}
s_+ \equiv s - \frac{1}{2}, \quad 
s_- \equiv s + \epsilon -\frac{1}{2}.
\end{equation}
This leads to the following expression.
\begin{eqnarray}
Z_{\rm NP}&=& \frac{T}{2 \pi i} \int_{-i \infty - \eta}^{i \infty - \eta} ds \,
e^{- T s} \, \sum_{n=1}^\infty \frac{(\alpha^2 K_+(s) K_-(s))^n}{n}
\nonumber \\
&=& - \frac{T}{2 \pi i} \int_{-i \infty - \eta}^{i \infty - \eta} ds
\, e^{- T s} \, \ln ( 1- \alpha^2 K_+(s) K_-(s)).
\end{eqnarray}
By partial integration we find that 
\begin{equation}
Z_{\rm NP} = - \frac{1}{2 \pi i} 
\int_{-i \infty - \eta}^{i \infty - \eta} ds \,
e^{- T s} \frac{\phi'(s)}{\phi(s)},
\label{parts}
\end{equation}
where
\begin{equation}
\phi(s) \equiv 1 - \alpha^2 K_+(s) K_-(s).
\label{eq:alpha}
\end{equation}
From Eq.(\ref{eqn:kpm}), we see that
$\phi(s)$ has both poles and zeros in $s$. Denoting the poles by
$s=E_0(N)$ and the zeros by $E(N)$, we find that
\begin{equation}
Z_{\rm NP} = \sum_{N} e^{-E(N) T} - \sum_N e^{-E_0(N) T}.
\label{eqn:sumsum}
\end{equation}
The poles of $\phi(s)$ are given by the poles of the
$\Gamma$-functions, $s=E_0^{(\pm)}(\epsilon, N_\pm)$ defined
as follows,
\begin{eqnarray}
E_0^{(+)}(\epsilon,N_+)&=&\frac{1}{2}+N_{+},\\
E_0^{(-)}(\epsilon,N_-)&=&-\epsilon+\frac{1}{2}+N_{-},
\end{eqnarray}
which are the perturbative zeroth order energies of
the $N_\pm$-th excited states around $q = q_\pm$, respectively.
The subtraction in Eq.(\ref{eqn:sumsum}) corresponds to the
fact that our $Z_{\rm NP}$ does not contain this zeroth order term
(and in fact vanishes for $g=0$).
Therefore the full partition function 
has only the first term of Eq.(\ref{eqn:sumsum}).
This means that the non-perturbative contributions to
the energy spectrum are given by the zeros of $\phi(s)$:
\begin{equation}
\phi(s)=0,
\label{eqn:phis0}
\end{equation}
This equation is identical to the result obtained by the
WKB approximation (this derivation is given in Appendix D).

Let us solve Eq.(\ref{eqn:phis0})
in a perturbation of $\alpha \ll 1$.
In the case when the poles of the $\Gamma$-functions of $K_+$ and
$K_-$ do not coincide, the solutions 
$s = E_0^{(\pm)}(\epsilon, N_\pm) + E_{\rm NP}^{(\pm)}(\epsilon,N_\pm)$ 
are found by a series expansion in $\alpha^2$:
\begin{equation}
E_{\rm NP}^{(\pm)}(\epsilon, N_\pm) = \alpha^2 
\frac{(-1)^{N_\pm+1}}{N_\pm!}
\left( -\frac{2}{g^2}\right)^{\pm \epsilon+2 N_\pm }
\Gamma \left(\mp \epsilon -  N_\pm \right) + O(\alpha^4).
\label{eqn:en+}
\end{equation}

\begin{figure}
\centerline{\epsfxsize=11cm\epsfbox{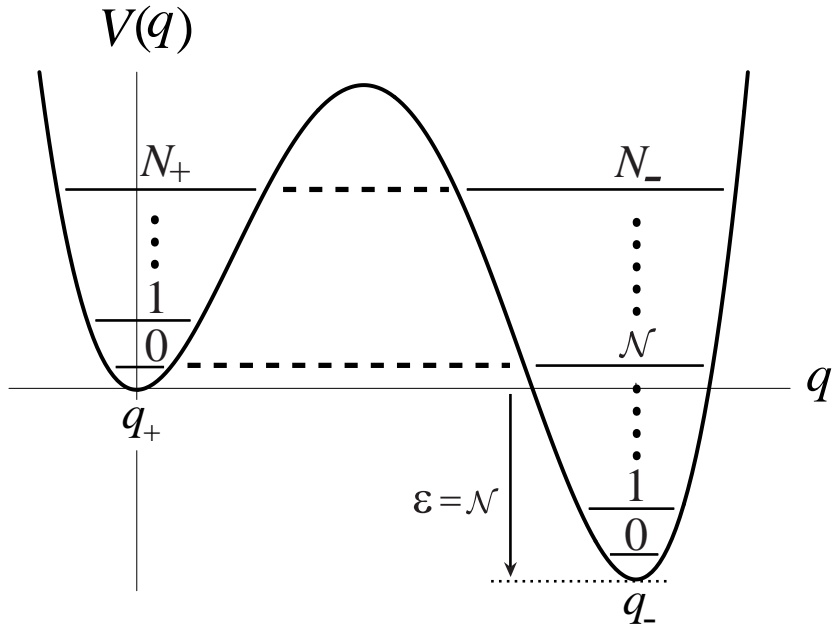}}
\caption{The potential $V(q)$ and the zeroth order eigenstates
when Eq.(\protect{\ref{eqn:npm}}) is satisfied.}
\label{fig:nfold}
\end{figure}

The result (\ref{eqn:en+}) may diverge
depending on $\epsilon$ and $N_\pm$.
In fact, the poles occur when the argument of the $\Gamma$-functions
is zero or an negative integer.
This happens when the zeroth order
energies of the states around $q_+$ and $q_-$ are degenerate,
which is allowed when $\epsilon={\cal N}$ $({\cal N} \in {\bf Z})$.
In fact, these divergences in $E_{\rm NP}^{(\pm)}(\epsilon, N_\pm)$ 
are caused by the confluence
of the corresponding poles of the $\Gamma$ functions
in $K_+$ and $K_-$.  In cases where this happens, we need to use
a different expansion scheme in $\alpha$, 
$E(N) = E_0(N) + \alpha \rho_1 + \alpha^2 \rho_2 + O(\alpha^3)$.
The zeroth order term $E(N)$ is chosen so that 
the arguments of the $\Gamma$ functions in $K_\pm$ are zero or a negative
integer $-N_\pm$ at the same time, which means
\begin{equation}
E_0(N_\pm) = \frac12 + N_+ =-\epsilon + \frac12 + N_- .
\label{eqn:npm}
\end{equation}
This situation is illustrated in Fig.\ref{fig:nfold}.
Straightforward calculation yields the following non-perturbative
correction $E_{\varsigma,\rm NP}(\epsilon, N_+)$ to the energy levels:
\begin{eqnarray}
&& E_{\varsigma,\rm NP}(\epsilon, N_+) = 
\varsigma \alpha 
\sqrt{\frac{1}{N_+! N_-!}\left(\frac{2}{g^2}\right)^{N_+ + N_-}}
\nonumber\\
&& \qquad + \frac{\alpha^2}{2}
\frac{1}{N_+! N_-!}\left(\frac{2}{g^2}\right)^{N_+ + N_-}
\left[2 \ln\left(-\frac{2}{g^2}\right)
-\psi(N_++1)-\psi(N_-+1)
\right],\nonumber\\
\label{eqn:dege}
\end{eqnarray}
where $\varsigma=\pm 1$ and $N_-=N_+ + \epsilon = N_+ + {\cal N}$.
The plus and the minus signs in the expression (\ref{eqn:dege})
correspond to two (approximate)
linear combinations of the perturbative states in the
left and right well with the same zeroth order energy.
This situation is analogous to the lifting of the degeneracy by
the instanton contribution for the symmetric double-well potential.
(In fact, in the limit $\epsilon \rightarrow 0$ the ordinary 
instanton result for the energy splitting between the ground state
and the first excited state is recovered.)

The non-perturbative contributions to the energy eigenvalue
obtained here contain both a real part and an imaginary part.
As discussed in the previous section the imaginary part
is related to the large order behaviour of the perturbative series.
We will confirm this in the following section by directly calculating
the perturbative coefficients and comparing them with
the prediction of the valley method.
The real part of the energy needs further consideration:
Normally the non-perturbative contribution is obscured by
the perturbative contribution.  
The quantities which receive only a non-perturbative contribution
provide a meaningful test of the non-perturbative effect, while
other quantities do not provide such a test.
We will discuss such a non-renormalization theorem in 
Section 6, and then using that result we will test the
real part of the energy expression in Section 7.

\subsection{$I\bar{I}$ valley and $\bar{I}I$ valley revisited}

Before proceeding to the various tests of the result in Section
\ref{multi;sec},
it is instructive to go back to the $I\bar{I}$ and $\bar{I}I$ valleys
and compare the result of Section
\ref{ibari;sec} with that of Section \ref{multi;sec}.
If we compare Eq.(\ref{ge-double;eqn}) with Eq.(\ref{eqn:dege}), and 
Eqs.(\ref{enpn+01;eqn}) 
and (\ref{enpn-0;eqn}) with Eq.(\ref{eqn:en+}),
it will be seen that there is a complete agreement between them.

In Section \ref{ibari;sec}, we did not evaluate
the non-perturbative
contribution to the perturbative ground state at $q=q_+$ in the case of
$\epsilon={\cal N} > 0,\,{\cal N}\in{\bf Z}$.
The reason is that in this case the ground state is degenerate with the
perturbative ${\cal N}$-th excited state at $q=q_-$, thus the mixing of
the states should be taken into account.  
Without knowledge of the mixing, the non-perturbative correction to
the energy levels cannot be uniquely determined.\footnote{In the case
of ${\cal N}$=0, ${\bf Z}_2$ symmetry exists in the system, which makes
it possible to determine the mixing.  
This is why we can obtain 
Eqs.(\ref{ge-double;eqn}) and (\ref{gw-double;eqn}).}

We, however, find that the result (\ref{znp-ibari2;eqn}) is consistent with
the result (\ref{eqn:dege}) if the non-perturbative
contribution of the perturbative ground state at $q=q_+$ satisfies 
the following relations:
\begin{eqnarray}
&& E_{\varsigma,\rm NP}(\epsilon, N_+=0) =
\varsigma \alpha 
\sqrt{\frac{1}{{\cal N}!}\left(\frac{2}{g^2}\right)^{\cal N}}
\nonumber\\
&& \qquad + \frac{\alpha^2}{2}
\frac{1}{{\cal N}!}\left(\frac{2}{g^2}\right)^{\cal N}
\left[2 \ln\left(-\frac{2}{g^2}\right)
-\psi(1)-\psi({\cal N}+1)
\right],
\label{enpn+02;eqn}
\end{eqnarray}
\begin{eqnarray}
&&|\Psi_{\varsigma}(q_+)|^2_{\rm NP}=\varsigma\alpha
\frac{|\Psi(0)|^2}{4}
\sqrt{\frac{1}{{\cal N}!}\left(\frac{2}{g^2}\right)^{\cal N}}
(\psi(1)-\psi({\cal N}+1))\nonumber\\
&&\quad+\alpha^2\frac{|\Psi(0)|^2}{2}\left(\frac{2}{g^2}\right)^{\cal N}
\frac{1}{{\cal N}!}
\left[-\ln\left(-\frac{2}{g^2}\right)\psi({\cal N}+1)
-\sum_{k=1}^{{\cal N}}\frac{\gamma}{k}+\sum_{k\ge l}^{{\cal N}}
\frac{1}{kl}\right.
\nonumber\\
&&\quad\left.
+\frac{\Gamma''(1)}{2}+\frac{1}{2}\left\{\ln\left(-\frac{2}{g^2}\right)
\right\}^2 \right]
\label{wfnp+02;eqn}
\end{eqnarray}
where $\varsigma=\pm 1$.
We also note that the above non-perturbative corrections for the energy and
the wave function are reproduced by the following 2$\times$2 
effective Hamiltonian:
\begin{eqnarray}
H_{\rm NP}
=
\left(
\begin{array}{cc}
W &Y \\
Y&X
\end{array}
\right),
\end{eqnarray}
where
\begin{eqnarray}
W&=&\alpha^2\frac{1}{{\cal N}!}\left(\frac{2}{g^2}\right)^{\cal N}
\left\{\ln\left(-\frac{2}{g^2}\right)-\psi({\cal N}+1)\right\},
\\
X&=&\alpha^2\frac{1}{{\cal N}!}\left(\frac{2}{g^2}\right)^{\cal N}
\left\{\ln\left(-\frac{2}{g^2}\right)-\psi(1)\right\},
\\
Y&=&\alpha\sqrt{\frac{1}{{\cal N}!}\left(\frac{2}{g^2}\right)^{\cal N}}.
\end{eqnarray}
The (1,1) component $W$ is interpreted as the matrix
element between the perturbative ground state at $q=q_+$ and itself, 
and the (2,2) component $X$ as that between the perturbative ${\cal N}$-th 
excited state at $q=q_-$and itself.   
The off-diagonal component $Y$ is the matrix element between the ground
state and the ${\cal N}$-th excited state. 
This result supports Eqs.(\ref{enpn+02;eqn}) and (\ref{wfnp+02;eqn}).

Finally, we would like to comment on the
non-perturbative corrections for the normalized wave function of the
ground state, Eqs.(\ref{gw-double;eqn}), (\ref{wfnp+01;eqn}),
(\ref{wfnp-0;eqn}) and (\ref{wfnp+02;eqn}). 
These corrections cannot be evaluated by the calculation in Section
\ref{multi;sec}, where we have considered
$Z=\lim_{T\rightarrow\infty}{\rm Tr}(e^{-HT})$,
since no information concerning the wave function
is contained in this partition function.
(See Eq.$(\ref{eqn:sumsum})$.)
According to the argument of Section \ref{dis;sec}, we can predict that the
perturbation series in $g^2$ for the normalized wave function of the ground
state should be non-Borel-summable.

\section{Large order behavior of the perturbative series} 
\label{lob;sec}

The expressions of the non-perturbative contribution to the energy
levels (\ref{eqn:en+}) and (\ref{eqn:dege})  
contain imaginary parts starting at the order $\alpha^2$, 
and are given by the following for the non-degenerate case:
\begin{eqnarray}
{\rm Im}\left(E^{(\pm)}_{\rm NP}(\epsilon, N)\right)=
-\alpha^2\frac{1}{N !}
\left(\frac{2}{g^2}\right)^{\pm\epsilon + 2N} \hspace*{-30pt}
\frac{\pi}{\Gamma(1 \pm\epsilon+N)},
\nonumber\\
\label{eqn:ime-n}
\end{eqnarray}
where we have omitted the subscripts $\pm$ of $N$ for brevity. 
For the degenerate case $\epsilon = {\cal N}$, we have
\begin{equation}
{\rm Im} \left(E_{\varsigma, \rm NP}({\cal N}, N) \right)=
-\alpha^2\frac{\pi}{N! (N+{\cal N})!}
\left(\frac{2}{g^2}\right)^{2 N + {\cal N}}.
\label{eqn:ime-d}
\end{equation}
We note that the result (\ref{eqn:ime-n}) reduces to (\ref{eqn:ime-d})
for $\epsilon = {\cal N}$:
\begin{equation}
{\rm Im}\left(E^{(+)}_{\rm NP}({\cal N}, N)\right) =
{\rm Im}\left(E^{(-)}_{\rm NP}({\cal N}, N+{\cal N})\right)=
{\rm Im} \left(E_{\varsigma, \rm NP}({\cal N}, N) \right).
\end{equation} 
The imaginary part of the energy is continuous in $\epsilon$ at
$\epsilon = {\cal N}$.

We recall the relation between the imaginary part of the non-perturbative 
contribution and the large order behaviour
of the perturbative series explained in Section 3.4.
The following relation holds:
\begin{equation}
E_{\rm P}^{(\pm)}(\epsilon,N,m) = - \frac{1}{\pi}\int_0^\infty dg^2
\frac{{\rm Im} (E_{\rm NP}^{(\pm)}(\epsilon, N))}{g^{2m+2}}  \label{eqn:disp}
\end{equation}
for the perturbative coefficients of the energy levels
defined by
\begin{equation}
E_{\rm P}^{(\pm)}(\epsilon,N) = 
\sum_{m=0}^\infty E_{\rm P}^{(\pm)}(\epsilon,N,m) g^{2m}.
\label{EPm}
\end{equation}
Substituting the result (\ref{eqn:ime-n}) into Eq.(\ref{eqn:disp}), we
obtain the following for the perturbative coefficient 
$E_{\rm P}^{(\pm)}(\epsilon,N,m)$, 
\begin{eqnarray}
E_{\rm P}^{(\pm)}(\epsilon,N,m) &=& A^{(\pm)}
(\epsilon, N) 3^m \Gamma(\pm \, \epsilon + 2N + m +1) 
\nonumber\\
&& \times \left[1 + O\left( {1 \over m}\right)\right],
\label{eqn:eaep}
\end{eqnarray}
where the coefficient $A^{(\pm)}(\epsilon, N)$ is defined as follows,
\begin{equation}
A^{(\pm)}(\epsilon, N) \equiv
-{3 \over \pi} {6^{\pm\epsilon + 2N} \over 
N! \, \Gamma(\pm \, \epsilon + 1 + N)}.
\label{eqn:aep}
\end{equation}
We note that the higher order corrections of $O(\epsilon g^2)$ 
and $O(g^2)$
that we have ignored in Eq.(\ref{eqn:ime-n}) contribute
only to the $O(1/m)$ correction,
or more specifically the $O(1/(\pm \epsilon + 2N + m))$ correction
in the result (\ref{eqn:eaep}).\footnote{An attempt
to calculate this type of correction is given in Ref.\cite{falsil}.}

The expression for $E_{\rm P}^{(-)}(\epsilon, 1)$
coincides with the expression obtained in \break
Ref.\cite{VWW,VWW2}.
In order to confirm our prediction, we 
have independently carried out the numerical and exact 
calculation of the perturbative coefficients 
$E_{\rm P}^{(\pm)}(\epsilon, N, m)$ 
for much wider parameter ranges and to much higher 
orders, using the methods described in Ref.\cite{Zin2,VWW,VWW2}.
Specifically, the method given 
in Ref.\cite{VWW} works for any excited state by replacing the
zeroth order wave function by the $n$-th order 
harmonic oscillator wave function, and is fastest in 
{\it exact} computer calculations. 
We have calculated the following five categories of the perturbative 
coefficients to the 200-th order:

\begin{itemize}
\item[(a)] $N=0$ ($-$) level (the perturbative ground state at $q = q_-$) for  
$\epsilon = 0$
to 10 with $\Delta\epsilon=0.2$ interval. 
Floating point calculation.
\item[(b)] $N=0$ ($+$) level (the perturbative ground state at $q = q_+$)
for  $\epsilon = 0$ to 20 
with $\Delta\epsilon=0.2$. Floating point calculation.
\item[(c)] $N=1, 2, 3$ ($-$) levels for  $\epsilon = 0$
to 10.5 with $\Delta\epsilon=0.5$. Exact calculation, 
\item[(d)] $N=1, 2, 3$ ($+$) levels for  $\epsilon = 0$ to 20 
with $\Delta\epsilon=1.5$. Exact calculation,
\item[(e)] $N=4$, 5, 6 ($-$) levels for $\epsilon=2.5$. 
Floating point calculation.
\end{itemize}
In the above, ``exact'' calculation indicates an algebraic
calculation carried out in Mathematica.

\begin{figure}
\centerline{\epsfxsize=10cm\epsfbox{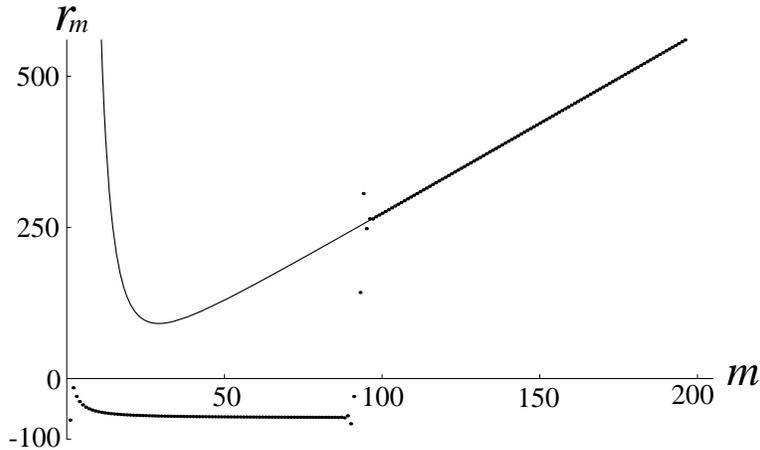}}
\caption{The plot of the ratio of successive perturbative 
coefficients $r_m$ (dots) and the fitted curve (solid line)
for the ground state at  $\epsilon=9.8$.
The fitting was done for $m= 150 \sim 200$.}
\label{fig:rm}
\end{figure}

We can compare these results with Eqs.(\ref{eqn:eaep}) 
and (\ref{eqn:aep}) with the following procedures.
First, in order to check the leading $m$-dependent terms 
in Eq.(\ref{eqn:eaep}), we take their ratio.
\begin{equation}
{E_{\rm P}^{(\pm)}(\epsilon, N, m) \over E_{\rm P}^{(\pm)}(\epsilon, N,m-1)}
=3(\pm \, \epsilon + 2 N + m).
\label{eq:predict}
\end{equation}
We use the result of the corresponding perturbative coefficients 
${\hat E}_m$ for $m=150$ to 200, and fit them as follows:
\begin{equation}
r_m = \frac{{\hat E}_m}{{\hat E}_{m-1}} = a_0 + a_1 m + \frac{a_2}{m} 
+ \frac{a_3}{m^2} + \frac{a_4}{m^3},
\label{eq:checka}
\end{equation}
for all the categories (a)--(e).
A plot of $r_m$ for the ground state for $\epsilon=9.8$
is given in Fig.\ref{fig:rm} as a sample.
The fit works to good accuracy for values from $m=150$ to $200$ as can be 
seen in this figure.  The same is true for all the cases in the above five
categories (a)--(e) we have examined so far.
The resulting coefficients are compared with their theoretical
predictions:
\begin{equation}
a_0 = 3(\pm \epsilon + 2N), \quad a_1=3.
\label{eqn:a01}
\end{equation}
In all the cases, the theoretical prediction (\ref{eqn:a01}) 
is confirmed with good accuracy, the maximum error being 2 \%.

\begin{figure}
\centerline{\epsfxsize=12cm\epsfbox{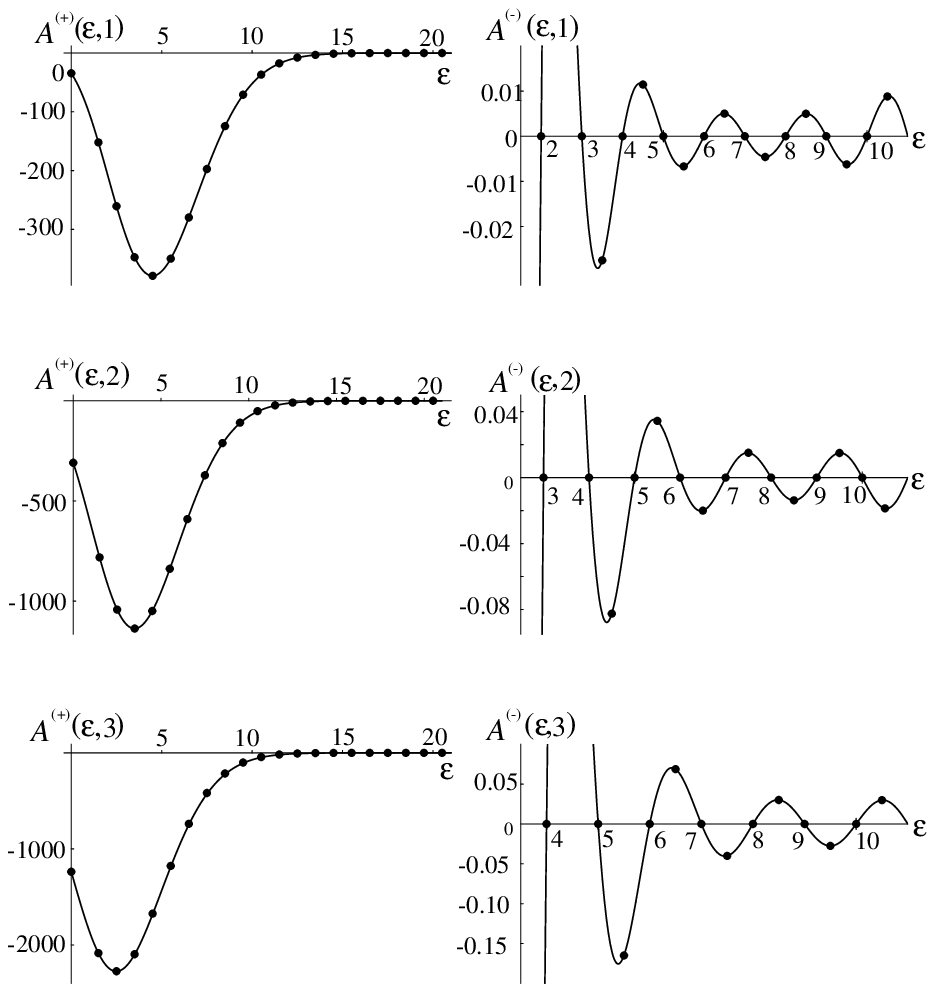}}
\caption{The comparison of the fitting to the perturbative coefficients 
(indicated by dots) and
the theoretical prediction (solid line) of
$A^{(\pm)}(\epsilon, N)$ for $N=1 \sim 3$.}
\label{fig:nall}
\end{figure}

Next, we examine $A^{(\pm)}(\epsilon, N)$ defined by 
Eq.(\ref{eqn:eaep}).
For this purpose we use the following quantity $b_m$:
\begin{equation}
b_m = (\pm\epsilon+2N+m+2) d_{m+1} - (\pm\epsilon+N+m+1) d_m,
\end{equation}
where
\begin{equation}
d_m \equiv \frac{{\hat E}_m}{3^m \Gamma(\pm \, \epsilon + 2N +m)}.
\label{eq:checkd}
\end{equation}
This quantity is defined so that it is free from
the correction term of order $1/(\epsilon + 2N + m+1)$ to ${\hat E}_m$. 
(This is not particularly crucial, since the order of our calculation is
high.)
Thus the numerical fit can be performed with the following formula:
\begin{equation}
b_m = c_0 + \frac{c_2}{m^2} +  \cdots + \frac{c_8}{m^8},
\end{equation}
and is compared with the theoretical prediction:
\begin{equation}
c_0 = A^{(\pm)}(\epsilon, N).
\end{equation}
The result for case (a) is plotted in Fig.\ref{fig:n0-p}.
The difference between Eq.(\ref{eqn:aep}) and the calculated value 
is at most 0.1 \%, which happens at $\epsilon=9.8$.  
For case (b), plotted in Fig.\ref{fig:aeps}, the error 
is at most 15 \% (at $\epsilon=20$).
Cases (c) and (d) are plotted in Fig.\ref{fig:nall}.
The maximum error is 65 \% for the case of $A^{(+)} (20.5, 3)$.
Since this error is somewhat large, we have carried out
a yet higher order calculation for this case.
The error decreases to  32 \% at the 364-th order,
and to 17 \% at the  478-th order, in both cases using
the last 50 coefficients.
This consistent decrease of the error for increasing $m$ is 
in excellent agreement with our large order predictions.
For the case (e) the maximum error is  0.15 \%.

\begin{figure}
\centerline{\epsfxsize=10cm\epsfbox{aepsilon.eps}}
\caption{The comparison of the fitting to the perturbative coefficients 
(indicated by dots) and
the theoretical prediction (solid line) of $A^{(-)}(\epsilon, 0)$.}
\label{fig:n0-p}
\centerline{\epsfxsize=10cm\epsfbox{n0-p-fig.eps}}
\caption{The comparison of the fitting to the perturbative coefficients 
(indicated by dots) and the theoretical prediction (solid line) of 
$A^{(+)}(\epsilon, 0)$.}
\label{fig:aeps}
\end{figure}

The reader may note that  $A^{(-)}(\epsilon, 0)$ is zero for 
any positive integer $\epsilon$ as seen in Fig.\ref{fig:n0-p}.
Similar zeros in $A^{(-)}(\epsilon, N)$ are also seen in
Fig.\ref{fig:nall}.
This is due to a new type of supersymmetry for these values
of $\epsilon$.  We will elaborate upon this in the next section.

\section{${\cal N}$-fold supersymmetry}
\label{nsusy;sec}

In the previous section, we have seen that
the Borel singularity of perturbation theory apparently 
disappears for the first ${\cal N}$ excited states at $q=q_-$ for
$\epsilon={\cal N}$ (${\cal N}=1,2,3,\cdots$).
Hereafter we will call these states isolated states, since these
states do not have degenerate partners at $q=q_+$ (see Fig.\ref{fig:nfold}).
In Table \ref{tb:1}, we have listed the perturbative coefficients
for several low-lying states for \({\cal N} = 1\).
The perturbative correction for the ground state is not only Borel
summable but in fact identically zero at every order.
We also notice that the perturbation calculation
cannot lift the degeneracy that exists at the zeroth order in \(g\):
\(E^{(-)}_{\rm P}(1, N, m)\) and \(E^{(+)}_{\rm P}(1, N-1, m)\) are
identical  for  all \(m\).
Interestingly  similar features prevail in the results of the other
values of \({\cal N}\).
The results for \({\cal N} = 2, 3 \) show similar characteristics as
seen in Table \ref{tb:2} and \ref{tb:3}.

\begin{table}
\begin{center}
\hspace*{-10mm}
\small
\renewcommand{\arraystretch}{1.5}
\begin{tabular}{|c||c|c|c|c|c|}
\hline
 m & $E^{(-)}_{\rm P}(0)$ & $E^{(-)}_{\rm P}(1), E^{(+)}_{\rm P}(0)$ 
& $E^{(-)}_{\rm P}(2), E^{(+)}_{\rm P}(1)$
& $E^{(-)}_{\rm P}(3), E^{(+)}_{\rm P}(2)$
& $E^{(-)}_{\rm P}(4), E^{(+)}_{\rm P}(3)$
\\  \hline  \hline
 0&$0.5$&$1.5$&$2.5$&$3.5$&$4.5$\\  \hline 
 1&$0$&$-3$&$-1.2\times10^{1}$&$-2.7\times10^{1}$&$-4.8\times10^{1}$\\  \hline 
 2&$0$&$-1.95\times10^{1}$&$-1.41\times10^{2}$&$-4.665\times10^{2}$
&$-1.098\times10^{3}$\\  \hline 
 3&$0$&$-2.70\times10^{2}$&$-3.330\times10^{3}$&$-1.5930\times10^{4}$
&$-4.9320\times10^{4}$\\ \hline 
 4&$0$&$-5.1791\times10^{3}$&$-1.0474\times10^{5}$&$-7.1373\times10^{5}$
&$-2.8885\times10^{6}$\\ \hline
 5&$0$&$-1.2110\times10^{5}$&$-3.8959\times10^{6}$&$-3.7318\times10^{7}$
&$-1.9624\times10^{8}$\\ \hline 
 6&$0$&$-3.2594\times10^{6}$&$-1.6258\times10^{8}$&$-2.1630\times10^{9}$
&$-1.4695\times10^{10}$\\ \hline 
 7&$0$&$-9.7888\times10^{7}$&$-7.3976\times10^{9}$&$-1.3523\times10^{11}$
&$-1.1804\times10^{12}$\\ \hline 
 8&$0$&$-3.2201\times10^{9}$&$-3.6075\times10^{11}$&$-8.9708\times10^{12}$
&$-1.0008\times10^{14}$\\ \hline 
 9&$0$&$-1.1464\times10^{11}$&$-1.8643\times10^{13}$&$-6.2466\times10^{14}$
&$-8.8621\times10^{15}$\\ \hline 
 10&$0$&$-4.3816\times10^{12}$&$-1.0131\times10^{15}$&$-4.5323\times10^{16}$
&$-8.1376\times10^{17}$\\ \hline 
 11&$0$&$-1.7875\times10^{14}$&$-5.7580\times10^{16}$&$-3.4085\times10^{18}$
&$-7.7085\times10^{19}$\\ \hline 
 12&$0$&$-7.7506\times10^{15}$&$-3.4088\times10^{18}$&$-2.6465\times10^{20}$
&$-7.5041\times10^{21}$\\ \hline 
 13&$0$&$-3.5604\times10^{17}$&$-2.0961\times10^{20}$&$-2.1154\times10^{22}$
&$-7.4855\times10^{23}$\\ \hline 
 14&$0$&$-1.7282\times10^{19}$&$-1.3358\times10^{22}$&$-1.7367\times10^{24}$
&$-7.6343\times10^{25}$\\ \hline 
 15&$0$&$-8.8443\times10^{20}$&$-8.8090\times10^{23}$&$-1.4620\times10^{26}$
&$-7.9464\times10^{27}$\\ \hline 
 16&$0$&$-4.7629\times10^{22}$&$-6.0040\times10^{25}$&$-1.2602\times10^{28}$
&$-8.4300\times10^{29}$\\ \hline 
 17&$0$&$-2.6943\times10^{24}$&$-4.2259\times10^{27}$&$-1.1113\times10^{30}$
&$-9.1043\times10^{31}$\\ \hline 
 18&$0$&$-1.5983\times10^{26}$&$-3.0695\times10^{29}$&$-1.0016\times10^{32}$
&$-1.0001\times10^{34}$\\ \hline 
 19&$0$&$-9.9264\times10^{27}$&$-2.2997\times10^{31}$&$-9.2230\times10^{33}$
&$-1.1167\times10^{36}$\\ \hline 
 20&$0$&$-6.4447\times10^{29}$&$-1.7765\times10^{33}$&$-8.6724\times10^{35}$
&$-1.2667\times10^{38}$\\ \hline 
\end{tabular}
\end{center}
\caption{The perturbative coefficients of the energy levels 
$E^{(\pm)}_{\rm P}(1, N, m)$ defined in Eq.(\ref{EPm}).
In the first row $E^{(\pm)}_{\rm P}(N)$ is a shorthand denotation
for  $E^{(\pm)}_{\rm P}(1, N, m)$.}
\label{tb:1}
\end{table}
\begin{table}
\renewcommand{\arraystretch}{1.5}
\begin{center}
\hspace*{-2pt}
\small
\begin{tabular}{|c||c|c|c|c|c|c|}
\hline
 m & $E^{(-)}_{\rm P}(0)$& $E^{(-)}_{\rm P}(1)$& 
$E^{(-)}_{\rm P}(2), E^{(+)}_{\rm P}(0)$& $E^{(-)}_{\rm P}(3),
E^{(+)}_{\rm P}(1)$
& $E^{(-)}_{\rm P}(4), E^{(+)}_{\rm P}(2)$ \\  \hline \hline
 0&$0.5$&$1.5$&$2.5$&$3.5$&$4.5$\\  \hline 
 1&$0$&$0$&$-6$&$-1.8\times10^{1}$&$-3.6\times10^{1}$\\  \hline 
 2&$0$&$0$&$-5.1\times10^{1}$&$-2.55\times10^{2}$&$-7.14\times10^{2}$\\
  \hline 
 3&$0$&$0$&$-9.09\times10^{2}$&$-7.227\times10^{3}$&$-2.7954\times10^{4}$\\
  \hline 
 4&$0$&$0$&$-2.2136\times10^{4}$&$-2.7102\times10^{5}$&$-1.4323\times10^{6}$\\
  \hline
 5&$0$&$0$&$-6.4877\times10^{5}$&$-1.1942\times10^{7}$&$-8.5387\times10^{7}$\\
  \hline 
 6&$0$&$0$&$-2.1620\times10^{7}$&$-5.8676\times10^{8}$&$-5.6267\times10^{9}$\\
  \hline 
 7&$0$&$0$&$-7.9445\times10^{8}$&$-3.1262\times10^{10}$
&$-3.9879\times10^{11}$\\  \hline 
 8&$0$&$0$&$-3.1605\times10^{10}$&$-1.7757\times10^{12}$
&$-2.9907\times10^{13}$\\  \hline 
 9&$0$&$0$&$-1.3451\times10^{12}$&$-1.0635\times10^{14}$
&$-2.3481\times10^{15}$\\  \hline 
 10&$0$&$0$&$-6.0762\times10^{13}$&$-6.6658\times10^{15}$
&$-1.9161\times10^{17}$\\  \hline 
 11&$0$&$0$&$-2.8970\times10^{15}$&$-4.3483\times10^{17}$
&$-1.6165\times10^{19}$\\  \hline 
 12&$0$&$0$&$-1.4520\times10^{17}$&$-2.9405\times10^{19}$
&$-1.4046\times10^{21}$\\  \hline 
 13&$0$&$0$&$-7.6282\times10^{18}$&$-2.0554\times10^{21}$
&$-1.2533\times10^{23}$\\  \hline 
 14&$0$&$0$&$-4.1917\times10^{20}$&$-1.4817\times10^{23}$
&$-1.1458\times10^{25}$\\  \hline 
 15&$0$&$0$&$-2.4051\times10^{22}$&$-1.0998\times10^{25}$
&$-1.0714\times10^{27}$\\  \hline 
 16&$0$&$0$&$-1.4391\times10^{24}$&$-8.3939\times10^{26}$
&$-1.0233\times10^{29}$\\  \hline 
 17&$0$&$0$&$-8.9697\times10^{25}$&$-6.5820\times10^{28}$
&$-9.9710\times10^{30}$\\  \hline 
 18&$0$&$0$&$-5.8182\times10^{27}$&$-5.2988\times10^{30}$
&$-9.9051\times10^{32}$\\  \hline 
 19&$0$&$0$&$-3.9242\times10^{29}$&$-4.3774\times10^{32}$
&$-1.0025\times10^{35}$\\  \hline 
 20&$0$&$0$&$-2.7497\times10^{31}$&$-3.7096\times10^{34}$
&$-1.0331\times10^{37}$\\  \hline 
\end{tabular}
\end{center}
\caption{The perturbative coefficients of the energy levels 
$E^{(\pm)}_{\rm P}(2, N, m)$.}
\label{tb:2}
\end{table}
\begin{table}
\begin{center}
\hspace*{-10mm}
\small
\renewcommand{\arraystretch}{1.5}
\begin{tabular}{|c||c|c|c|c|c|c|}
\hline
 m & $E^{(-)}_{\rm P}(0)$& $E^{(-)}_{\rm P}(1)$& $E^{(-)}_{\rm P}(2)$ 
& $E^{(-)}_{\rm P}(3), E^{(+)}_{\rm P}(0)$ 
& $E^{(-)}_{\rm P}(4), E^{(+)}_{\rm P}(1)$ \\ \hline \hline
 0&$0.5$&$1.5$&$2.5$&$3.5$&$4.5$\\ \hline 
 1&$-1$&$2$&$-1$&$-1.0\times10^1$&$-2.5\times10^{1}$\\  \hline 
 2&$1.5$&$0$&$-1.5$&$-1.05\times10^{2}$&$-4.125\times10^{2}$\\  \hline 
 3&$-4$&$8$&$-4$&$-2.290\times10^{3}$&$-1.36\times10^{4}$\\  \hline 
 4&$1.3125\times10^1$&$0$&$-1.3125\times10^1$&$-6.7676\times10^{4}$
&$-5.9131\times10^{5}$\\  \hline
 5&$4.8\times10^1$&$9.6\times10^{1}$&$-4.8\times10^1$
&$-2.3888\times10^{6}$&$-3.009\times10^{7}$\\  \hline 
 6&$1.8769\times10^2$&$0$&$-1.8769\times10^2$&$-9.5181\times10^{7}$
&$-1.7008\times10^{9}$\\  \hline 
 7&$-7.68\times10^2$&$1.536\times10^{3}$&$-7.68\times10^2$
&$-4.1525\times10^{9}$&$-1.0385\times10^{11}$\\  \hline 
 8&$3.2477\times10^3$&$0$&$-3.2477\times10^3$&$-1.9476\times10^{11}$
&$-6.7357\times10^{12}$\\  \hline 
 9&$-1.4080\times10^4$&$2.8160\times10^{4}$&$-1.4080\times10^4$
&$-9.7049\times10^{12}$&$-4.5906\times10^{14}$\\  \hline 
 10&$6.2247\times10^4$&$0$&$-6.2247\times10^4$&$-5.0965\times10^{14}$
&$-3.263\times10^{16}$\\  \hline 
 11&$-2.7955\times10^5$&$5.5910\times10^{5}$&$-2.7955\times10^5$
&$-2.8048\times10^{16}$&$-2.4059\times10^{18}$\\  \hline 
 12&$1.2718\times10^6$&$0$&$-1.2718\times10^6$&$-1.6112\times10^{18}$
&$-1.8330\times10^{20}$\\  \hline 
 13&$-5.8491\times10^6$&$1.1698\times10^{7}$&$-5.8491\times10^6$
&$-9.6315\times10^{19}$&$-1.4386\times10^{22}$\\  \hline 
 14&$2.7148\times10^7$&$0$&$-2.7148\times10^7$&$-5.9791\times10^{21}$
&$-1.1606\times10^{24}$\\  \hline 
 15&$-1.2701\times10^8$&$2.5402\times10^{8}$&$-1.2701\times10^8$
&$-3.8483\times10^{23}$&$-9.6071\times10^{25}$\\  \hline 
 16&$5.9828\times10^8$&$0$&$-5.9828\times10^8$&$-2.5649\times10^{25}$
&$-8.1491\times10^{27}$\\  \hline 
 17&$-2.8353\times10^9$&$5.6706\times10^{9}$&$-2.8353\times10^9$
&$-1.7687\times10^{27}$&$-7.0762\times10^{29}$\\  \hline 
 18&$1.3508\times10^{10}$&$0$&$-1.3508\times10^{10}$&$-1.2610\times10^{29}$
&$-6.2853\times10^{31}$\\  \hline 
 19&$-6.4664\times10^{10}$&$1.2933\times10^{11}$&$-6.4664\times10^{10}$
&$-9.2892\times10^{30}$&$-5.7075\times10^{33}$\\  \hline 
 20&$3.1087\times10^{11}$&$0$&$-3.1087\times10^{11}$&$-7.0678\times10^{32}$
&$-5.2964\times10^{35}$\\  \hline 
\end{tabular}
\end{center}
\caption{The perturbative coefficients of the energy levels 
$E^{(\pm)}_{\rm P}(3, N, m)$.}
\label{tb:3}
\end{table}

The degeneracy reminds us of the non-renormalization theorem
of supersymmetric theories.
This holds that quantum correction at any finite order of perturbation
theory cannot 
induce supersymmetry breaking if it is unbroken at the tree level,
and the present model can actually be cast into a supersymmetric
quantum mechanics  at \({\cal N} = 1\) \cite{Witten}.
The degeneracy between the perturbative coefficients,
\(E^{(-)}_{\rm P}(1, N, m)\) and \(E^{(+)}_{\rm P}(1, N-1, m)\),
is then understood as that between
the bosonic and fermionic levels of the perturbatively unbroken supersymmetry.
In this section we will show that this explanation also applies 
to other  \(\cal N\). 
We will reveal a new symmetry of the model
that reduces to the ordinary supersymmetry at \({\cal N} = 1\) and can
be understood as one of its possible extensions.
We dub it \(\cal N\)-fold supersymmetry.

Considerations based on this  symmetry will further reveal that the
perturbative series of the isolated levels is convergent with a finite
convergence radius. 
The perturbative series of these levels
has a much stronger property than expected by the lack of the Borel
singularity. 
To begin with, we briefly review the supersymmetric quantum 
mechanics, focusing upon the specific realization of the
non-renormalization theorem. 

\subsection{Supersymmetric quantum mechanics}
We will start by clarifying the supersymmetry that underlies our model
at \({\cal N} = 1\).
To see this, let us define the supercharges \cite{Witten} by
\begin{equation}
 Q^{\dagger} \equiv D \psi^{\dagger}, \quad  Q= D^\dagger \psi,\label{QQdag}
\end{equation}
where
\begin{equation}
D =  p - i W(q), \quad  D^\dagger = p + i W(q), \label{Ddag}
\end{equation}
\(p = -i (d/dq) \) is the canonical momentum of \(q\),
and \(W(q)\) is defined by 
\begin{equation}
 W(q) = q( 1 -gq ).
\end{equation}
In the definitions (\ref{QQdag}), $\psi$ and $\psi^{\dagger}$ are the  
annihilation and creation operators for a fermionic degree of freedom.
They satisfy 
\( \left\{ \psi^{\dagger}, \psi \right\} = 1\) and \( \psi^{\dagger 2} =
\psi^2 = 0\). 
As a result, \(Q^2\) and \(Q^\dagger{}^2\) vanish:
\begin{equation}
Q^2 =  Q^\dagger{}^2 = 0.
\end{equation}
The supersymmetric Hamiltonian \( {\bf H}\) is given by 
\begin{equation}
{\bf H} = {1\over 2} \left\{ Q^{\dagger}, Q \right\}  = 
{1\over 2} \left( p^2 + W^2(q) \right) + W'(q) 
\left( \psi^{\dagger} \psi - {1\over 2} \right).
\label{eq:susyH}
\end{equation}
All the states are labeled by the quantum number of the fermion number
operator $N_{f}=\psi^{\dagger}\psi$. 
The operator $Q^{\dagger}$ transforms a bosonic state $|{\rm b}\rangle$
(with $N_f=0$) to a fermionic one $|{\rm f}\rangle$ (with $N_f=1$), and
$Q$ does vice versa.  
By construction,  \(Q\) and \(Q^\dagger\) commute with \({\bf H}\) 
\begin{equation}
\left[ {\bf H}, Q \right] = \left[ {\bf H}, Q^\dagger \right] = 0. 
\label{susy}
\end{equation}
Therefore, the Hamiltonian is invariant under the transformations
generated by the supercharges. 

The relation between this supersymmetric Hamiltonian and our
Hamiltonian
becomes obvious if we introduce a matrix representation of $\psi$ 
\begin{eqnarray}
\psi^{\dagger}
=\left(
\begin{array}{cc}
0&1 \\
0&0
\end{array}
\right),\quad
\psi
=\left(
\begin{array}{cc}
0&0 \\
1&0
\end{array}
\right).
\end{eqnarray}
In this representation, the bosonic state $|{\rm b}\rangle$ and 
the fermionic one $|{\rm f}\rangle$ are written as 
\begin{eqnarray}
|{\rm b}\rangle=
\left(
\begin{array}{c}
0 \\
\Psi^{(-)}
\end{array}
\right),
\quad
|{\rm f}\rangle=
\left(
\begin{array}{c}
\Psi^{(+)}\\
0
\end{array}
\right),
\end{eqnarray}
and the supersymmetric Hamiltonian is decomposed into two components:
\begin{eqnarray}
{\bf H}=
\left(
\begin{array}{cc}
H_+&0 \\
0&H_-
\end{array}
\right).                                                                 
\end{eqnarray}
Here we denote the fermionic component by $H_+$ and the bosonic by $H_-$
which are given by
\begin{equation}
H_- = {1\over 2} D^{\dagger}D,  \quad H_+ = {1\over 2} D D^{\dagger},
\end{equation}
or more explicitly
\begin{equation}
H_\pm = -{ 1\over 2} {d^2 \over dq^2} + {1\over 2}q^2 ( 1 - gq )^2 
\mp  ( g q - {1\over 2}).
\end{equation}
One can readily check   \(H_+  - 1/2 \) is indeed the Hamiltonian of 
our model with \({\cal N} = 1\), i.e., the one derived from the 
action (\ref{eqn:action})-(\ref{eqn:pot}) with \(\epsilon = 1\). 
The  Hamiltonian  \(H_- - 1/2 \) becomes \(H_+  - 1/2 \)
as a result of the mirror transformation with respect 
to the \(q = 1/2g\) axis.
Therefore, the dynamics of our model with
${\cal N}=1$ are same as that of the supersymmetric Hamiltonian ${\bf H}$.  

\begin{figure}
\centerline{\epsfxsize=10cm\epsfbox{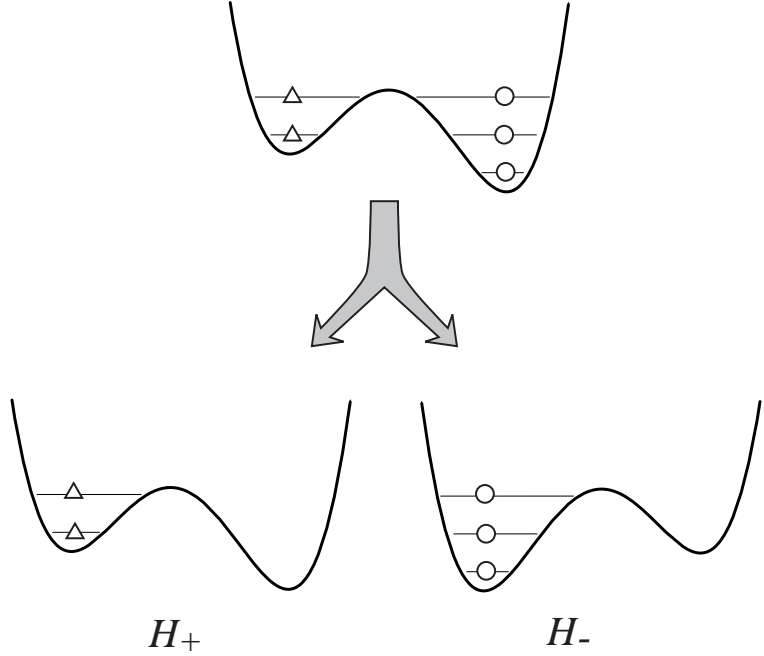}}
\caption{The correspondence between the states in the 
left and right wells of the potential (\ref{eqn:pot}) with
$\epsilon=1$ and 
the states in the left wells of $H_\pm - 1/2 $.}
\label{fig:crspnd}
\end{figure}

In considering  the eigenvalues  of \({\bf H}\),
we will work on the decomposed form of the equation, i.e., 
\begin{equation}
H_\pm \Psi^{(\pm)} = E^{(\pm)} \Psi^{(\pm)}. \label{eq:eig}
\end{equation}
The perturbative energy levels \(E_{\rm P}^{(\pm)}\) calculated in the
previous section
were those of the well-local states in the left well of \(H_+ - 1/2 \) 
and of the states in the right well.
It is convenient to identify the latter states  with the well-local ones in
the left well at \(q \sim 0\) of \(H_- - 1/2 \) 
(see Fig.\ref{fig:crspnd}).
These states correspond to ``bosonic" states, while the well-local states
in  the left well of \(H_+ - 1/2 \) correspond to ``fermionic" states.

At \( g=0\), \({\bf H}\) reduces to that of the supersymmetric 
harmonic oscillator.
The eigenvalues \( E^{(\pm)} \) and eigenfunctions \( \Psi^{(\pm)}\) can
be solved immediately. 
They are labeled by an integer \(N (=  0, 1, 2, ... ) \) representing
the level of excitation. 
The solution for \(\Psi^{(\pm)}\) is 
\begin{equation}
\Psi_N^{(\pm)}(q) =  H_N(q) e^{- q^2 /2}
\end{equation}
by the use of the \(N\)-th Hermite polynomial
$H_N(q)\equiv(-1)^Ne^{q^2}\frac{d^N}{dq^N}e^{-q^2}$;  
the corresponding eigenvalues are 
\(E_N^{(- )} = N \) and \(E_N^{(+)} = N + 1\).
The ground state \(\Psi_0^{(-)}\) is of zero-point energy 
and the supersymmetry is not broken.
The degeneracy between the fermionic and bosonic spectra due
to the unbroken supersymmetry is manifest 
in the pairs of \(E_{N+1}^{(-)}\) and \( E_{N}^{(+)} \).

Now let us examine a case where $g$ is a nonzero quantity.
When one applies  perturbation theory to
Eq.(\ref{eq:eig}) by expanding the wave functions and
the eigenvalues in \(g\), 
one gets 
\begin{equation}
E_N^{(\pm)} = - {1\over 2} + \sum_{m = 0}^\infty E_{\rm P}^{(\pm)}
(1, N, m) g^{2m},
\end{equation}
where  \(E_{\rm P}^{(\pm)}(1, N, m)\) is exactly what we have calculated in
the previous section, Eq.(\ref{EPm}).
If we recall 
the identification of the state in the right well of \(H^{(+)}\)
and the one in the left of \(H^{(-)}\),
it will be seen that
the perturbative coefficients of \(E_N^{(-)}\) are given by those of
\(E_{\rm P}^{(-)}(1, N) \) calculated in the previous section.
The disappearance of the correction to
\(E_0^{(-)}\) and the persistence of the degeneracy 
between  \(E_{N+1}^{(-)}\) and
\(E_N^{(+)}\) are now  seen in the context of the unbroken supersymmetry 
of \({\bf H}\) in perturbation.

A direct demonstration of the non-renormalization theorem 
can be obtained for the present case,
by considering the function obtained from \( D \Psi_{\rm G} = 0\),
\begin{equation}
\Psi_{\rm G}(q) = \exp \left( - \int_0^q dq' W(q') \right) = 
\exp \left( - {q^2 \over 2} + g{q^3\over 3}\right).
\end{equation}
Since this function starts with \(\Psi_0^{(-)}(q)\) 
in the expansion in \(g\)
and  obviously satisfies \( H_- \Psi_{\rm G}= 0\),
it mimics the zero-energy state of \(H_-\), whose existence is the
indicator of unbroken supersymmetry.

The function \(\Psi_{\rm G}\) is not normalizable, and cannot,
therefore, be accepted as the physical zero-energy state.
Thus, the supersymmetry is dynamically broken at any non zero values of \(g\)
\cite{Witten}. 
The perturbation theory, however, cannot detect it, for the
unnormalizability does not 
appear at any finite order  in \(g\) in the expansion of \(\Psi_{\rm G}\).
(There can be no zero-energy state in \(H_+\) either, for the spectrum
of \(H_+\) is necessarily the same as that of \(H_-\).)

Since the perturbation cannot break the supersymmetry, the
degeneracy of  the excitations must persist in the perturbation.
Thanks to the simplicity of the quantum mechanics, we can check this directly.
The explicit contents of the supersymmetry relation (\ref{susy}) are the
two equations 
\begin{eqnarray}
D H_-  & = & H_+ D  \label{susycont1}, \\
D^\dagger H_+  & = & H_- D^\dagger. \label{susycont2}
\end{eqnarray} 
We will focus on Eq.(\ref{susycont1}) in the following discussion.
This holds that if \(\Psi^{(-)}\) is an eigenfunction
of \(H_-\) with the eigenvalue \(E^{(-)}\) 
then \(D \Psi^{(-)}\) is that of \(H_+\) with the same eigenvalue.
The operators \(D\) and \(D^\dagger\) are written as
\begin{equation}
D = (-i) e^{(g q^3 / 3)} a e^{-(g q^3/ 3)},
\quad 
D^\dagger = i e^{- (g q^3 / 3)} a^\dagger e^{(g q^3/3)} \label{eq:conj},
\end{equation}
where
\begin{equation}
a \equiv {d \over dq} + q
\end{equation}
is the annihilation operator in the harmonic oscillator 
up to an unimportant factor $\sqrt{2}$, \( [a, a^\dagger ] = 2\).
The operator \(D\) has \(a\) as the leading term in its expansion in \(g\);
\(D = a + O(g) \).
Thus if the state \(\Psi^{(-)}\) is for the \((N+1)\)-th excitation 
and starts with
\(H_{N+1}(q) e^{-q^2/2}\) in the perturbative expansion, \(D\Psi^{(-)}\) starts
with \(H_{N}(q) e^{-q^2/2}\) and represents the \(N\)-th excitation of \(H_+\).
This observation suggests the degeneracy in the perturbative energy levels.

To be more specific we can rewrite the two Hamiltonians as
\begin{eqnarray}
 H_+ &=& (1/2) e^{(g q^3 /3)}\, a (a^\dagger - 2 g q^2) \, e^{-(g q^3
   /3)},\label{perH+}\\
H_- &=& (1/2) e^{(g q^3 /3)} \, (a^\dagger - 2 g q^2) a \, e^{-(g q^3
  /3)}.\label{perH-}
\end{eqnarray} 
In conducting the perturbative analysis of  
\(E_{N+1}^{(-)}\) we need not necessarily
expand \(\Psi^{(-)}_{N+1}\) in \(g\) directly.
Since  the asymptotic expansion is unique, we may
adopt another convenient expansion scheme for the eigenfunction as long as
it is well-defined.
Looking at Eqs.(\ref{eq:conj}), (\ref{perH+}), and (\ref{perH-}) we
can expand
\(\Phi^{(-)} \equiv e^{- (g q^3 /3)} \Psi^{(-)} \) in \(g\),
\begin{equation}
\Phi_{N+1}^{(-)}(q) = H_{N+1}(q) e^{- q^2 / 2} + \sum_{m=1}^\infty g^m
\Phi_{N+1}^{(m)}(q).
\end{equation}
The eigenvalue equation for \(E_{N+1}^{(-)}\) now reads
\begin{equation}
{1\over 2} \left( a^\dagger - 2g q^2 \right) a \Phi_{N+1}^{(-)} 
= E_{N+1}^{(-)} \Phi_{N+1}^{(-)}. 
\label{eigphi}
\end{equation}
By multiplying \(a\) on both sides of Eq.(\ref{eigphi}), 
we obtain the following:
\begin{equation}
{1\over 2} a \left( a^\dagger - 2g q^2 \right) \left[ a
  \Phi_{N+1}^{(-)} \right] 
= E_{N+1}^{(-)} \left[ a \Phi_{N+1}^{(-)} \right].
\label{eigaphi}
\end{equation}
The function \( [ a \Phi_{N+1}^{(-)} ] \) 
starts with \(H_N(q) e^{ -(q^2/2)}\) when expanded in $g$.\linebreak
Eq.(\ref{eigaphi}) shows that \( \Psi^{(+)} \equiv e^{ (g q^3 /3)} [ a
\Phi_{N+1}^{(-)} ]\) has the same perturbative series as
\(E_{N+1}^{(-)}\) to the eigenvalue in \(H_+\). 
This proves that the \((N+1)\)-th excitation of \(H_-\) and  the
\(N\)-th of \(H_+\) have the same perturbative series.

We can now turn to the other values of \({\cal N}\).
The first 20 coefficients of  \(E^{(\pm)}_{\rm P}\)
for \({\cal N} = 2\) are listed in Table \ref{tb:2} and 
those for \({\cal N} = 3\) are listed in Table \ref{tb:3}.
The vanishing coefficients at \({\cal N} = 2\)
for the two isolated states draw our attention first.\footnote{Supersymmetry
provides an explanation for the vanishing coefficients 
for only the first excited state \cite{VWW}.}
The perturbative degeneracy remains,
but this time between \(E^{(-)}_{\rm P}({\cal N}, N+{\cal N}, m)\)
and \(E^{(+)}_{\rm P}({\cal N}, N, m)\).
In the next section, we will present the general proofs of these properties of
the perturbative coefficients.

\subsection{Non-renormalization theorem }
We will start with the degeneracy,
and define a pair of two Hamiltonians \(H_\pm({\cal N}) \)
with \({\cal N}\) as a parameter,
\begin{equation}
H_\pm({\cal N}) \equiv  -{ 1\over 2} {d^2 \over dq^2} + {1\over 2}q^2
( 1 - gq )^2 \mp {\cal N} \left( g q - {1\over 2}\right). \label{eq:H+-}
\end{equation}
The Hamiltonians \(H_{\pm}({\cal N})\) are related to the action
(\ref{eqn:action})-(\ref{eqn:pot}) again with \(\epsilon = {\cal N}\).
We denote the wave function and the eigenvalue 
of the $N$-th excited state in the left well
of $H_{\pm}({\cal N})$ by $\Psi_N^{(\pm)}$ and $E_N^{(\pm)}$, 
as in the case of the previous subsection.
The degeneracy between the \(N+{\cal N}\)-th excited level at the right
well and the \(N\)-th at the left in Table \ref{tb:2} and
\ref{tb:3} is realized between the eigenvalues \(E^{(\pm)}\) of the left
wells of \(H_\pm({\cal N})\).
The Hamiltonians are also rewritten as 
\begin{eqnarray}
H_+({\cal N}) &=& {1\over 2} D D^\dagger - ({\cal N} - 1) \left( gq -
  {1\over 2} \right), \label{He+}\\
H_-({\cal N}) &=& {1\over 2} D^\dagger D + ({\cal N} - 1) \left( gq -
  {1\over 2} \right). \label{He-}
\end{eqnarray}
As in the supersymmetric case, these two Hamiltonians transform to
each other by mirror transformation and
thus describe the same quantum mechanics.

The point of the  explanation concerning the degeneracy in \({\cal N} = 1\) 
was Eq.(\ref{susycont1}), and the fact that the operator \(D\) reduces
the level of excitation by \(1\).
What we have now is degeneracy  between \(E_{N+{\cal N}}^{(-)}\) and
\(E_{N}^{(+)}\). 
Correspondingly, we find \(H_\pm({\cal N})\) satisfies
\begin{eqnarray}
D^{\cal N} H_-({\cal N}) &=&  H_+( {\cal N} ) D^{\cal N}, \label{Nf1}\\
D^{\dagger\,{\cal N}} H_+({\cal N}) & = & H_-({\cal N}) D^{\dagger\,{\cal N}} 
\label{Nf2}. 
\end{eqnarray}
The proof is as follows. 
We first note that the equation
\begin{eqnarray}
H_+({\cal N}) D &=& D H_+( {\cal N} -2 ) -i  ({\cal N} -1) g,
\label{eq:rel1}
\end{eqnarray}
is proved from Eqs.(\ref{Ddag}) and (\ref{He+}).
We examine Eq.(\ref{Nf1}). The other will be proved in a similar way.
We start with \(H_+({\cal N}) D^{\cal N}\) and move each \(D\) to
the left using Eq.(\ref{eq:rel1}). 
After moving all \(D\)'s we obtain
\begin{equation}
H_+({\cal N}) D^{\cal N} = D^{\cal N} H_+( {\cal N} - 2 {\cal N} ) -i g 
D^{{\cal N}-1}
\sum_{k = 0}^{{\cal N} -1}({\cal N} - 1 - 2k).
\end{equation}
The terms in the summation over \(k\) on the right-hand side  cancel each other
 and \(H_+(-{\cal N} ) = H_-({\cal N}) \).
This proves Eq.(\ref{Nf1}).

Once we have obtained Eq.(\ref{Nf1}),
the perturbative degeneracy is proved in a similar manner to 
the ordinary supersymmetric case.
In terms of \(a \) and \( a^\dagger\),
\begin{equation}
H_\pm({\cal N}) = e^{(gq^3/3)} h_\pm e^{-(gq^3/3)},
\label{eq:hconj}
\end{equation}
where
\begin{eqnarray}
h_+ &=& (1/2)  a (a^\dagger - 2 g q^2) - 
({\cal N} -1)\left( gq -{1\over 2}\right),
\\
h_- &=& (1/2)  (a^\dagger - 2 g q^2) a + 
({\cal N} -1)\left( gq -{1\over 2}\right).
\end{eqnarray}
Substituting Eqs.(\ref{eq:conj}) and (\ref{eq:hconj}) into
Eq.(\ref{Nf1}) leads
\begin{equation}
a^{\cal N} h_- = h_+ a^{\cal N}.
\end{equation}
This establishes that the perturbative series of the \((N+{\cal N})\)-th level
 in \(h_-\) is the same  as that of the \(N\)-th level
in \(h_+\).
In turn it establishes the perturbative degeneracy
between \(E_{N+{\cal N}}^{(-)}\) and \(E_{N}^{(+)}\).
The wave function $\Psi^{(-)}_{N}$ of the $(N+{\cal N})$-th state of
$H^{(-)}({\cal N})$ is related to the wave function $\Psi^{(+)}_{N}$ of
the $N$-th excited state of $H^{(+)}({\cal N})$:
\begin{eqnarray}
D^{\cal N}\Psi^{(-)}_{N+{\cal N}}=\Psi^{(+)}_{N}.
\end{eqnarray}

On the contrary, the isolated states should satisfy
\begin{equation}
D^{\cal N} \Psi_N^{(-)} = 0 \quad (N=0,1,2, \cdots,{\cal N}-1)
\label{Nvac},
\end{equation}
which is the analogue of the ground state condition of the
supersymmetric case.
Since \(D^{\cal N}\) is rewritten as 
\begin{equation}
D^{\cal N} = e^{ -( q^2/2 ) + (g q^3/3) } 
\left( -i {d \over dq} \right)^{\cal N}
 e^{ ( q^2/2 ) - (g q^3/3) }, 
\end{equation}
the solutions for Eq.(\ref{Nvac}) have the form
\begin{equation}
\Psi_N^{(-)} = f(q) \exp\left(-{ q^2\over 2}  + g { q^3\over 3}\right), 
\label{eq:factrization}
\end{equation}
where \(f(q) \) is a polynomial of \(q\) of order \({\cal N} -1\) at most.
Again the unnormalizability of \(\Psi_{\cal N}^{(-)}\) disappears once
it is expanded in \(g\), as in the  case of \({\cal N} = 1\).
Plugging Eq.(\ref{eq:factrization}) into 
\begin{equation}
H_-({\cal N}) \Psi_N^{(-)} = E \Psi_N^{(-)}, \label{Nfeig}
\end{equation}
we obtain
\begin{equation}
f'' + 2 ( g q^2 - q ) f' + 
\left[ - 2 g ({\cal N} - 1) q + 2 E + {\cal N} - 1 \right] f = 0.
\label{eq:qtop}
\end{equation}
Let us write the polynomial \(f\) explicitly as
\begin{equation}
f(q) = \sum_{n=0}^{{\cal N} -1} a_n q^n.
\end{equation}
Only  polynomials of the order of \(({\cal N} -1)\) at most 
can be a consistent solution for (\ref{eq:qtop}):
Otherwise the terms \( 2 g [ q^2 f' - ({\cal N} -1) q f] \) in
(\ref{eq:qtop}) forces \(f(q)\) to have much higher powers.\footnote{We 
note that the equation (\ref{eq:qtop}) is rewritten
in the form 
\( [- (1/2) (J^-)^2 - g J^+ + J^0] f = E f \) where
\(J^- = d/dq\), \(J^0 = q (d/dq) - (1/2) ({\cal N} -1)\), 
and \(J^+ = q^2 (d/dq) - ({\cal N} -1 ) q\) are the SL(2,Q) generators for
the \({\cal N}\)-dimensional representation.
This is the eigenvalue problem equation of the quantum top, known in 
 connection with the quasi-exactly-solvable problem \cite{turbiner}.
This  also accounts for the fact that the solution \(f\) is obtained in
polynomials of the order \({\cal N} - 1\). }

It is convenient to write the resulting equation for \(a_n\) in a matrix form
\begin{equation}
{\bf M}_{{\cal N}}(E) \cdot {\bf a} = 0, \label{eq:ma}
\end{equation}
where \( {\bf a}^{\rm T} = ( a_0, a_1, ..., a_{{\cal N} -1} ) \).
The (\(i,j\))-element of \( {\bf M}_{{\cal N}}(E)\) is given by
\begin{equation}
{\bf M}_{{\cal N}}(E)_{ij} = 2g(i-{\cal N}) \delta_{i, j+1} + 
(2 E + {\cal N} - 1 - 2i) \delta_{i,j}
+ (i + 2)(i+1) \delta_{i, j-2},
\end{equation}
where \(i\) and \(j\) run from \(0\) to \({\cal N} -1\).
The matrix \( {\bf M}_{{\cal N}}(E)\)
has the explicit form 
\begin{equation}
\begin{array}{l}
{\bf M}_{{\cal N}}(E) \equiv \\
\\
\left(
\renewcommand{\arraystretch}{1.3}
\begin{array}{cc}
\begin{array}{cccc}
2 E + {\cal N} - 1 & 0 & 2 & 0\\
2 g (1- {\cal N} ) & 2E + {\cal N} - 3  & 0 & 6\\
0 & 2 g ( 2 - {\cal N}) & 2E + {\cal N} - 5 & 0 \\
0 & 0 & 2 g ( 3 - {\cal N}) & 2E + {\cal N} - 7
\end{array}
& \cdots\\
\vdots & \ddots 
\end{array}\right) 
\end{array}
\end{equation}
at the upper left corner and
\begin{equation}
\begin{array}{l}
{\bf M}_{{\cal N}}(E) \equiv \\
\\
\hspace*{-1pt}\left(
\renewcommand{\arraystretch}{1.3}
\begin{array}{cc}
\ddots & \vdots \\
\cdots & 
\begin{array}{cccc}
2E-{\cal N}+ 7 & 0 & ({\cal N}-2)({\cal N}-3) & 0\\
-6g& 2E-{\cal N}+5 & 0 & ({\cal N}-1)({\cal N}-2)\\
0 &-4g & 2E-{\cal N}+3 &0\\
0 &  0 & -2g & 2 E-{\cal N}+1 
\end{array}
\end{array}\right) 
\end{array}
\end{equation}
at the lower right.

The condition that there exists a non-trivial solution for \(\bf a\) is 
\begin{equation}
 \det {\bf M}_{\cal N}(E) =0. 
\label{eq:detm}
\end{equation}
Writing this condition explicitly for several lowest values of \({\cal N}\),
we obtain 
\begin{eqnarray}
{\cal N} =1: && E=0, 
\nonumber\\
{\cal N} =2: && \left( E-\frac12 \right) \left( E+\frac12 \right) =0, 
\nonumber\\
{\cal N} =3: && E \left( E-1 \right) \left( E+1 \right) +2 g^2=0, 
\nonumber\\
{\cal N} =4: && \left( E-\frac32\right) \left( E-\frac12\right) 
\left(E+\frac12\right)\left(E+\frac32\right) + 12 E g^2 =0, 
\nonumber\\
{\cal N} =5: && E(E-2)(E-1)(E+1)(E+2) + 6 (7 E^2 - 52) g^2=0. 
\nonumber\\
\label{eq:energyform}
\end{eqnarray}
These algebraic equations for \(E\) of the order \({\cal N}\) determine
the perturbative eigenvalues of the \({\cal N}\) isolated states.
The solutions are consistent with those we expect as the energy levels in
perturbation theory. 
They reproduce the energy levels of the first \({\cal N}\) lowest states
of the harmonic oscillator at \( g=0\), which are 
\(E = - ({\cal N}/2) + (1/2), -({\cal N}/2) + (3/2),
..., ({\cal N}/2) - (1/2)\).
The perturbative corrections are identically zero at any \(g^2 \) for 
\({\cal N} = 2\) as well as \({\cal N} = 1\).
We have also checked that the equation at other values of 
${\cal N}$ (\({\cal N}\ge
3\)) correctly regenerates the perturbative coefficients obtained
in the previous section once it is solved in powers of \(g^2\).

The very existence of the algebraic equation (\ref{eq:detm}) for
\(E\) tells us that the perturbative series for the isolated energy levels are
{\em convergent}, not simply Borel summable.

Finally, we will combine the relations (\ref{Nf1}) and (\ref{Nf2}) into 
a compact form,  which suggests the name ``\(\cal N \)-fold
supersymmetry'',
and will present an interesting relation
between the Hamiltonians $H_{\pm}({\cal N})$ 
and a new Hamiltonian that will be introduced naturally as a direct
extension of the supersymmetric Hamiltonian.

We will define the  \(\cal N\)-fold supercharge in \(2\times 2\) 
matrix notation by
\begin{equation}
Q_{ \cal N}^{\dagger}  = \left(
\renewcommand{\arraystretch}{1.3}
  \begin{array}{cc}
 0 &  D^{\cal  N} \\
  0 & 0 
 \end{array}
\right),
\quad \quad
Q_{ \cal N} =\left(
  \begin{array}{cc}
 0 & 0 \\
 ( D^\dagger)^{ \cal N} & 0 
 \end{array}
\right).
\end{equation}
These obviously satisfy
\begin{equation}
Q_{\cal N}^2 = Q_{\cal N}^\dagger {}^2 = 0,
\end{equation}
and induce transformations between the ``bosonic'' state $|{\cal B}\rangle$ and
the \linebreak ``fermionic'' state $|{\cal F}\rangle$:
\begin{eqnarray}
 |{\cal B}\rangle\equiv
\left(
\begin{array}{c}
0 \\
\Psi^{(-)}
\end{array}
\right),\quad
 |{\cal F}\rangle\equiv
\left(
\begin{array}{c}
\Psi^{(+)} \\
0
\end{array}
\right).
\end{eqnarray}
The relations (\ref{Nf1}) and (\ref{Nf2}) now become compact as
\begin{equation}
\left[ H_{\cal N}, Q_{\cal N }\right] 
= \left[ H_{\cal N}, Q_{\cal N}^\dagger \right] = 0 \label{HNQN}
\end{equation}
where \(H_{\cal N}\) is defined by
\begin{equation}
H_{\cal N}= \left(\begin{array}{cc}
 H_+({\cal N}) & 0 \\
 0 & H_-({\cal N}) 
\end{array}\right),
\end{equation}
in terms of the Hamiltonians 
\(H_\pm({\cal N})\) in Eq.(\ref{eq:H+-}).
The supersymmetric Hamiltonian is constructed from the supercharges $Q$
and $Q^{\dagger}$: 
\begin{equation}
{\bf H}_{ \cal N} = {1\over 2} 
\left\{ Q_{ \cal N}, Q_{\cal  N}^\dagger \right\} 
= {1\over 2 }
\left(
\renewcommand{\arraystretch}{1.3}
  \begin{array}{cc}
 D^{\cal N} D^{\dag {\cal N}} & 0 \\
 0 & D^{\dag {\cal N}} D^{\cal N} 
 \end{array}
\right),
\end{equation}
which we call the ${\cal N}$-fold supersymmetric Hamiltonian.
The Hamiltonian \({\bf H}_{\cal N}\) commutes with \( Q_{\cal N}\) and
\(Q_{\cal N}^\dagger\). 
We can define the states 
\begin{equation}
|{\cal B}_0\rangle \equiv \left( \begin{array}{c}
0\\\Psi_N^{(-)}
\end{array}\right), \quad N = 0, 1, ..., {\cal N} -1,
\end{equation}
using the perturbative isolated states $\Psi_N^{(-)}$
of \(H_-({\cal N})\).
These states satisfy
\begin{equation}
{\bf H}_{\cal N} |{\cal B}_0\rangle = 0,
\label{eq:hb0}
\end{equation}
and play the role of the \(\cal N\) vacua of
\({\bf H}_{\cal N}\). 
The symmetry generated by \(Q_{\cal N}\) and \(Q_{\cal N}^\dagger\) is
\(\cal N\)-fold of the ordinary supersymmetry in the sense of the number of
perturbative vacua and the change in the level of excitation.

By the definition of \({\bf H}_{\cal N}\) and Eq.(\ref{HNQN}), 
it is obvious that
\begin{equation}
\left[{\bf H}_{ \cal N}, H_{\cal N} \right] = 0.
\end{equation}
Therefore \({\bf H}_{\cal N}\) is a function of \(H_{\cal  N}\).
By counting the highest orders of $d/dq$ in both operators, we find that
${\bf H}_{\cal N}$ must be a polynomial of $H_{\cal N}$ of order ${\cal N}$.
Since the perturbative isolated states satisfy Eqs.(\ref{eq:detm}) and
(\ref{eq:hb0}), 
we conjecture that
\begin{eqnarray}
{\bf H}_{\cal N}=\frac12\det {\bf M}_{\cal N}(H_{\cal N}).
\label{conj}
\end{eqnarray} 
In the above, we have fixed the overall constant by comparing the
highest powers of $d/dq$ on both sides.
We have checked this conjecture explicitly by hand to the value
\( {\cal N} = 3\), and algebraically on a computer to \({\cal N} = 10\).

The non-renormalization theorem in the \(\cal N\)-fold supersymmetric quantum 
mechanics guarantees the absence of any perturbative correction to 
the energy of the ground states in \({\bf H}_{\cal N}\) and the
persistence of the degeneracy in perturbation theory. 

\section{Numerical verification of the prediction of the energy
spectrum}\label{num;sec}

In this section we will numerically verify the prediction 
with regard to the energy spectrum
made from the valley method. 
The result (\ref{eqn:en+}) obtained in Section 4
is for the case of no perturbative degeneracy, or
even for the isolated states in the presence of degeneracy.
The result (\ref{eqn:dege}) is for the perturbatively degenerate states.
In principle we can use these to check the validity 
of the valley method by a
comparison with exact numerical energy levels.
There is, however, a great difficulty for general values of
\(\epsilon\).
The valley result represents only the non-perturbative part and needs 
to be combined with the perturbative predictions.
This is important for numerical studies, 
since the perturbative part has in general a much larger numerical value than 
does the non-perturbative.
The problem is that we know only the leading Borel singularity of
the perturbation series.
After canceling it with the imaginary part of the 
non-perturbative contribution we obtained, we are still left with non-leading
Borel singularities in the perturbation theory.
This prohibits us from making numerical predictions for the energy
levels.

The model with  \(\epsilon = {\cal N} \) has
a great advantage in this respect.
For the isolated states we can obtain the perturbative
contribution exactly by solving Eq.(\ref{eq:detm}).
Remember that the equation generates all the perturbative coefficients
and, thus, it contains complete information on the perturbative
contribution.
Correspondingly, the valley prediction for these states derived 
from Eq.(\ref{eqn:en+}) is free from any imaginary element.
Regarding  the perturbatively degenerate states, their energy splittings
have no perturbative corrections because of 
the \(\cal N\)-fold supersymmetry.

In this section we will carry out the numerical calculation for
the energy spectra in the case of \( {\cal N} = 0, 1, 2, 3\),
and compare them with the valley predictions.
We label its energy levels by a single integer \(k\),  \(E_k\)
(\(k = 0, 1, 2, ...\)), 
counting from bottom.
That is, the energy of the ground state, the first excited state, and
so on,
are denoted by \(E_0\), \(E_1\), ..., respectively.
They thus include the isolated states localized at the lower right well.
We will first make the correspondence with the result in the previous sections 
and identify the quantities on which we will perform the comparison.

For  \(\cal N\) isolated states, \( k = 0, 1, ..., {\cal N} -1 \),
we use \(E^{(\rm P)}_k \) for denoting the perturbative contribution
to the energy, which is the solution of Eq.(\ref{eq:detm}).
In fact they are zero at \({\cal N} = 1\) and just \(- 1/2\) or \(1/2\) at
\({\cal N } = 2\).
In the case of \({\cal N} = 3\), the equation reads
\begin{equation}
E ( E -1 ) (E + 1) + 2 g^2 = 0.
\label{gforN3}
\end{equation}
This equation has three real roots up to \(g^2 = \sqrt 3/9\) and
each of them represents \(E^{(\rm P)}_k\).
We will first solve the equation numerically at various values of \(g\),
and then we will define the non-perturbative correction  by
\begin{equation}
\Delta E_k \equiv E_k - E^{(\rm P)}_k. \label{delEk}
\end{equation}
We can compare this with the valley prediction
\begin{equation}
\Delta E^{\rm (V)}_k = \alpha^2 {(-1)^{k + 1} \over k!} 
\Gamma({\cal N} - k) \left( -{2\over g^2} \right)^{-{\cal N} + 2 k},
\end{equation}
which is derived from Eq.(\ref{eqn:en+}) 
by the insertions  \(\epsilon = {\cal N} \) and 
\(N_- = k\). 

For the perturbatively degenerate energy levels, 
\( k = {\cal N}, {\cal N} + 1, ... \),
we will consider the energy splitting defined by the following equation:
\begin{equation}
\Delta E_{k\, k+1} \equiv E_{k+1} - E_k.
\end{equation}
The valley prediction in relation to this splitting is
\begin{equation}
\Delta E^{\rm (V)}_{ k\, k+1} = 2 \alpha \sqrt{
{1 \over [(k-{\cal N})/2]! [(k + {\cal N})/2]! }
\left({2 \over g^2} \right)^k 
}, \label{eq:7-3}
\end{equation}
as derived from Eq.(\ref{eqn:dege}) by the insertion of  
\(N_\pm = ( k \mp {\cal N}) /2 \).

\begin{figure}
\centerline{\epsfxsize=9cm\epsfbox{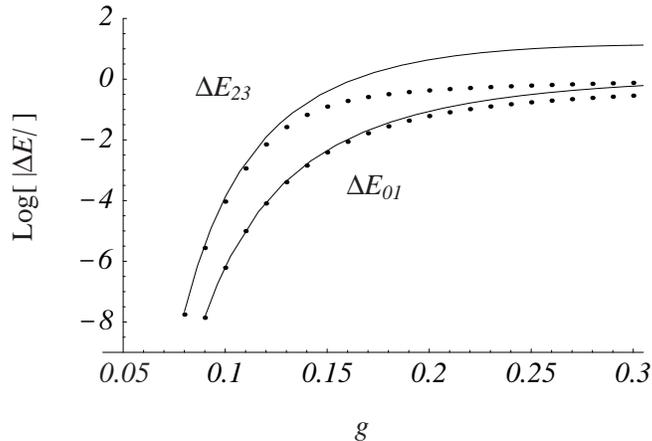}}
\caption{
The comparison in absolute values on the logarithmic scale between
the instanton prediction and the exact numerical result for
the energy splitting at ${\cal N} = 0$.
The solid lines stand for the instanton predictions, while
the dots  show the corresponding exact numerical results.}
\label{fg:7-1}
\end{figure}

\begin{figure}
\centerline{\epsfxsize=9cm\epsfbox{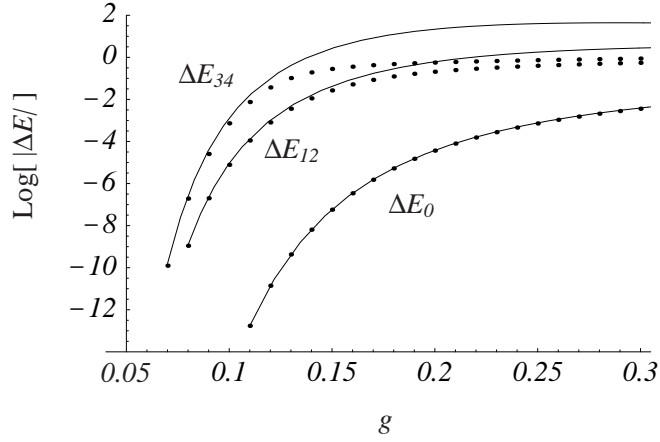}}
\caption{The comparison 
between the valley prediction and the exact numerical result for
the non-perturbative corrections to the isolated energy level and splitting at 
${\cal N} = 1$.
The solid lines stand for the valley prediction, while
the dots  show the corresponding exact numerical result.}
\label{fg:7-2}
\end{figure}

\begin{figure}
\centerline{\epsfxsize=9cm\epsfbox{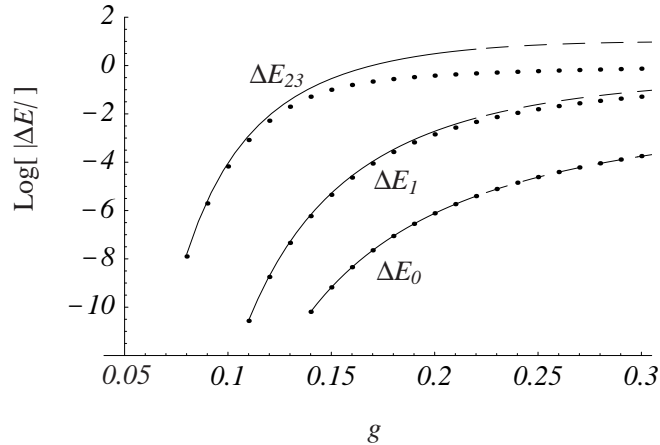}}
\caption{The comparison
between the valley prediction and the exact numerical result at
${\cal N} = 2$.
We use  dashed lines to represent the valley prediction 
at \(g\) larger than \(0.22\),  
as the valley prediction has no justification 
for those values, as explained in the text.}
\label{fg:7-3}
\end{figure}

\begin{figure}
\centerline{\epsfxsize=9cm\epsfbox{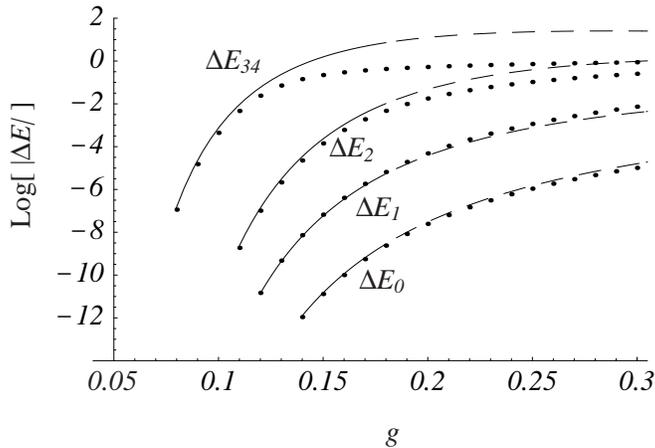}}
\caption{The comparison between the valley prediction and the exact
 numerical result at ${\cal N} = 3$.
We use dashed lines to represent the valley prediction 
at \(g\) larger than \(0.18\),  
as the valley prediction has no justification 
for those values, as explained in the text.
The critical value of  \(g\) for the equation  (\ref{gforN3}) to cease to
have real solutions is well beyond the scope of this figure.}
\label{fg:7-4}
\end{figure}

The result of the comparison is plotted in Figs.\ref{fg:7-1}-\ref{fg:7-4}.
The valley prediction monotonically
approaches to the exact values as \(g\) decreases for all levels.
This fact clearly shows that
the results (\ref{eqn:en+})
and (\ref{eqn:dege}) obtained on the basis of the valley method
are indeed the leading terms
of the non-perturbative correction to the energy spectrum.

There are various sources of corrections to the leading term that
could account for the discrepancy in the figures.
First is the fact that the potential (\ref{eqn:pot}) ceases to have
two minima at \( g \) larger than 
\(\sqrt{\sqrt 3/(18 {\cal N})}\), that are \(g > 0.31\) at \({\cal N} = 1\),
\(0.22\) at \(2\), and \(0.18\) at \(3\), respectively.
For those values of \(g\), the valley-instanton is no longer
a solution of the valley equation and 
the valley  prediction becomes unreliable.
Thus it is natural for the valley prediction to have a relatively large 
discrepancy at these values as above.
Even for the smaller \(g\), there must be higher-order corrections from
fluctuations in the valley background. 
The inclusion of those corrections will generate an $O(g^2)$ correction
relative to the leading term.

In Tables  \ref{tb:7-1}-\ref{tb:7-3}, we list the specific values of
$\Delta E_k$, $\Delta E_{k\, k+1}$, $\Delta E^{(\rm V)}_k$, 
$\Delta E^{(\rm V)}_{k\, k+1}$, and the
relative discrepancy \(\delta\) between them,
\begin{equation} \delta \equiv
 \left| { \Delta E^{(\rm V)} - \Delta E
\over \Delta E} \right|.
\end{equation}
Because of  the accumulation of round-off error in
the calculation of the energy,
it is difficult to carry out the numerical calculation for
smaller \(g\) than shown in the figures and tables.
The discrepancy is not necessarily 
small, even at the lowest \(g\).
Nevertheless the discrepancy decreases as \(g\) decreases
and the behavior of the decrease seems consistent with
the expected dependence on \(g^2\).
The legitimacy of our valley prediction as the leading term of
the non-perturbative contribution is solid.
\begin{table}
\begin{center}
\begin{tabular}{|c|c|c|c|} \hline 
g    &  $\Delta E_0$           &  $\Delta E_{0}^{(\rm V)} $     
&  $\delta $\\ \hline
0.12 &  1.389 $\times{10}^{-11}$ &1.408$\times{10}^{-11}$ &0.0137 \\ 
0.14 & 6.431$\times{10}^{-9}$ &6.544$\times{10}^{-9}$  & 0.0176 \\
0.16 &3.442$\times{10}^{-7}$  &3.523$\times{10}^{-7}$  & 0.0236  \\
0.18 &5.255$\times{10}^{-6}$  &5.417$\times{10}^{-6}$  & 0.0309    \\
0.20 &3.680$\times{10}^{-5}$  &3.826$\times{10}^{-5}$  & 0.0395    \\ 
0.22 &1.548$\times{10}^{-4}$  & 1.624$\times{10}^{-4}$ & 0.0495   \\
0.24 &4.604$\times{10}^{-4}$  & 4.881$\times{10}^{-4}$ & 0.0603 \\
0.26 &1.073$\times{10}^{-3}$  &  1.149$\times{10}^{-3}$  &  0.0710  \\
0.28 &  2.098$\times{10}^{-3}$  &2.266$\times{10}^{-3}$  &   0.0805\\
0.30 &  3.603$\times{10}^{-3}$  & 3.920$\times{10}^{-3}$ &  0.0879  \\
\hline
g    &  $\Delta E_{12}$           &  $\Delta E_{12}^{(\rm V)} $     
&  $\delta $\\ \hline
0.08 &1.103$\times{10}^{-9}$&1.222$\times{10}^{-9}$&0.108\\
0.10 &7.801$\times{10}^{-6}$&9.220$\times{10}^{-6}$&0.182\\
0.12 &8.066$\times{10}^{-4}$&1.042$\times{10}^{-3}$&0.292\\
0.14 &1.123$\times{10}^{-2}$&1.651$\times{10}^{-2}$&0.470\\
0.16 &5.175$\times{10}^{-2}$&9.275$\times{10}^{-2}$&0.792\\
0.18 &1.238$\times{10}^{-1}$&2.873$\times{10}^{-1}$&1.32\\
0.20 &2.070$\times{10}^{-1}$&6.185$\times{10}^{-1}$&1.99\\
0.22 &2.880$\times{10}^{-1}$&1.053 &2.66\\
0.24 &3.627$\times{10}^{-1}$&1.534&3.23\\
0.26 &4.308$\times{10}^{-1}$&2.006&3.66\\
0.28 &4.930$\times{10}^{-1}$&2.429&3.93\\
0.30 &5.504$\times{10}^{-1}$&2.783&4.06\\
\hline
g    &  $\Delta E_{34}$           &  $\Delta E_{34}^{(\rm V)} $     
&  $\delta $\\ \hline
0.07&1.246$\times{10}^{-10}$&1.589$\times{10}^{-10}$&0.275\\
0.09&2.550$\times{10}^{-5}$&3.985$\times{10}^{-5}$&0.563\\
0.11&7.517$\times{10}^{-3}$&1.607$\times{10}^{-2}$&1.14\\
0.13&1.037$\times{10}^{-1}$&4.118$\times{10}^{-1}$&2.97\\
0.15&2.814$\times{10}^{-1}$&2.705&8.61\\
0.17&4.238$\times{10}^{-1}$&8.455&18.9\\
0.19&5.241$\times{10}^{-1}$&17.12&31.7\\
0.21&6.027$\times{10}^{-1}$&26.50&43.0\\
0.23&6.709$\times{10}^{-1}$&34.54&50.5\\
0.25&7.334$\times{10}^{-1}$&40.14&53.7\\
\hline
\end{tabular}
\end{center}
\caption{The values of non-perturbative part of the energy spectrum
at ${\cal N} = 1$;
$\Delta E $ is from the exact numerical calculation,
$\Delta E^{\rm (V)} $ the valley prediction, and $\delta$ is
their relative discrepancy
\(\delta = |( \Delta E^{(V)} - \Delta E) / \Delta E| \).}
\label{tb:7-1}
\end{table}
\begin{table}
\begin{center}
\begin{tabular}{|c|c|c|c|} \hline
g    &  $\Delta E_0$           & $\Delta E^{(V)}_{0} $ &  $\delta $\\ \hline
0.14 &-6.438$\times{10}^{-11}$ &-6.413$\times{10}^{-11}$ &0.00387\\
0.16 &-4.530$\times{10}^{-9}$  &-4.510$\times{10}^{-9}$ &0.00445\\
0.18 &-8.825$\times{10}^{-8}$  &-8.776$\times{10}^{-8}$ &0.00569\\    
0.20 &-7.706$\times{10}^{-7}$  &-7.651$\times{10}^{-7}$ &0.00713\\            
0.22 &-3.967$\times{10}^{-6}$  &-3.932$\times{10}^{-6}$ &0.00877\\
0.24 &-1.421$\times{10}^{-5}$  &-1.406$\times{10}^{-5}$&0.0106\\
0.26&-3.933$\times{10}^{-5}$   &-3.884$\times{10}^{-5}$&0.0126\\
0.28&-9.018$\times{10}^{-5}$   &-8.884$\times{10}^{-5}$&0.0148\\
0.30&-1.795$\times{10}^{-4}$   &-1.764$\times{10}^{-4}$&0.0170\\
 \hline
g    &  $\Delta E_1$           & $\Delta E^{(V)}_{1} $ &  $\delta $\\ \hline
0.12 &1.788$\times{10}^{-9}$ &1.956$\times{10}^{-9}$ &0.0942\\
0.14 &5.883$\times{10}^{-7}$ &6.678$\times{10}^{-7}$ &0.135\\
0.16 &2.317$\times{10}^{-5}$ &2.753$\times{10}^{-5}$ &0.188\\
0.18 &2.659$\times{10}^{-4}$  &3.344$\times{10}^{-4}$ &0.258\\    
0.20 &1.423$\times{10}^{-3}$  &1.913$\times{10}^{-3}$ &0.344\\            
0.22 &4.658$\times{10}^{-3}$  &6.714$\times{10}^{-3}$ &0.441\\
 \hline
g    &  $\Delta E_{23}$  & $\Delta E^{(V)}_{23} $ &  $\delta $\\ \hline
0.08 &1.260$\times{10}^{-8}$&1.527$\times{10}^{-8}$&0.212\\
0.10 &6.689$\times{10}^{-5}$&9.220$\times{10}^{-5}$&0.378\\
0.12 &5.211$\times{10}^{-3}$&8.687$\times{10}^{-3}$ &0.667\\
0.14 &5.139$\times{10}^{-2}$ &1.179$\times{10}^{-1}$ &1.29\\
0.16 &1.576$\times{10}^{-1}$ &5.797$\times{10}^{-1}$ &2.68\\
0.18 &2.765$\times{10}^{-1}$ &1.596 &4.77\\    
0.20 &3.826$\times{10}^{-1}$  &3.093 &7.08\\            
 \hline
\end{tabular}
\end{center}
\caption{The values of non-perturbative part of the energy spectrum
at ${\cal N} = 2$.
Since the numerical agreement of \(\Delta E^{(\rm V)}_0\) with
the exact value is excellent, we think it permissible to write them up to
\(g = 0.3\), although the valley-instanton ceases to be a solution of
the valley equation at \(g \ge 0.2 \).}
\label{tb:7-2}
\end{table}
\begin{table}
\begin{center}
\begin{tabular}{|c|c|c|c|} \hline
g    &  $\Delta E_0$           & $\Delta E^{(V)}_{0} $ &  $\delta $\\ \hline
0.15 &1.305$\times{10}^{-11}$ &1.483$\times{10}^{-11}$ &0.137\\
0.16 &1.002$\times{10}^{-10}$ &1.155$\times{10}^{-10}$ &0.152\\
0.17 &5.553$\times{10}^{-10}$ &6.508$\times{10}^{-10}$ &0.172\\
0.18 &2.383$\times{10}^{-9} $ &2.843$\times{10}^{-9}$ &0.193\\ 
 \hline
g    &  $\Delta E_1$           & $\Delta E^{(V)}_{1} $ &  $\delta $\\ \hline
0.13 &-4.738$\times{10}^{-10}$ &-4.324$\times{10}^{-10}$ &0.0874\\
0.14 &-7.289$\times{10}^{-9}$  &-6.544$\times{10}^{-9}$ &0.102\\
0.15 &-6.639$\times{10}^{-8}$  &-5.859$\times{10}^{-8}$ &0.117\\
0.16 &-4.068$\times{10}^{-7}$  &-3.523$\times{10}^{-7}$ &0.134\\    
0.17 &-1.836$\times{10}^{-6}$  &-1.558$\times{10}^{-6}$ &0.151\\            
0.18 &-6.527$\times{10}^{-6}$  &-5.417$\times{10}^{-6}$ &0.170\\
 \hline
g    &  $\Delta E_{2}$  & $\Delta E^{(V)}_{2} $ &  $\delta $\\ \hline
0.11 &1.868$\times{10}^{-9}$&2.362$\times{10}^{-9}$&0.265\\
0.12 &1.019$\times{10}^{-7}$&1.358$\times{10}^{-7}$&0.333\\
0.13 &2.144$\times{10}^{-6}$&3.028$\times{10}^{-6}$&0.412\\
0.14 &2.257$\times{10}^{-5}$&3.407$\times{10}^{-5}$&0.510\\
0.15 &1.420$\times{10}^{-4}$&2.315$\times{10}^{-4}$&0.630\\    
0.16 &6.040$\times{10}^{-4}$&1.075$\times{10}^{-3}$&0.780\\  
0.17 &1.900$\times{10}^{-3}$&3.731$\times{10}^{-3}$&0.964\\
0.18 &4.732$\times{10}^{-3}$&1.032$\times{10}^{-2}$&1.18\\             
 \hline
g    &  $\Delta E_{34}$  & $\Delta E^{(V)}_{34} $ &  $\delta $\\ \hline
0.08 &1.140$\times{10}^{-7}$&1.559$\times{10}^{-7}$&0.367\\
0.09 &1.524$\times{10}^{-5}$&2.301$\times{10}^{-5}$&0.509\\
0.10 &4.415$\times{10}^{-4}$&7.528$\times{10}^{-4}$ &0.705\\
0.11 &4.657$\times{10}^{-3}$&9.275$\times{10}^{-3}$ &0.991\\
0.12 &2.402$\times{10}^{-2}$&5.910$\times{10}^{-2}$ &1.46\\
0.13 &7.170$\times{10}^{-2}$&2.378$\times{10}^{-1}$&2.32\\    
0.14 &0.143$\times{10}^{-1}$&6.878$\times{10}^{-1}$ &3.81\\            
 \hline
\end{tabular}
\end{center}
\caption{The values of non-perturbative part of the energy spectrum
at ${\cal N} = 3$.}
\label{tb:7-3}
\end{table}

The excellence of the agreement of
the valley prediction and the exact result
for the ground state energy in both
\({\cal N} = 1\) and \(2\) is very striking.
This may indicate a new sort of non-renormalization theorem, 
even in the non-perturbative sector.

Finally, we note that these non-zero values of \(\Delta E\) are 
the quantitative measure of the dynamical breaking of
the \(\cal N\)-fold supersymmetry. 

\section{Discussions} 

\begin{enumerate}
\item   First we would like to outline the results of this paper.
We have examined the geometrical structure of the configuration space of
the path integral, and shown that in one-dimensional quantum mechanics:

\begin{enumerate}
\item There are valleys in the configuration space which connect the
      vacuum and the instanton-like configuration. (Section \ref{valcon;sec}) 
\item The perturbative and non-perturbative contributions in the path integral
      can be separated in terms of the valley. (Section \ref{ibari;sec})
\item In the separation, the imaginary part of the
      non-perturbative contribution and the Borel 
      singularity of the perturbative
      contribution cancels trivially. (Section \ref{ibari;sec})
\item Bogomolny's trick is proved to be sound,
      thus enabling us to go beyond the dilute-gas approximation. (Section
      \ref{bog;sec} and Section \ref{multi;sec})
\item Based on the above separation, the large order behavior of the
      perturbation series is predicted. 
      The prediction reproduces the perturbation theory
      excellently. (Section \ref{dis;sec} and  Section \ref{lob;sec})   
\item A new type of supersymmetry in quantum mechanics is found.  
      Non-renormalization theorems of the ${\cal N}$-fold supersymmetry
      hold. (Section \ref{nsusy;sec})
\item The energy spectrum calculated by our valley calculus reproduces 
      those derived directly from the Schr\"{o}dinger equation.  
      (Section \ref{num;sec})
\end{enumerate}
 \item The asymmetric double-well potential was also treated by Balitsky and
        Yung \cite{BY}. 
        Their calculation, however, utilized a complex solution, which
        requires the deformation of the path-integral in
        the infinite dimensional configuration space. 
        This procedure is difficult to justify, and furthermore it
        obscures the analytic continuation in the complex $g^2$-plane.
        On the other hand, our analysis involves only the real
        configurations, and the analytic continuation is well-defined in
        the finite dimensional collective coordinate space.

\item 
       To parameterize the valley, we have used $R$, which corresponds to
        the relative distance between $I$ and $\bar{I}$.   
       But as was denoted in the footnote to Section \ref{ibari;sec},
       the definition of $R$ is not simple.  
       The simplest variable to parameterize the valley 
       is the eigenvalue $\lambda$ in the
       valley equation (\ref{eqn:valley}).
       In further research on valleys, one might use $\lambda$
       rather than $R$.
 \item 
       Our analyses have been done in the topologically trivial
       sector, but the extension to the topologically non-trivial sector is
       straightforward. 
       For example, we can construct a valley in the one instanton sector
       which consists of the $n+1$ instantons and the $n$ anti-instantons. 
       The valley equation is the same as in the topological trivial
       sector, and the only change is the boundary condition of $q$ at
       $\tau=\pm\infty$.  
\item 
        In Section \ref{dis;sec}, we have predicted the large order
        behavior of the perturbative series for the square of the absolute
        value  of the normalized wave function at $q=q_{\pm}$
        (see Eq.(\ref{eqn:psicorr})).  
        We have not tested this prediction, 
        but we note that the prediction is consistent with the
        result in Section \ref{nsusy;sec}.  
        In Section \ref{nsusy;sec}, we have shown that if
        $\epsilon={\cal N}$ the perturbative ground state (or more
        generally first ${\cal N}$ perturbative excited states) at $q=q_-$
        can be constructed exactly.   
        The state is normalizable at any finite order of the
        perturbation of $g$, but if we sum up the contributions at all
        orders, the state becomes unnormalizable. 
        This unnormalizability indicates the non-Borel-summability of
        the the perturbative series for the
        square of the absolute value  of the {\em normalized}
        wave function at $q=q_{-}$.    
 \item 
       In Section \ref{num;sec}, we have found in
       Figs.\ref{fg:7-2}-\ref{fg:7-4} that for the first
       $({\cal N}-1)$-th energy levels our valley calculus agrees 
       excellently with the numerical calculation.    
       In spite of the fact that we have only taken into
       account the lowest order corrections in the background of
       the valley, and neglected the higher order corrections,
       the agreement is better than expected.
       This may suggest another
       non-renormalization theorem for the higher order corrections on
       the valley calculus.
       Indeed, for the instanton calculus in the supersymmetric QCD, 
       similar types of the non-renormalization theorem are known to hold 
       \cite{NSVZ}. 
\item 
       The techniques of this paper may be applied to the model in
       quantum field theory. In quantum field theory, there are three
       types of singularities in the Borel plane, which are known as the
       IR renormalon, UV renormalon, and the instanton singularities
        \cite{review,thooft}.     
       The former two singularities are outside the scope of this paper, but
       the instanton singularity is manageable by a straightforward
       extension of our method.  
       A promising application of our method is to 
       processes with huge numbers of particles. 
       The number of Feynman diagrams of these processes is large
        and the instanton type singularity is closely related with this fact
       \cite{parisi,thooft}.
  \item One of the interesting processes with high
        multiplicities is the electroweak baryon and lepton number
        violating process at high energy scattering \cite{thooft2,RS}. 
        The process is the tunneling effect,
        and the height of the tunneling barrier is $O(10)$ TeV.
        The amplitude of the process at that energy is naively expected
        to be enhanced, and the process inevitably contains a
        huge number of particles (massive gauge and Higgs bosons)
        \cite{AG,R,E}. 
        A plausible calculation does not yet exist. 
        
        Based on the optical theorem, Porrati and later Khoze and
        Ringwald proposed a method to treat a huge number of particles
        \cite{porrati,KR}.   
        The problem with their method is that the optical theorem 
        itself does not
        distinguish the baryon number violating process (tunneling
        effect) and the baryon number preserving one (the perturbative
        effect).
        The extension of our method to the electroweak theory  
        might make it possible to separate them, although the detailed
        analysis of the geometrical structure of the theory is needed.
        The valley instanton in the electroweak theory has already been
        constructed in Ref.\cite{AHSW}.    
       
        More detailed arguments about this subject are given in
        Ref.\cite{msato}.

\item 
        The extension of our method to string theory is an interesting
                possibility.
        Unfortunately, it is difficult at present because of the lack of
        a satisfactory non-perturbative 
        formulation of string theory (e.g. string field theory).
        But the consideration
        of the large order behavior of string theory
        \cite{shenker} and the existence of the solitonic object
        \cite{polchinski} suggests that a geometrical structure similar
                to that of our model exists in the configuration space of 
                the non-perturbative
        formulation of the string theory.    
\end{enumerate}

\vspace*{12pt}
\centerline{\large\bf Acknowledgment}
\vspace*{12pt}
M.~Sato wishes to acknowledge suggestions by T.~Eguchi and thank
K.~Fujikawa, M.~Kato and H.~Kawamura for discussions.  
H.~Aoyama's work was supported in part by the Grant-in-Aid 
for Scientific Research 10120218 and 10640259.
H.~Kikuchi's work was supported in part by the Grant-in-Aid 
for Scientific Research 09226232.
M.~Sato and S.~Wada were supported in part by
Grants-in-Aid for JSPS fellows.
Numerical computation in this work was in part supported by
the computing facility at the Yukawa Institute for Theoretical
Physics, Kyoto University.
We would like to thank J.~Constable, Faculty of Integrated Human
Studies, Kyoto University, for careful reading of the manuscript.

\newpage
\appendix
\section{Some notes on the valley method}

In this appendix we will note some of the important properties 
of the valley method, which have not been explicitly 
described in previous references.

Firstly, we note its relation to the ordinary collective 
coordinate method.
In our context, this corresponds to the cases when there
are non-trivial solutions of equations of motion.
That is, cases where $F(\tau) \equiv 0$.
These cases can be treated by a limiting procedure
and the valley method reduces to the usual collective coordinate
method. This is done in the following manner:  
When $F(\tau)=0$, both the numerator and the denominator
of the Jacobian $\Delta[\varphi_\alpha]$ 
of Eq.(\protect{\ref{eqn:jacob}}) are zero.
This quantity can be obtained by introducing a small
perturbation to the action
that renders $F(\tau) \ne 0$ and making the perturbation vanish at the end.
Let us take $S[q] \rightarrow S[q] + \rho W[q]$.
Since the eigenvalue $\lambda$ is zero $\rho\ne 0$, 
$\lambda$ is of order $\rho$ and the
valley solutions are expanded as $q = q_0 + \rho q_1 + \cdots$,
$F = \rho F_1 + \cdots$.  The $\rho$-expansion of
Eq.(\ref{eqn:valleysf}) and Eq.(\ref{eqn:valleyf}) yields the following:
\begin{eqnarray}
\frac{\delta S[q_0]}{\delta q_0(\tau)} &=& 0,\\
\int d\tau'\frac{\delta^2 S[q_0]}{\delta q_0(\tau)\delta q_0(\tau')}
F_1(\tau')&=&0.
\end{eqnarray}
The first equation is the original equation of motion.
The second gives $F_1 \propto \dot q_0$, which proportional
constant is to be determined at the higher-order equations. 
The gradient vector $G(\tau)$ is independent from this proportional 
constant:
\begin{equation}
G(\tau) = \frac{F_1(\tau)}{\displaystyle \sqrt{\int F_1^2 d\tau'}}
= \frac{\dot q_0(\tau)}{\displaystyle \sqrt{\int \dot q_0^2 d\tau'}}, 
\end{equation}
By choosing the valley parameter $\alpha$ as the
translation coordinate of $q_0(\tau)$,
the Jacobian is found to be the following:
\begin{equation}
\Delta[q_\alpha] = 
\int \dot q_0(\tau) G(\tau)d\tau 
= \sqrt{\int \dot q_0^2 d\tau'}.
\end{equation}
This is the Jacobian of the ordinary collective coordinate
method \cite{Coleman}.

The second is the extension to the multi-dimensional valley,
which will be needed when there are multiple dangerous eigenvalues.
The definition of the $D$-dimensional valley is as follows:
\begin{eqnarray}
\frac{\delta S[q]}{\delta q(\tau)} &=& \sum_{i=1}^D F_i(\tau), 
\\
\int d\tau'\frac{\delta^2 S[q]}{\delta q(\tau)\delta q(\tau')}
F_i(\tau')&=&\lambda_i F_i(\tau) \quad (i=1, \cdots, D),
\end{eqnarray}
The Gaussian integration is performed in the subspace defined
by\linebreak $\int \phi(\tau) G_i(\tau) d\tau =0$ for $i=1, \cdots, D$
\cite{AK},
where \(G_j\) is the normalized function of \(F_i\).
The Jacobian of the collective coordinate volume element
\( \prod_i da_i \) is given by the determinant of 
\begin{equation}
C_{ij} \equiv 
 \int d\tau' {\partial q(\tau')\over \partial\alpha_i} G_j(\tau'),
\end{equation}
where the valley is parametrized by the 
\(D\) collective coordinates \(\alpha_i\). 

\section{The determinant of the valley-instanton} 
\def\fc{F}
\def\fd{\bar F}
\def\dotfc{\dot{F}}
\def\dotfd{\dot{\bar F}}

In this appendix we will calculate the determinant
for a valley-instanton, utilizing the fact that it has a zero eigenvalue,
as is proven in Section 2.2.
We will consider the valley-instanton located at $\tau=\tau_{I}$
in the region $\tau \in [-T/2,T/2]$, and 
calculate the normalized determinant, using the following formula
\cite{Coleman} for $\omega_\pm (T/2 \pm \tau_{I}) \gg 1$.
\begin{equation}
\frac{\det'(-\partial_\tau^2 + V^{\prime\prime})}
{\det(-\partial_\tau^2 + \omega_+^2)}
=
\frac{\psi (T/2)} {\lambda\psi_0(T/2)},
\end{equation}
where the eigenvalues are determined with the use of the 
Dirichlet boundary conditions at $\pm T/2$.
The eigenvalue $\lambda$ is the one that appears in the
valley equations and goes to zero in the limit
$T \rightarrow \infty$.
The functions $\psi(\tau), \psi_0(\tau)$ are the solutions of 
following differential equations:
\begin{eqnarray}
(-\partial_\tau^2 + V^{\prime\prime}(q))\psi(\tau)&=&0,\\
(-\partial_\tau^2 + \omega_+^2)\psi_0(\tau)&=&0,
\end{eqnarray}
with boundary conditions, $\psi(-T/2)=\psi_0(-T/2)=0$
and $\dot\psi(-T/2)=\dot\psi_0(-T/2)=1$.

The function $\psi$ is given by the following:
\begin{equation}
\psi_0(\tau) = \frac{1}{2\omega_+}\left(
e^{\omega_+ (\tau + T/2)} - e^{-\omega_+ (\tau + T/2)}\right),
\end{equation}
which yields $\psi_0(T/2) \sim e^{\omega_+ T}/(2\omega_+)$.

For the construction of the function $\psi(\tau)$,
we can utilize the auxiliary coordinate $F(\tau)$,
which is the eigenfunction with zero eigenvalue,
\begin{equation}
\left(-\partial_\tau^2 + V^{\prime\prime}(q)\right)F=0.
\label{eq:feq}
\end{equation}
with converging boundary conditions at $\tau=\pm\infty$.
It has the following asymptotic behavior:
\begin{equation}
\fc(\tau) \sim \fc_\mp e^{\mp\omega_\mp(\tau - \tau_{I})},
\end{equation}
for $\tau \rightarrow \pm T/2$ as was shown in Section 2.2.
We denote the other independent solution of Eq,(\ref{eq:feq}) by
$\fd(\tau)$, in which we choose to have the following Wronskian:
\begin{equation}
W=\fc(\tau) \dotfd(\tau) - \dotfc(\tau) \fd(\tau)
 = 2\omega_+\omega_-.
\end{equation}
This implies that the asymptotic behavior of $\fd(\tau)$ 
for $\tau \rightarrow \pm T/2$ is the following:
\begin{equation}
\fd(\tau)\sim \fd_\mp e^{\pm\omega_\mp(\tau- \tau_{I})},
\end{equation}
with $\fd_\pm = \mp {\omega_\mp}/{\fc_\pm}$.
Using these solutions, the function $\psi(\tau)$ is 
given as follows,
\begin{equation}
\psi(\tau) = \frac{1}{2\omega_+} \left(
\frac{\fc(\tau)}{\fc(-T/2)}-\frac{\fd(\tau)}{\fd(-T/2)}
\right).
\label{eq:plainpsi}
\end{equation}
which satisfies the required boundary conditions at $\tau=-T/2$.

Next we will evaluate the lowest eigenvalue $\lambda$,
\begin{equation}
\left(-\partial_\tau^2 +
  V^{\prime\prime}(q)\right)\bar\psi(\tau)=\lambda\bar\psi(\tau),
\end{equation}
with $\bar\psi(\pm T/2)=0$.
We can construct the eigenfunction $\bar\psi(\tau)$ 
as a perturbation series in $\lambda$ as follows,
\begin{eqnarray}
\bar\psi(\tau) &=&
\psi(\tau) + \lambda \int_{-T/2}^{+T/2} d\tau' G(\tau, \tau') \bar\psi(\tau') 
\nonumber\\
&=& \psi(\tau) + \lambda \int_{-T/2}^{+T/2} d\tau'
G(\tau, \tau') \psi(\tau') + O(\lambda^2),
\label{eq:bpcons}
\end{eqnarray}
where the function $\psi(\tau)$ given by Eq.(\ref{eq:plainpsi})
is used so that the boundary condition $\bar\psi(-T/2)=0$ is 
satisfied.
The Green function $G(\tau, \tau')$ is defined by,
\begin{equation}
G(\tau, \tau') = \frac{1}{W} (\fc(\tau) \fd(\tau') -
\fc(\tau') \fd(\tau)) \theta(\tau - \tau').
\label{eq:gdef}
\end{equation}
and satisfies the following,
\begin{equation}
\left(-\partial_\tau^2 +  V^{\prime\prime}(q)\right)
G(\tau, \tau') = \delta(\tau-\tau').
\end{equation}
The boundary condition $\bar\psi(T/2)=0$ determines the eigenvalue $\lambda$.
Straightforward calculation shows that
the product of the second term of the Green function 
(\ref{eq:gdef}) and the $F(\tau')$ term in $\psi(\tau')$ makes
the dominant contribution to $\bar\psi(T/2)$.
Therefore we find that
\begin{eqnarray}
\psi\left(\frac{T}{2}\right)
&\simeq& \frac{\lambda}{W}
\frac{\fd(T/2)}{2\omega_+\fc(-T/2)}\int_{-T/2}^{T/2}d\tau' F^2(\tau').
\end{eqnarray}
Combining the results obtained so far, we find that
\begin{equation}
\frac{\det'(-\partial_\tau^2 + V^{\prime\prime})}
{\det(-\partial_\tau^2 + \omega_+^2)}
\simeq
\frac{e^{(\omega_- - \omega_+)(T/2 - \tau_{I})}}
{2\omega_-\fc_+\fc_-}
\int_{-\infty}^{\infty}d\tau F^2(\tau).
\label{eq:appa}
\end{equation}

\section{The relation between the valley parameter and the separation
of valley-instantons}

As has been shown by the explicit numerical integration of the valley
equations carried out in Section 2, 
the solution approaches to the well-separated 
pair of the (anti-)valley-instantons
as the parameter \(\lambda\) in the equation approaches zero from below.
In this appendix we will clarify the relation between \(\lambda\) 
and the separation \(R\).
For the sake of simplicity we will confine ourselves to the case 
of the $I \bar I$ valley.
Obtaining a similar result for the $\bar I I$ valley is a
simple matter.

Assuming that \(\lambda\) is small enough that the solution consists of
a well-separated pair of valley-instanton and anti-valley-instanton,
we can locate the valley-instanton at \(\tau = 0\) and
the anti-valley-instanton at \(\tau = R\).
At the symmetric point \(\tau_0 (= R/2) \), the time derivatives of 
the configuration
vanish simultaneously; \(\dot F(\tau_0) = \dot q(\tau_0) = 0 \). 
In the  asymptotic region to the far left of the valley-instanton,   
\(\tau<0 \) and $ |\tau| \gg 1/\omega_+$,
the solution is well described by  linearizing the
valley equations (\ref{eqn:valleysf2})-(\ref{eqn:valleyf2}).
By requiring that it becomes a vacuum at \(q_+\) as 
\(\tau \rightarrow -\infty \), the solution is described by two parameters 
\(F_+\) and \(Q_-\),
\begin{eqnarray}
F & \simeq & F_+ e^{\kappa_+ \tau} \\
q - q_+ &\simeq& \left(Q_+ - {F_+ \over \lambda}\right) e^{\omega_+ \tau} + 
{F_+ \over \lambda} e^{\kappa_+ \tau},
\end{eqnarray}
where \(\kappa_\pm \equiv \sqrt{ \omega_\pm -\lambda}\).
Of two degrees of freedom in \(F_+\) and \(Q_+ \),
one is related to the overall translation and the
other should be fixed so that \(\dot q\) and \(\dot F\) 
are zero  at \( \tau_0 \).
We already know the asymptotic form of the valley-instanton as 
Eqs.~(\ref{eqn:approxq})-(\ref{eqn:approxF}). 
Thus
\begin{equation}
Q_+ \rightarrow {1\over g},  \qquad  F_+ \rightarrow - 6 \epsilon g
\end{equation}
at the limit of \(\lambda \rightarrow 0\).

Similarly  the solution is   well described by 
linearized equations around  \(\tau_0\), for the configuration is
almost a vacuum at \(q_-\).
Since \(\dot F(\tau_0) =\dot q(\tau_0) = 0 \),
we can conclude the \(I \bar I\) valley behaves as
\begin{eqnarray}
F &\simeq& F_- e^{ - \kappa_- \tau} + F_- e^{\kappa_- (\tau - 2 \tau_0)} 
\label{eq:F-}\\
q - q_- &\simeq& \left(Q_- - {F_- \over \lambda}\right) e^{-\omega_- \tau} + 
{F_- \over \lambda} e^{-\kappa_- \tau} 
\nonumber\\
&&+\left(Q_- - {F_- \over \lambda} \right) e^{\omega_- (\tau- 2\tau_0) } + 
{F_- \over \lambda} e^{\kappa_- (\tau - 2 \tau_0)}.
\label{eq:q-}
\end{eqnarray}
When \(\lambda \rightarrow -0 \), \(\tau_0\) goes to the positive infinity
and 
\begin{equation}
Q_- \rightarrow -{1\over g} \qquad F_- \rightarrow - 6 \epsilon g, 
\label{eq:QF-}
\end{equation}
obtained again by comparing (\ref{eq:F-}) and (\ref{eq:q-}) with the
asymptotic forms for positive \(\tau\) in (\ref{eqn:approxq}) and 
(\ref{eqn:approxF}).

We use the conservation of \(H_V\) (\ref{eqn:hami}) 
to relate \(\lambda\) to \(\tau_0\).
At \(\tau = -\infty \) \(H_V\) is equal to \(- V(q_+) \),
while at \(\tau_0\)
\begin{equation}
H_V \simeq  - V(q_-) - {1\over 2} \omega_-^2 \left[ q(\tau_0) - q_- \right]^2
 + {1\over \lambda} \omega_-^2 F(\tau_0) \left[ q(\tau_0) - q_- \right]
- {1\over 2\lambda} F(\tau_0)^2.
\end{equation} 
Equating the two expressions for \(H_V\) and using (\ref{eq:F-})
and (\ref{eq:q-}) to relate \(F(\tau_0)\) and
\(q(\tau_0)\) to \(F_-\), \(Q_-\), and \(\tau_0\), we obtain
\begin{equation}
\lambda = - {2 \over V(q_+) - V(q_-)} \left[2 \omega_-^2 F_- Q_-
+ F_-^2 (\omega_- \tau_0 - 1) \right] e^{-2\omega_- \tau_0}.
\end{equation}
Although \(F_-\) and \(Q_-\) may have  an O(\(\lambda\))  correction,
they are exponentially small compared to the values in (\ref{eq:QF-}) and
are negligible.
Thus we obtain
\begin{equation}
\lambda = - \left[ 24 \omega_-^2 + 36 \epsilon g^2 (\omega_- R - 2) \right]
e^{-\omega_- R}.
\end{equation}
The valley parameter \(\lambda\) decreases simply as \( e^{-\omega_- R}\) 
at \(\epsilon = 0\), while
it has an exotic  dependence \(\omega_- R e^{- \omega_- R}\)
at non-zero \(\epsilon\).

\section{WKB evaluation}

In this appendix, we will re-derive the result (\ref{eqn:phis0}) 
by the WKB approximation.
The system we will consider obeys the Schr\"odinger equation  
\begin{eqnarray}
\left[-\frac{1}{2}\frac{d^2}{dq^2}+V(q)\right]\Phi(q)=E\Phi(q).
\end{eqnarray}
Around the upper minimum $q_+$ of the potential, this becomes
\begin{eqnarray}
\left[-\frac{1}{2}\frac{d^2}{dq^2}+\frac{1}{2}q^2\right]\Phi(q)=E\Phi(q),
\end{eqnarray}
and a solution which vanishes at $q\rightarrow -\infty$ is
\begin{eqnarray}
\Phi(q)=AD_{\nu}(-\sqrt{2}q),
\label{wvf-l:eqn}
\end{eqnarray}
where $E=\nu+\frac{1}{2}$ and $A$ is a constant.
The function $D_\nu$ is the parabolic cylinder function \cite{grads}.
And around the lower minimum $q_-$ of the potential, the Schr\"odinger equation
becomes
\begin{eqnarray}
\left[-\frac{1}{2}\frac{d^2}{dq^2}+\frac{1}{2}\left(q-\frac{1}{g}\right)^2
-\epsilon\right]\Phi(q)=E\Phi(q),  
\end{eqnarray}
and a solution which vanishes at $q\rightarrow\infty$ is
\begin{eqnarray}
\Phi(q)=\tilde{A}D_{\nu+\epsilon}(\sqrt{2}(q-1/g)),
\label{wvf-r:eqn}
\end{eqnarray}
where $\tilde{A}$ is a constant.
We must now connect these solutions with that in the forbidden region.

In the forbidden region, the usual semi-classical expression
for the wave function is available:
\begin{eqnarray}
\Phi(q)=\frac{C_1}{k(q)^{1/2}}\exp\left[-\int^{q}_{q_1}k(x)dx\right]
+\frac{C_2}{k(q)^{1/2}}\exp\left[\int^{q}_{q_1}k(x)dx\right],
\label{wvf-f:eqn}
\end{eqnarray}
where
\begin{eqnarray}
k(q)=\sqrt{q^2(1-gq)^2-2\epsilon gq-2E}.
\end{eqnarray}
and $q_i$ ($i=1,2$) are the tunneling points $V(q_i)=E$.
$C_i$ ($i=1,2$) are constant.
When $\epsilon g^2\ll 1$, these are given by
\begin{eqnarray}
gq_1&=&\epsilon g^2+\sqrt{2Eg^2+(\epsilon g^2)^2},\\
gq_2&=&1+\epsilon g^2-\sqrt{2(E+\epsilon)g^2+(\epsilon g^2)^2}.
\end{eqnarray}
The integral in the wave function can be evaluated by expanding about $g^2$:
\begin{eqnarray}
\int_{q_1}^{q}k(x)dx&=&\frac{1}{g^2}
\int_{gq_1}^{gq}[w^2(1-w)^2-2\epsilon g^2w-2g^2E]^{1/2}
dw\nonumber\\
&=&\frac{1}{g^2}
\int_{gq_1}^{gq}w(1-w)dw-\int_{z_1}^{z}\frac{\epsilon w+E}{w(1-w)}dw+\cdots\\
&=&\frac{1}{g^2}
\left[\frac{1}{2}w^2-\frac{1}{3}w^3-g^2E\ln\left(\frac{w}{1-w}\right)
+g^2\epsilon\ln(1-w)\right]_{gq_1}^{gq}+\cdots\nonumber
\end{eqnarray}
In the region near the upper minimum, it becomes
\begin{eqnarray}
\int_{q_1}^{q}k(x)dx
&=&\left[\frac{1}{2}q^2-\frac{1}{3}gq^3-
E\ln\left(\frac{gq}{1-gq}\right)+\epsilon\ln(1-gq)\right]\nonumber\\
&&-\left[\frac{1}{2}\left(\epsilon g+\sqrt{2E+\epsilon^2g^3}\right)^2-
\frac{1}{3}g\left(\epsilon g+\sqrt{2E+\epsilon^2g^3}\right)^3\right.\nonumber\\
&&-\left.E\ln\left(\frac{\epsilon g^2+\sqrt{2Eg^2+(\epsilon g^2)^2}}
{1-\left(\epsilon g^2+\sqrt{2Eg^2+(\epsilon g^2)^2}\right)}\right)
\right.\nonumber\\
&&\left.
+\epsilon\ln\left(1-\left(\epsilon g^2+\sqrt{2Eg^2+(\epsilon g^2)^2}
\right)\right)\right]\nonumber\\
&=&\left[\frac{1}{2}x^2-E\ln(gq)-E-E\ln(\sqrt{2E}g)\right]+\cdots.
\end{eqnarray}
In the final expression, 
we have neglected the terms which vanish as $g\rightarrow 0$,
so the wave function becomes
\begin{eqnarray}
\Phi(q)&=&C_1q^{-1/2}\exp\left(-\frac{1}{2}q^2+E\ln\left(\frac{q}{\sqrt{2E}}
\right)+E\right)\nonumber\\
&+&C_2q^{-1/2}\exp\left(\frac{1}{2}q^2-E\ln\left(\frac{q}{\sqrt{2E}}
\right)-E\right)+\cdots\nonumber\\
&=&C_1e^{-q^2/2+E}(2E)^{-E/2}q^{\nu}+C_2e^{q^2/2-E}(2E)^{E/2}q^{-\nu}+\cdots.
\nonumber\\
\end{eqnarray}
We are here using
$k(q)=q+\cdots$.
This wave function must be connected with the one given by
Eq.(\ref{wvf-l:eqn}).
As the wave function (\ref{wvf-l:eqn}) 
has the following asymptotic form for $q\gg 1$: 
\begin{eqnarray}
\Phi(q)=A\left[e^{-q^2/2}(-\sqrt{2}q)^{\nu}
-\frac{\sqrt{2\pi}}{\Gamma(-\nu)}e^{\pm\nu\pi i}e^{-q^2/2}
(-\sqrt{2}q)^{-\nu-1}+\cdots\right],
\end{eqnarray}
so we obtain the following matching condition 
\begin{eqnarray}
\frac{\sqrt{2\pi}}{\Gamma(-\nu)}e^{\pm\nu\pi i}2^{-\nu-1/2}
=\frac{C_2}{C_1}e^{-(2\nu+1)}(2\nu+1)^{-(\nu+1/2)}.
\end{eqnarray}

In a similar way, by connecting the wave function (\ref{wvf-f:eqn})
with the one given by (\ref{wvf-r:eqn}) near the lower minimum of the
potential, we obtain another matching condition: 
\begin{eqnarray}
\frac{\sqrt{2\pi}}{\Gamma(-\nu-\epsilon)}e^{\pm(\nu+\epsilon)\pi i}
2^{-(\nu+1/2+\epsilon)}=\frac{C_1}{C_2}e^{-1/3g^2}(2\nu+1)^{-(\nu+1/2)}
g^{-(4\nu+2+2\epsilon)}.\nonumber\\
\end{eqnarray}
Finally, if we eliminate the undetermined ration of the coefficients
$C_1/C_2$ in the matching conditions, the master equation
(\ref{eqn:phis0})
for the non-perturbative corrections of the energies is reproduced: 
\begin{eqnarray}
\alpha^2\Gamma\left(-E+\frac{1}{2}\right)
\Gamma\left(-E+\frac{1}{2}-\epsilon\right)
\left(-\frac{2}{g^2}\right)^{2E+\epsilon}=1,
\end{eqnarray}
where
\begin{eqnarray}
\alpha=\frac{e^{-1/6g^2}}{g\pi^{1/2}}.
\end{eqnarray}

\newcommand{\J}[4]{{\sl #1} {\bf #2} (19#3) #4}
\newcommand{\MPL}{Mod.~Phys.~Lett.}
\newcommand{\NP}{Nucl.~Phys.}
\newcommand{\PL}{Phys.~Lett.}
\newcommand{\PR}{Phys.~Rev.}
\newcommand{\PRL}{Phys.~Rev.~Lett.}
\newcommand{\AP}{Ann.~Phys.}
\newcommand{\CMP}{Commun.~Math.~Phys.}
\newcommand{\CQG}{Class.~Quant.~Grav.}
\newcommand{\PRP}{Phys.~Rept.}
\newcommand{\SPU}{Sov.~Phys.~Usp.}
\newcommand{\RMPA}{Rev.~Math.~Pur.~et~Appl.}
\newcommand{\SPJ}{Sov.~Phys.~JETP}
\newcommand{\MP}{Int.~Mod.~Phys.}
\newcommand{\JMP}{J.~Math.~Phys.}

\end{document}